\documentclass[12pt,preprint]{aastex}

\newcommand{\av}{$A_V$}

\newcommand{\eg}{{\it e.g.}}
\newcommand{\etal}{et~al.}

\newcommand{\jhk}{$JHK_{\rm s}$}

\newcommand{\ks}{$K_{\rm s}$}

\newcommand{\lsun}{$L_{\sun}$}

\newcommand{\mum}{$\mu$m}

\begin{document}

\title{The Spitzer c2d Survey of Large, Nearby, Interstellar
Clouds: \\
VI. Perseus Observed with MIPS}

\author{L.\ M.\ Rebull\altaffilmark{1}, 
K.\ R.\ Stapelfeldt\altaffilmark{2}, 
N.\ J.\ Evans II\altaffilmark{7},
J.\ K.\ J{\o}rgensen\altaffilmark{8},
P.\ M.\ Harvey\altaffilmark{7},
T.\ Y.\ Brooke\altaffilmark{6},
T.\ L.\ Bourke\altaffilmark{8},
D.\ L.\ Padgett\altaffilmark{1},
N.\ L.\ Chapman\altaffilmark{3},
S.-P.\ Lai\altaffilmark{3,4,5},
W.\ J.\ Spiesman\altaffilmark{7},
A.\ Noriega-Crespo\altaffilmark{1},
B.\ Mer\'in\altaffilmark{12},
T.\ Huard\altaffilmark{8},
L.\ E.\ Allen\altaffilmark{8},
G.\ A.\ Blake\altaffilmark{9},
T.\ Jarrett\altaffilmark{10},
D.\ W.\ Koerner\altaffilmark{11},
L.\ G.\ Mundy\altaffilmark{3},
P.\ C.\ Myers\altaffilmark{8},
A.\ I.\ Sargent\altaffilmark{6},
E.\ F.\ van Dishoeck\altaffilmark{12},
Z.\ Wahhaj\altaffilmark{11},
K.\ E.\ Young\altaffilmark{7,13}}

\altaffiltext{1}{Spitzer Science Center/Caltech, M/S 220-6, 1200
E.\ California Blvd., Pasadena, CA  91125
(luisa.rebull@jpl.nasa.gov)}
\altaffiltext{2}{Jet Propulsion Laboratory, MS 183-900, California Institute
of Technology, Pasadena, CA 91109}
\altaffiltext{3}{Department of Astronomy, University of Maryland, College Park,
MD 20742}
\altaffiltext{4}{Institute of Astronomy and Department of Physics,
National Tsing Hua University, Hsinchu 30043, Taiwan}
\altaffiltext{5}{Academia Sinica Institute of Astronomy and Astrophysics, P.O.
Box 23-141, Taipei 106, Taiwan}
\altaffiltext{6}{Division of Physics, Mathematics, and Astronomy, MS 105-24,
California Institute of Technology, Pasadena, CA 91125}
\altaffiltext{7}{Department of Astronomy, University of Texas at Austin,
1 University Station C1400, Austin, TX 78712}
\altaffiltext{8}{Harvard-Smithsonian Center for Astrophysics, 60 Garden Street,
MS42, Cambridge, MA 02138}
\altaffiltext{9}{Division of Geological and Planetary Sciences, MS 150-21, 
California Institute of Technology, Pasadena, CA 91125}
\altaffiltext{10}{Infrared Processing and Analysis Center,
California Institute of Technology, Pasadena, CA 91125}
\altaffiltext{11}{Department of Physics and Astronomy, Northern Arizona
University, NAU Box 6010, Flagstaff, AZ 86011-6010}
\altaffiltext{12}{Leiden Observatory, PO Box 9513, NL 2300 RA Leiden, The
Netherlands}
\altaffiltext{13}{Department of Physical Sciences, Nicholls State University,
Thibodaux, Louisiana 70301}

\begin{abstract}

We present observations of 10.6 square degrees of the Perseus
molecular cloud at 24, 70, and 160 \mum\ with the {\it Spitzer
Space Telescope} Multiband Imaging Photometer for Spitzer
(MIPS).  The image mosaics show prominent, complex extended
emission dominated by illuminating B stars on the East side of
the cloud, and by cold filaments of 160 \mum\ emission on the
West side.  

Of 3950 point sources identified at 24 \mum, 1141 have 2MASS
counterparts.  A quarter of these populate regions of
the \ks\ vs.\ \ks$-$[24] diagram that are distinct from stellar
photospheres and background galaxies, and thus are likely to be cloud
members with infrared excess.  Nearly half (46\%) of these 24
\mum\ excess sources are distributed outside the IC 348 and NGC
1333 clusters.  NGC 1333 shows the highest fraction of stars
with flat or rising spectral energy distributions (28\%), while
Class II SEDs are most common in IC 348.  These results are
consistent with previous relative age determinations for the two
clusters.  A significant number of IRAS PSC objects are not
recovered by {\it Spitzer}/MIPS, most often because the IRAS
objects were confused by bright nebulosity.  There is no
evidence for 24 \mum\ source variability to 10\% between the
$\sim$3-6 hours of our two observation epochs.  

The intercluster region contains several tightly clumped 
($r\sim$0.1 pc) young stellar aggregates whose members exhibit a 
wide variety of infrared spectral energy distributions characteristic 
of different circumstellar environments.  One possible explanation is 
a significant age spread among the aggregate members, such that some
have had time to evolve more than others.  Alternatively, if the aggregate 
members all formed at roughly the same time , then remarkably rapid 
circumstellar evolution would be required to account for the association 
of Class I and Class III sources at ages $\lesssim$ 1 Myr.

We highlight important results for the HH 211 flow, where the
bowshocks are detected at both 24 and 70 \mum; and for the
debris disk candidate BD +31$\arcdeg$643, where the MIPS data
shows the linear nebulosity to be an unrelated interstellar
feature.  Our data, mosaics, and catalogs are available at the
Spitzer Science Archive for use by interested members of the
community.

\end{abstract}


\section{Introduction}
\label{sec:intro}

The Spitzer Space Telescope Legacy program ``From Molecular
Cores to Planet-Forming Disks'' (c2d; Evans \etal\ 2003)
selected five large star-forming clouds for mapping with the
Infrared Array Camera (IRAC, 3.6, 4.5, 5.8, and 8 \mum; Fazio
\etal\ 2004) and the Multiband Imaging Photometer for Spitzer
(MIPS, 24, 70, and 160 \mum; Rieke \etal\ 2004).  These clouds
were selected to be within 350 pc, to have a substantial mass of
molecular gas, and to have a range of cloud properties, thereby
allowing studies of star formation in isolation, in groups, and
in clusters.  The goals of this aspect of the c2d project (see
Evans \etal\ 2003 for more information) include determining the
stellar content of the clouds, the distributions of the youngest
stars and substellar objects, and the properties of their disks
and envelopes.  All of these Spitzer cloud studies represent the
first unbiased mid-infrared surveys across entire clouds at this
sensitivity and spatial resolution, where in the past only
targeted, small-field-of-view observations have been possible. 

Perseus is one of five nearby star forming clouds mapped with IRAC
and MIPS by c2d, also including Chamaeleon II, Lupus I, III \&
IV, Ophiuchus and Serpens (see Evans \etal\ 2003 for an
overview).  Previous papers in this series presented IRAC
observations of Serpens (Harvey \etal\ 2006), Perseus
(J{\o}rgensen \etal\ 2006, hereafter J06) and Chamaeleon II
(Porras \etal\ 2007) and MIPS observations of Chamaeleon II
(Young \etal\ 2005) and Lupus I, III and IV (Chapman \etal\
2007).

The Perseus molecular cloud is a source of active star formation
with few high-mass stars, none earlier than early B.  While
nowhere near as chaotic as the Orion star forming region, it is
also not as quiescent as the Taurus molecular cloud, and so
provides an ``intermediate'' case study.  IC 348 and NGC 1333
are the two densest and most famous star-forming clusters in
this region, containing numerous stars $<$1-2 Myr old.  Ongoing
star formation is certainly occurring throughout the cloud,
including in named regions such as L1448 and B5; there are
deeply embedded Class 0 objects found outside the two major
clusters (see, \eg, J06).  

The Perseus cloud is large enough that different parts of it may
be at significantly different distances; see the discussion in
Enoch \etal\ (2006).  Following that paper, and J06, we take the
distance to Perseus to be 250 pc, though we acknowledge that
there may indeed be a substantial distance gradient across the
cloud, or there might even be multiple pieces located at several
different distances.  

The Perseus cloud is rich in both point sources and complex
extended emission.  IRAS observations revealed more than 200
apparent point sources and complex ISM structures, including a
large ring thought to be an \ion{H}{2} region, G159.6$-$18.5,
excited by HD~278942 (Andersson \etal\ 2000).   Ridge \etal\
(2006) argue that the ring is {\em behind} the main Perseus
cloud.  Hatchell \etal\ (2005) find via a survey at 850 and 450
\mum\ a high degree of point-source clustering and filamentary
structures throughout the cloud.  Enoch \etal\ (2006) also find
strong point-source clustering in their 1.1 mm continuum survey.
The Spitzer/IRAC study of Perseus by J06 concluded (a) that there 
are significant numbers of stars being formed outside of the two
main clusters (IC 348 and NGC 1333); (b) the fraction of Class I,
Class II, and ``flat spectrum'' young stellar objects (YSOs)  
differs between the two rich clusters and the extended cloud
population; and (c) that deeply embedded Class 0 objects are
detected, with very red [3.6]$-$[4.5] colors (but not similarly
red [5.8]$-$[8] colors). 

MIPS observations at 24 \mum\ ($\sim6\arcsec$ resolution), 70
\mum\ ($\sim20\arcsec$ resolution), and 160 \mum\
($\sim40\arcsec$ resolution) can elucidate many aspects of the
ongoing star formation in Perseus.  Although the emission from
stellar photospheres is falling rapidly at 24 \mum, emission
from circumstellar material makes many of the young cluster
members still quite bright at 24 \mum, so MIPS finds the young
stars easily.  Emission from the cloud itself becomes increasingly
prominent in the MIPS 24, 70, and 160 \mum\ bands, allowing
dusty molecular material in the temperature range 120-20 K to be 
probed.  MIPS reveals complex extended emission throughout the Perseus 
region at all three of its wavelengths.

This paper presents MIPS data covering more than 10.5 square
degrees in Perseus.  We also use information obtained from the
IRAC data for Perseus from J06.  As a result of observational
constraints (see \S\ref{sec:obs} below) the IRAC data cover only
about one-third the area of the MIPS data (IRAC covers 3.86
square degrees), so we use the IRAC data where possible, but
large areas of our map have no IRAC data at all.   

The goals of this paper are to present the MIPS data in a format
similar to that found in the other papers in the series, to
discuss some of the high-level conclusions drawn from these
data, and to highlight some of the interesting objects we have
found.  Because this paper is part of a series, there is synergy
with both the papers that have gone before and those to come.
Because the IRAC data, where they exist, are usually important
for understanding the objects seen in the Perseus map, there are
extensive references to J06; for example, SEDs for some objects
discussed there were deferred to this paper for presentation. 
Similarly, there are references to future work throughout this
paper.  One such future paper will present a complete list of
YSO candidates associated with Perseus using the combined IRAC
and MIPS data, which is beyond the scope of this paper.  As part
of the c2d ancillary data, there has been a paper on the Bolocam
1.1 mm continuum survey of Perseus (Enoch \etal\ 2006) and one
on the JCMT/SCUBA sub-mm maps from the COMPLETE team as compared
to the Spitzer data (J{\o}rgensen \etal\ 2006b).  

This paper can be broadly divided into three major parts.  First we
give the details of the observations, reductions, and source extraction 
(\S\ref{sec:obs}).  This is followed by a presentation of the 
ensemble MIPS results for the entire Perseus cloud (\S\ref{sec:pall}).
Finally, \S\ref{sec:indobj} gives a focused discussion of noteworthy
stellar aggregates and individual young stars.  The main results of our 
study are summarized in \S\ref{sec:concl}.

\section{Observations, Data Reduction, and Source Extraction}
\label{sec:obs}

\begin{figure*}[tbp]
\epsscale{0.75}
\plotone{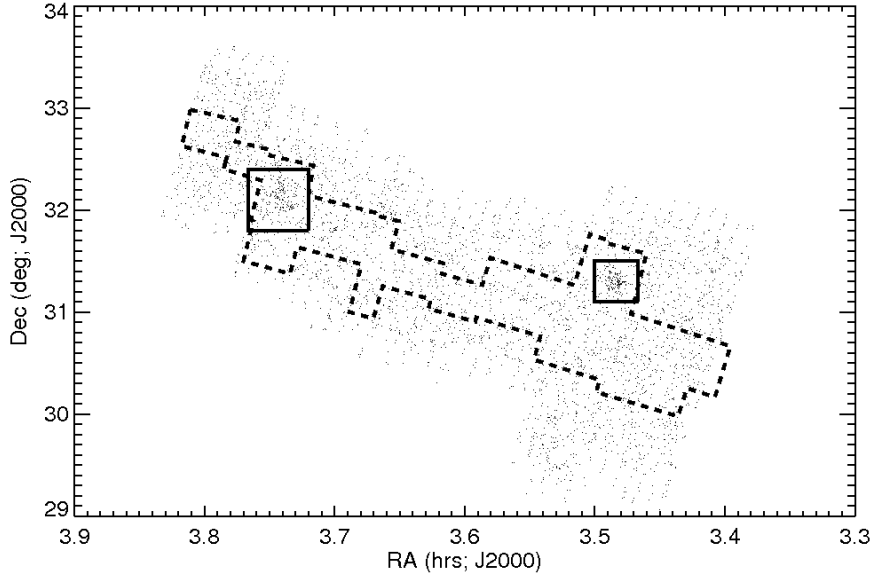}
\caption{Location of MIPS coverage (small points are MIPS-24
detections), with the region of IRAC coverage (dashed line)
indicated.  The smaller squares (solid lines) indicate the
regions defined to be IC 348 (left) and NGC 1333 (right).  The
definition of IC 348 includes the new objects recently found by
Lada \etal\ (2006). }
\label{fig:where}
\end{figure*}

\begin{deluxetable}{llll}
\tablecaption{Summary of observations (program 178)}
\label{tab:observations}
\tablewidth{0pt}
\tablehead{
\colhead{field} & \colhead{map center} & \colhead{first epoch
AORKEY} & \colhead{second epoch AORKEY} }
\startdata
per1  & 3h47m05.0s,+32d38m23.0s & 5780992 & 5787648 \\
per2  & 3h44m30.7s,+32d06m08.1s & 5781248 & 5787904 \\
per3  & 3h42m34.5s,+31d55m33.0s & 5781504 & 5788160 \\
per4  & 3h40m39.0s,+31d37m53.0s & 5781760 & 5788416 \\
per5  & 3h37m42.0s,+31d16m41.0s & 5782016 & 5788672 \\
per6  & 3h33m34.0s,+31d08m57.0s & 5782272 & 5788928 \\
per7  & 3h31m10.6s,+30d51m12.0s & 5782528 & 5789184 \\
per8  & 3h29m10.0s,+31d11m00.0s & 5782784 & 5789440 \\
per9  & 3h30m54.0s,+30d00m48.0s & 5783040 & 5789696 \\
per10 & 3h28m26.0s,+30d39m55.0s & 5783296 & 5789952 \\
per11 & 3h26m11.0s,+30d32m03.0s & 5798656 & 5790208 \\
\enddata
\end{deluxetable}

\subsection{Observations}

The MIPS observations of Perseus were conducted on 18-20 Sep
2004 and covered $\sim$10.5 square degrees; they were designed
to cover the $A_V$=2 contour, which then by extension completely
covered the c2d IRAC map of Perseus.  The center of this large
map is roughly at $\alpha, \delta$ (J2000) = 3$^h$37$^m$,
31$\arcdeg$11$\arcmin$30, or galactic coordinates
$l,b$=160$\arcdeg$, $-$19.5$\arcdeg$, or ecliptic coordinates
(J2000) 59.5$\arcdeg$, +11.5$\arcdeg$.

These observations were part of Spitzer program id 178; ``AORKEYs''
labeling the datasets in the Spitzer Archive are given in Table~1.
Fast scan maps were obtained at two
separate epochs.  At each epoch, the spacing between adjacent
scan legs was 240$\arcsec$.  The second epoch observation was
offset 125$\arcsec$ in the cross-scan direction from the first,
to fill in the 70 \mum\ sky coverage which was  incomplete at
each individual epoch.  Furthermore, the second epoch scan was
also offset 80$\arcsec$ from the first along the scan direction,
to maximize 160 $\mu$m map coverage in the combined epoch mosaic.  
These mapping parameters
resulted in every part of the map being imaged at two epochs at
24 \mum\ and only one epoch at 70 and 160 \mum, with total
integration times of 30 sec, 15 sec, and 3 sec at each point in
the map (respectively).  The 160 \mum\ maps have some coverage
gaps, and suffer from saturation (particularly in NGC 1333
and IC 348).  Figure~\ref{fig:where} shows the region of 3-band
coverage with MIPS, and the 4-band coverage with IRAC.  At about
10.5 square degrees, the MIPS observations cover a much larger
area than the IRAC observations, which cover only $\sim$4 square
degrees.  The three main reasons for this apparent mismatch are
entirely instrumental: (1) we are constrained by the ecliptic
latitude to observe with scan legs in a particular orientation
(the maximum rotation is about 10$\arcdeg$), (2) we are limited
in the available choices for scan leg lengths, and (3) MIPS
covers large areas very efficiently, so we can easily cover
large areas with MIPS in much less time than for IRAC. 

The c2d MIPS observations were designed for even coverage (all
to the same exposure time depth), independent of the GTO
observations of IC 348 (Lada \etal\ 2006) and NGC 1333 (R.\
Gutermuth \etal\ in prep.).  The GTO observations of these
regions are {\em not} included in the discussion here, primarily
to enable discussion of a catalog obtained to uniform survey
depth.  This is different from the c2d IRAC observations of
Perseus discussed in J06, where the GTO observations provided
the images for one of the two epochs.

The two observation epochs were separated by 3 to 6 hrs
to permit asteroid removal in this relatively low ecliptic
latitude  (+11--12$\arcdeg$) field.  Indeed, by comparing the 24
\mum\ maps obtained at the two epochs, at least 100 asteroids
are easily visible, ranging in flux density from at least as
faint as 0.6 to as bright  as 30 mJy.  Some asteroids are
clearly visible even at 70 \mum.  The asteroids in this and
other c2d cloud maps will be discussed further in K.\
Stapelfeldt \etal, in preparation.

We started with the SSC-pipeline-produced basic calibrated data
(BCDs), version S11.4.  For a description of the pipeline, see
Gordon \etal\ (2005).  As in Chapman \etal\ (2006), each MIPS
channel was then processed differently and therefore is
discussed separately below.   Mosaics and source catalogs from
these data were delivered back to the SSC for distribution; see
http://ssc.spitzer.caltech.edu/legacy/ for additional
information.  Multiple deliveries were made; the data discussed
here were part of the 2005 data delivery.

Figures~\ref{fig:24mosaic}, \ref{fig:70mosaic}, and
\ref{fig:160mosaic} show the individual mosaics by channel, and
\ref{fig:3color} shows a 3-color image with all three channels
included.  (For an indication of where ``famous'' regions are,
please see Fig.~\ref{fig:wherehilite}, which indicates several
objects highlighted for discussion below.)  There is substantial
extended emission in all three MIPS channels throughout the MIPS
maps.  In the 70 and 160 \mum\ channels, the MIPS instrument
uses internal stimulator flashes to calibrate the data (for more
information, see the Spitzer Observer's Manual, available at the
SSC website\footnote{http://ssc.spitzer.caltech.edu/}).  For
most of a scan leg, the correct calibration can be obtained via
an interpolation.  On the ends of scan legs, it necessarily must
use extrapolation solutions.  When the ends of the scan legs run
across particularly bright emission, as they do in some cases
here (particularly in the ``ring'' of bright emission), the
absolute calibration is not as good as it is in darker regions. 
Fluxes obtained in these regions have larger errors than the
rest of the map.

The IC 348 and NGC 1333 regions are encompassed by the overall
Perseus cloud map dataset.  To explore difference between the
memership of these two clusters and the more broadly distributed
Perseus young stellar population, we have chosen to consider
the IC 348 and NGC 1333 stellar populations separately from the
rest of the cloud; see Figure~\ref{fig:where}.  We have defined 
the regions belonging
to the clusters based on the surface density of 24 \mum\
sources. The region we define to be IC 348 is given by a box
bounded by the coordinates $\alpha$=55.8$\arcdeg$ to
56.5$\arcdeg$  (3.720$^{h}$ to 3.767$^{h}$, or
03$^h$43$^m$12.0$^s$ to 03$^h$46$^m$00.0$^s$),
$\delta$=31.8$\arcdeg$ to 32.4$\arcdeg$ (or
+31$\arcdeg$48$\arcmin$00.0$\arcsec$ to
+32$\arcdeg$24$\arcmin$00.0$\arcsec$).  This region is larger
than what has historically been assumed to encompass the
cluster, and large enough (0.36 square degrees) to include the
$\sim$300 likely cluster members (with the new likely members)
found by Lada \etal\ (2006), but not so large that it includes
substantial numbers of likely field members.  (Note that this
box includes most of the new members found by Cambresy \etal\
2006, but does not include their farthest southwest part of the
cluster.)  The region we define to be NGC 1333 is given by a box
bounded by the coordinates $\alpha$=52$\arcdeg$ to 52.5$\arcdeg$
(3.467$^{h}$ to 3.500$^{h}$, or 03$^h$28$^m$00.0$^s$ to
03$^h$30$^m$00.0$^s$), $\delta$=31.1$\arcdeg$ to 31.5$\arcdeg$
(or +31$\arcdeg$06$\arcmin$00.0$\arcsec$ to
+31$\arcdeg$30$\arcmin$00$\arcsec$); this region is 0.17 square
degrees.  In both cases, for ease of comparison, these regions
are the same as those used in J06.

\subsection{MIPS-24}

\begin{figure*}[tbp]
\plotone{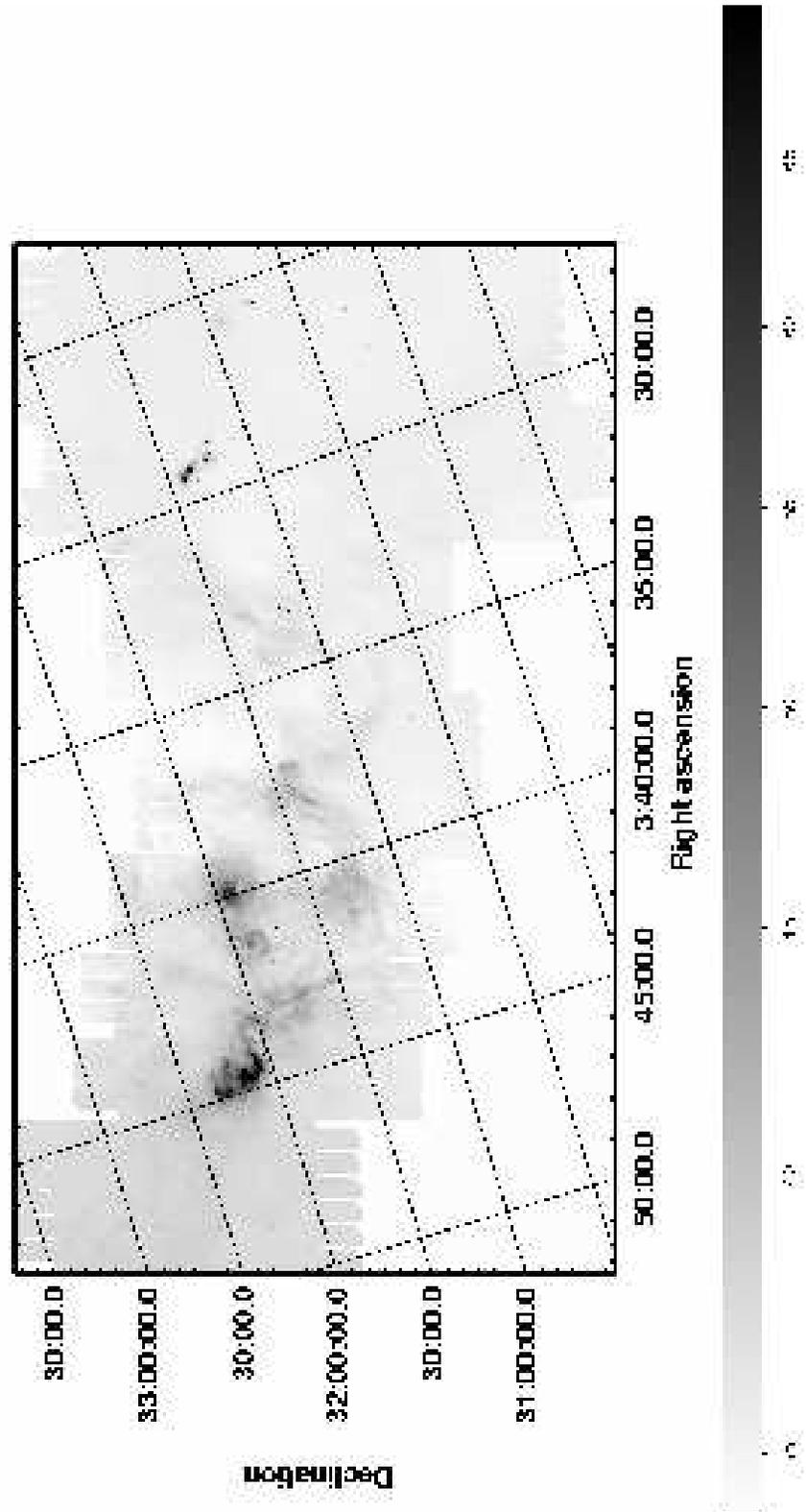}
\caption{Mosaic of Perseus map at 24 \mum.  The reverse
greyscale colors correspond to a logarithmic stretch of surface
brightnesses. }
\label{fig:24mosaic}
\end{figure*}

As discussed in Chapman \etal\ (2006), standard c2d pipeline
processing on the S11 BCDs (2005 delivery) was used for MIPS-24
(Evans \etal\ 2005; see also Young \etal\ 2005). In summary, the
c2d reduction starts with the pipeline-produced BCDs, and then
further processes them to remove artifacts, e.g., ``jailbars''
near bright sources.  A mosaic was then constructed from the
entire data set using the SSC MOPEX software (Makovoz \&
Marleau 2005); see Figure~\ref{fig:24mosaic}.  Sources were
extracted from the mosaic and bandmerged into the catalog along
with 2MASS (Skrutskie \etal\ 2006) \jhk\ and IRAC-1,2,3,4
measurements.  For more details on this process, please see
Chapman \etal\ (2006).  The uncertainty on the flux densities
derived at 24 \mum\ is estimated to be 15\%.

The high-quality catalog we assembled consisted of all
detections at MIPS-24 from the last 2005 delivery where the c2d
catalog detection quality flag (see the 2005 c2d Delivery
Document, available online linked from
http://ssc.spitzer.caltech.edu/legacy/) was `A' or `B,' which
translates to a signal-to-noise ratio of $>$5, and where the
object was detected at both epochs.  While resulting in a
shallower survey than would be possible using other combinations
of flags, this ensures that no asteroids are included in the
catalog.  The extraction pipeline flags some objects as extended
(``imtype'' flag); no filter was imposed on these extended
objects to create the catalog we used, but only 8 of the objects
in our catalog are flagged as extended.

There were 3950 total point sources detected at 24 \mum\ meeting
our criteria, ranging from 0.603 to 3530 mJy (comparable to the
saturation limit; see below).  The source surface density is
about 370 sources per square degree.  The zero point used to
convert these flux densities to magnitudes was 7.14 Jy, based on
the extrapolation from the Vega spectrum as published in the
MIPS Data Handbook.  About 30\% of these objects had
identifiable 2MASS counterparts at $K_s$.  

The faint limit of the catalog of 24 \mum\ sources is
a function of the nebular brightness across the field, but what
might be less obvious is that the saturation limit for point
sources with MIPS-24 is also a function of location in
the cloud because the total flux density registered by the
detector is that due to the point source itself plus any
surrounding extended nebular emission.  Because the extended
emission at 24 \mum\ varies from $\sim$450 MJy sr$^{-1}$ in IC
348 to $\sim$100 MJy sr$^{-1}$ in NGC 1333 to $\lesssim$1 MJy
sr$^{-1}$ in the darker parts of the cloud, the completeness of
the 24 \mum\ catalog at both the bright and faint limits is a
function of location in the cloud.  For example, as will be seen
in the source counts discussion (\S\ref{sec:sourcecounts}),
there are fewer faint sources in the clusters than in the field,
and fewer bright sources in NGC 1333 than in IC 348. This may
indeed be entirely due to the brightness of the background. An
additional issue when considering completeness is the
resolution; the resolution of MIPS-24 ($\sim6\arcsec$,
2.55$\arcsec$ pixel size) is poorer than IRAC or 2MASS
($\sim2\arcsec$). Source multiplicity and confusion may also
affect the completeness of the catalog, particularly in dense
regions such as the clusters.

\subsection{MIPS-70}
\label{sec:mips70}

\begin{figure*}[tbp]
\plotone{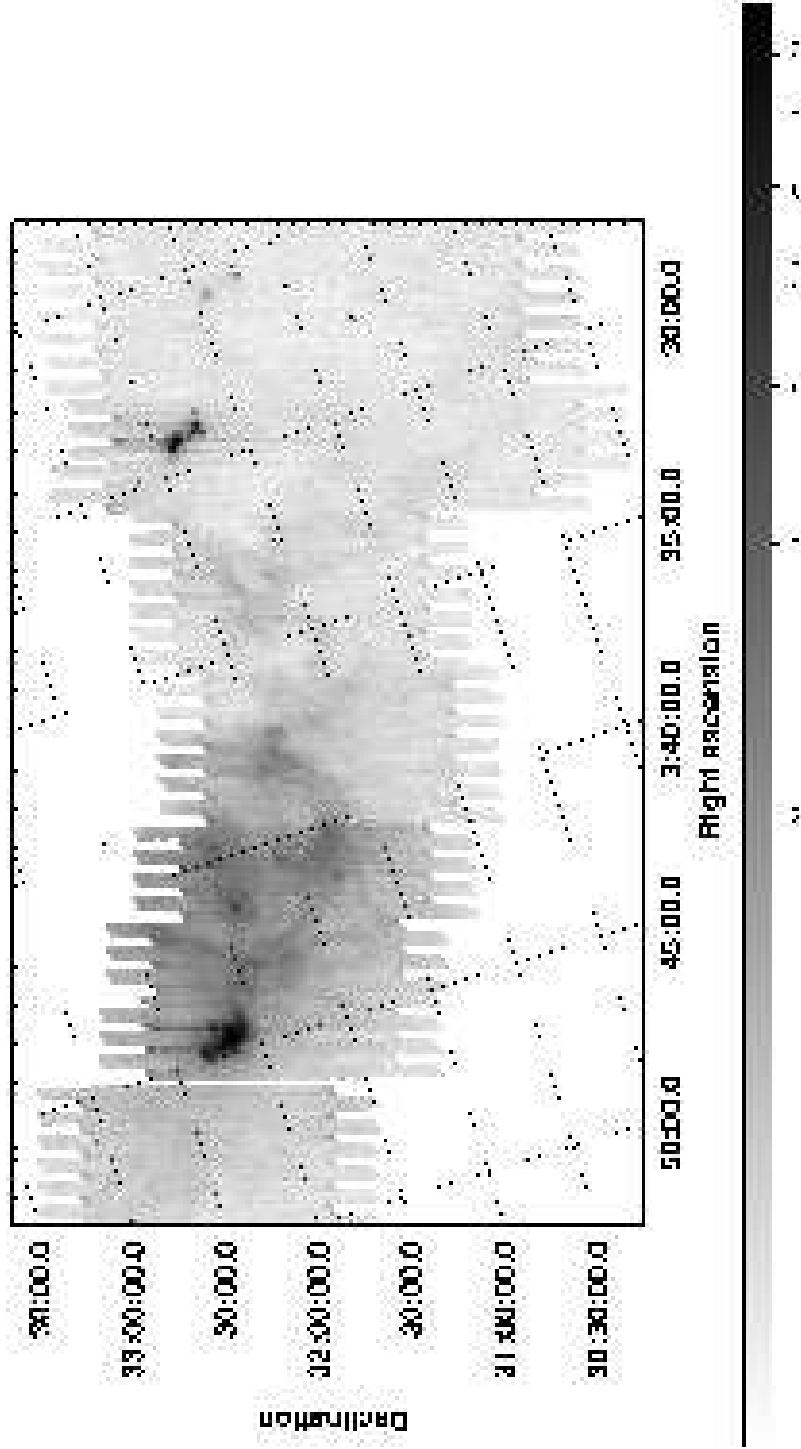}
\caption{Mosaic of Perseus map at 70 \mum.  The reverse
greyscale colors correspond to a logarithmic stretch. }
\label{fig:70mosaic}
\end{figure*}

To reduce the MIPS-70 data, we started with the automated
pipeline-produced BCDs.  The SSC produces two sets of BCDs; one
is simply calibrated, and the other has spatial and temporal
filters applied that attempt to remove instrumental signatures
in an automated fashion. These filtered BCDs do not conserve
flux for extended emission, nor for bright point sources, but
they do conserve flux for the fainter point sources.  We started
with both the filtered and unfiltered S11 BCD products.  Then,
we mosaicked these individual BCDs into one filtered and one
unfiltered mosaic using the MOPEX software.  We resampled the
pixels to be 4$\arcsec$ square (smaller than the native pixel
scale of $\sim10\arcsec$) to better enable source extraction.  
The unfiltered mosaic is presented in Figure~\ref{fig:70mosaic}
to better show the extended emission.  

We defined the point response function (PRF) from clean and
bright point sources selected from this large mosaic, and then
using this, performed point source detection and extraction 
using the APEX-1 frame option of MOPEX.  The initially-produced
source list was cleaned for instrumental artifacts via
manual inspection of the 70 \mum\ image and comparison to the 24
\mum\ image; e.g., if there was some question as to whether a
faint object seen at 70 \mum\ was real or an instrumental
artifact, and comparison to the 24 \mum\ image revealed a 24
\mum\ source, then the 70 \mum\ object was retained as a real
source.   This catalog also has the same limitations as was
found at 24 \mum; the brightness of the nebulosity drowns out
the faintest objects in the clusters, and contributes to
saturation of the brightest point-source objects (particularly
in the clusters) as well.  And, the resolution at 70
\mum\ ($\sim20\arcsec$) is coarser than it is at 24 \mum\
($\sim6\arcsec$), which complicates source matching to \ks\ and
source extraction in confused regions such as the clusters. For
all of these reasons, the 70 \mum\ catalog is not necessarily
complete and unbiased, particularly in the regions of bright ISM
and/or the faintest end.

Based on a comparison of PRF and aperture photometry fluxes, we
empirically determined the limit between where PRF fitting
photometry on the filtered image was more appropriate (fainter
than $4\times10^{3}$ mJy), and where aperture photometry with an
aperture radius of 32$\arcsec$ on the unfiltered image was more
appropriate (brighter than this level).  There were 19 objects
brighter than $4\times10^{3}$ mJy, for which we used aperture
photometry with a 32$\arcsec$ aperture and a 17\% aperture
correction.  The estimated uncertainty on our point source
fluxes is 20\%.  No color corrections were applied. Sources that
were determined by APEX to be extended were not fitted, so
sources that appear at initial manual inspection to be
point-like but are actually resolved in the 70 \mum\ maps do not
appear in our catalog. 

There were 139 total point sources detected at 70 \mum, ranging
from 64 mJy\footnote{An A0 photosphere with a 70 \mum\ flux
density of this value would have \ks$\sim$2.9 mag.} to 97
Jy\footnote{Although the instrument's published saturation
limits are about half of this value, some useful information can
still be extracted from the first few reads of the BCDs. Note
that the measured value of the flux of this bright object prior
to the aperture correction is $\sim$80 Jy.  This value may not
necessarily be well-calibrated, as it is indeed quite bright.};
the surface density is about 13 sources per square degree. 
About 90\% of the 70 \mum\ objects had identifiable counterparts
at 24 \mum.  The zero point used to convert the flux densities
to magnitudes was 0.775 Jy, based on the extrapolation from the
Vega spectrum as published in the MIPS Data Handbook.

\subsection{MIPS-160}
\label{sec:mips160}

\begin{figure*}[tbp]
\plotone{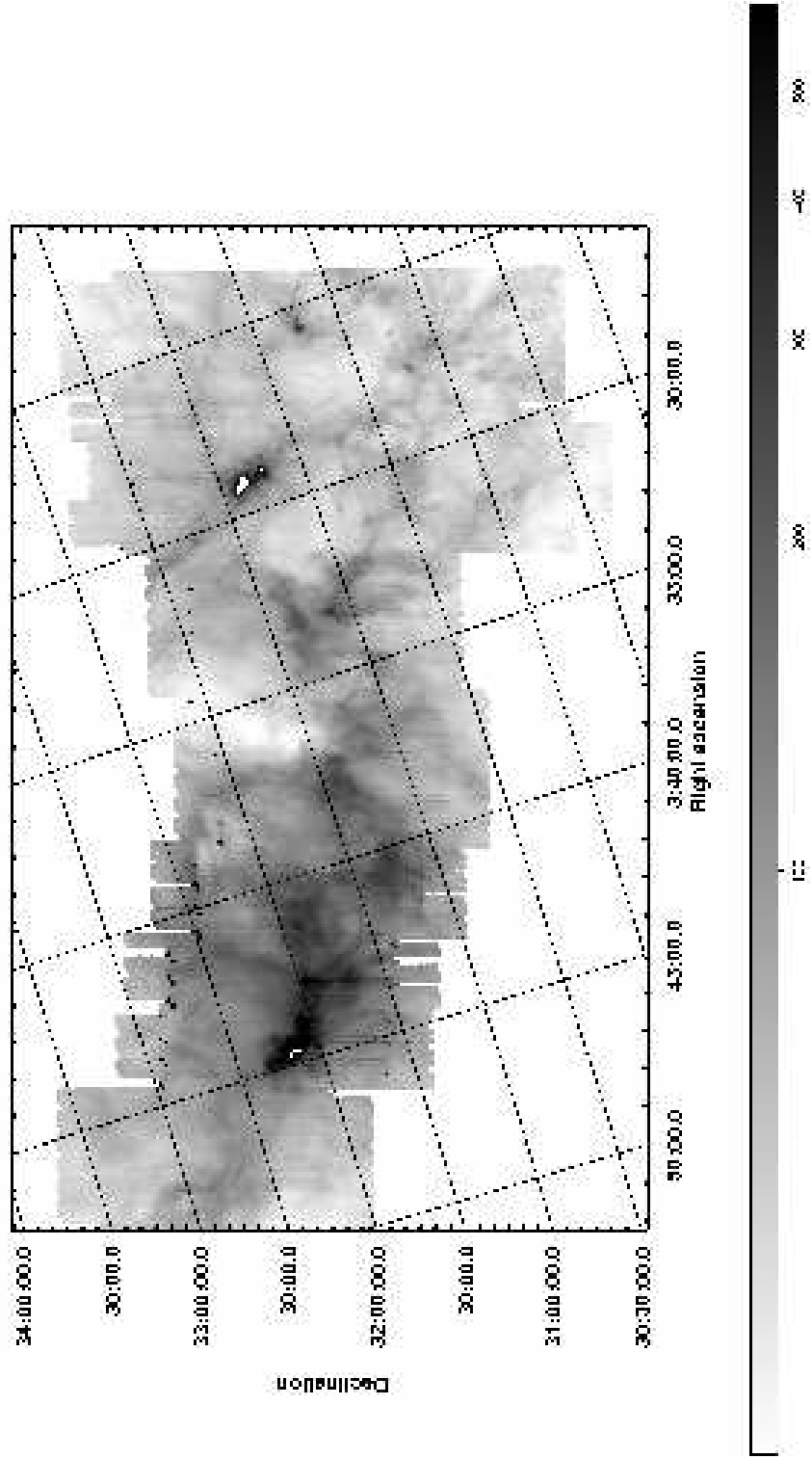}
\caption{Mosaic of Perseus map at 160 \mum.  The reverse
greyscale colors correspond to a logarithmic stretch. }
\label{fig:160mosaic}
\end{figure*}

For the 160 \mum\ data, we started with the S11.4 SSC BCDs.  To
create the mosaics that appear in the Figures in this paper
(such as Figure~\ref{fig:160mosaic}), we used MOPEX to mosaic
the unfiltered BCDs and resampled the mosaics to have 8$\arcsec$
pixels (half the native pixel scale). In order to ``fill in''
the gaps in the map caused by incomplete coverage, we used a
two-dimensional 3$\times$3 (native) pixel boxcar median to
interpolate across missing (``NaN'') pixels. No attempt was made
to interpolate across saturated regions, nor was any further
filtering applied to remove instrumental signatures.  

In order to obtain point source flux densities, we started with
the unfiltered 160 \mum\ pipeline products and created mosaics
with pixels of similar size to the native pixel scale.  
Photometry was then performed with APEX-1-frame PRF fitting on
this mosaic.  Because of the complexity of the extended emission
in this channel, APEX can be confused. APEX's initial source
list was cleaned by hand, and only flux densities from those
point sources considered reliable were retained.  Not all
emission peaks seen on the mosaic were retained; as for 70 \mum,
only unresolved point sources are included.  Similarly,
saturated sources, or close multiples, or low signal-to-noise
sources, are not included.  As with the other MIPS wavelengths,
in regions of particularly bright ISM (at 160 \mum, the ISM is
bright enough that this restriction is no longer limited just to
the clusters), point sources may be lost due to saturation or
(at the faint end) to simply being drowned out by the
nebulosity.  And, finally, source confusion is a particularly
difficult problem here, where the resolution is $\sim40\arcsec$.

There are 28 total point sources in our catalog detected at 160
\mum\ (see Figure~\ref{fig:whereall3}), ranging from 0.67 to 18
Jy.  About 60\% of the 160 \mum\ objects had identifiable
counterparts at 24 and 70 \mum; many of the 160-only objects are
lacking shorter-wavelength counterparts because the shorter
wavelengths are saturated.  The zero point used to convert the
flux densities to magnitudes was 0.159 Jy, based on the
extrapolation from the Vega spectrum as published in the MIPS
Data Handbook.

No color corrections were applied, nor was any attempt made to
remove the contribution from the blue filter leak.  As discussed
in the Spitzer Observer's Manual and the MIPS Data Handbook
(available on the SSC website), stars fainter than $J\sim$5.5
mag will not produce a detectable leak signal above the
confusion level.   Of the 12 objects detected at both $J$ and
160 \mum, all of them have  $J$ fainter than 5.5.  We conclude
that the leak is not important for  our objects.  We estimate an
overall 160 \mum\ flux uncertainty of 20\%.

\subsection{Bandmerged Source Catalogs and Statistics}

\begin{figure*}[tbp]
\includegraphics[angle=90]{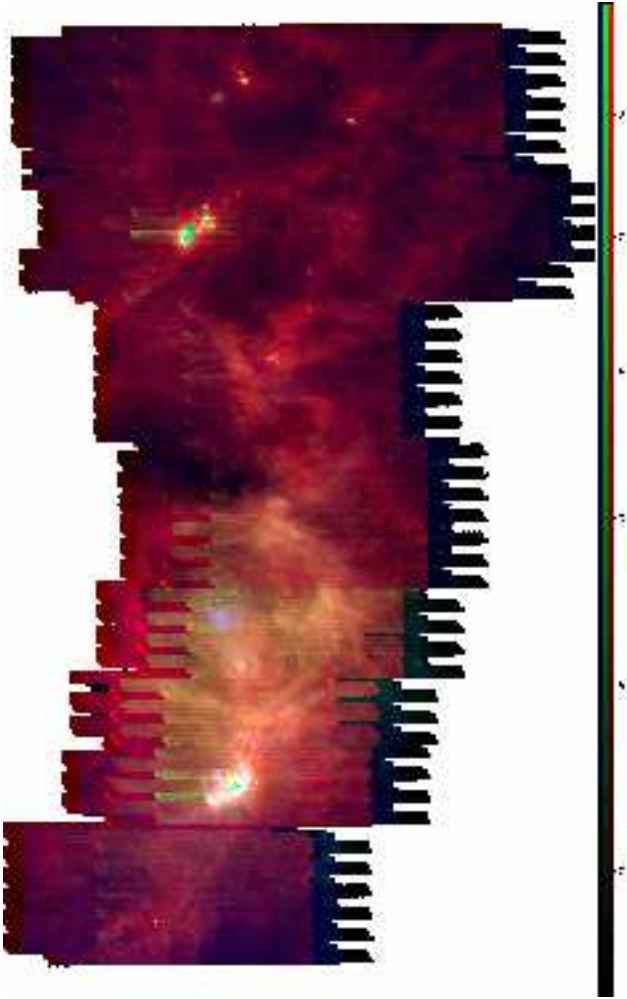}
\caption{Three-color mosaic of Perseus map at 24 (blue), 70
(green) and 160 (red) \mum.   }
\label{fig:3color}
\end{figure*}

A multi-wavelength mosaic using MIPS data can be found in
Figure~\ref{fig:3color}.  The extended emission is particularly
dramatic and complex when compared across all of the MIPS
wavelengths.

The c2d project mapped this cloud in IRAC as well; the IRAC map
(see J06) covers less than half the area covered by MIPS (see
Fig.~\ref{fig:where}), so there are large parts of the MIPS map
which do not have IRAC coverage.  The IRAC source extractions
discussed in J06 were bandmerged with the MIPS sources (see
Chapman \etal\ 2006 or Harvey \etal\ 2006 for much more
discussion on this multi-wavelength bandmerging process) and
these were included in our catalog.  

Table~2 presents some statistics on the ensemble catalog
spanning $J$-band through 160 \mum.  Whereas only 37\% of the
MIPS-24 objects in the entire catalog have an IRAC match at some
band, 92\% of the MIPS-24 objects {\em in the region covered by
IRAC} have a match in at least one IRAC band. There are near-IR
\jhk\ data covering this entire region from 2MASS, but 2MASS is
relatively shallow; only 29\% of the MIPS-24 sources have
\ks-band counterparts.

We wish to remind our readers of the discussion in
\S\ref{sec:obs} above, namely, that due to bright nebulosity,
particularly in the clusters, both the faintest sources and the
brightest sources (due to saturation) may not be present in the
catalog.  For example, about half of the 70 \mum\ objects that
lack 24 \mum\ counterparts and about half of the 160 \mum\
objects lacking 70 and 24 \mum\ counterparts are lacking those 
counterparts solely because the shorter wavelength counterpart
is saturated.

\begin{deluxetable}{lrrrrr}
\tablecaption{Statistics of MIPS point source detections}
\label{tab:statistics}
\tablewidth{0pt}
\tablehead{
\colhead{item} & \colhead{overall} & \colhead{IC 348} 
& \colhead{NGC 1333} & \colhead{rest of cloud} &
\colhead{IRAC coverage region}}
\startdata
24 \mum\ &             3950   & 252 & 166 & 3532 & 1602 \\ 
70 \mum\ &              139   &  11 &  21 &  107 & 85 \\ 
160 \mum\ &              28   &   0 &   1 &   27 & 15 \\ 
24 \mum\ \& 70 \mum\ &     121  &   9 &  13 &   99 & 85 \\ 
24 \mum, 70 \mum, \& 160 \mum\  &  16 &   0 &    0 &   16 &i 13 \\ 
24 \mum\ \& 2MASS K  &  1141  & 146 &  78 &  917 & 557 \\ 
24 \mum\ \& any IRAC band &  1476  & 205 & 146 & 1125 & 1476 \\ 
70 \mum\ \& any IRAC band &    91  &  10 &  16 &   65 & 83 \\ 
160 \mum\ \& any IRAC band &   19  &   0 &   1 &   19 & 15 \\ 
\enddata
\end{deluxetable}

The c2d IRAC study by J06 selected YSO candidates using a preliminary
SED classification scheme that required a source to be detected in
one or more IRAC photometric bands.  Perseus YSO candidates (``YSOc'') 
selected in this fashion were statistically described by J06 and are
not reproduced here.  A new result presented here for those objects 
discussed in J06 is that
$\sim$80\% have MIPS-24 counterparts, $\sim$13\% have MIPS-70
counterparts, and $\sim$2\% have MIPS-160 counterparts.  In the
region of Perseus not covered by IRAC, we are limited to using
$K-[24]$ colors to select YSO candidates.  Further discussion of
the global properties of these candidates appears below.  An
upcoming paper by S.-P.\ Lai \etal\ will synthesize data from
IRAC, MIPS, and groundbased surveys and present a list of all
YSO candidates in this cloud based on Spitzer data. 
A separate study by M.\ Enoch \etal\ will include very cold Spitzer
objects associated with millimeter emission.  J{\o}rgensen
\etal\ (2006b) discus a comparison between the JCMT/SCUBA
sub-millimeter and Spitzer data in Perseus.

Finally, in the present paper, we give only a statistical
picture of the cloud population compared to the extragalactic
source backgrounds.  As with the other c2d papers in this
series, our comparison extragalactic source background is data
from the Spitzer Wide-area InfraRed Extragalactic Survey (SWIRE;
Lonsdale \etal\ 2003), which were retrieved for the ELAIS N1
field (Surace \etal\ 2005), and processed through the same steps
as for the molecular cloud data.  For further discussion of
these steps, see Harvey \etal\ (2007, submitted).  

\subsection{Summary of Observations}

This paper presents 10.5 square degrees of Perseus as observed
with MIPS at 24 \mum\ ($\sim6\arcsec$ resolution), 70 \mum\
($\sim20\arcsec$ resolution), and 160 \mum\ ($\sim40\arcsec$
resolution).  It references bandmerged complementary data
obtained with IRAC (3.6, 4.5, 5.8, and 8 \mum) and 2MASS
($JHK_s$) over this same region.  There were 3950 point sources
detected at 24 \mum, 139 point sources at 70 \mum, and 28 point
sources at 160 \mum.  The sensitivity limits for all three bands
are a function of location on the sky because the sky brightness
changes substantially at all three bands; it is brightest in the
clusters IC348 and NGC 1333.  The rest of the analysis of the
ensemble of point sources in Perseus separates out the objects
found in the clusters from the rest of the cloud, and compares
them (where possible) to similar observations of one of the SWIRE
fields, expected to be populated entirely by galaxies. 

\section{Global Results Across the Perseus Cloud}
\label{sec:pall}

\subsection{Source Counts}
\label{sec:sourcecounts}

\begin{figure*}[tbp]
\epsscale{1}
\plotone{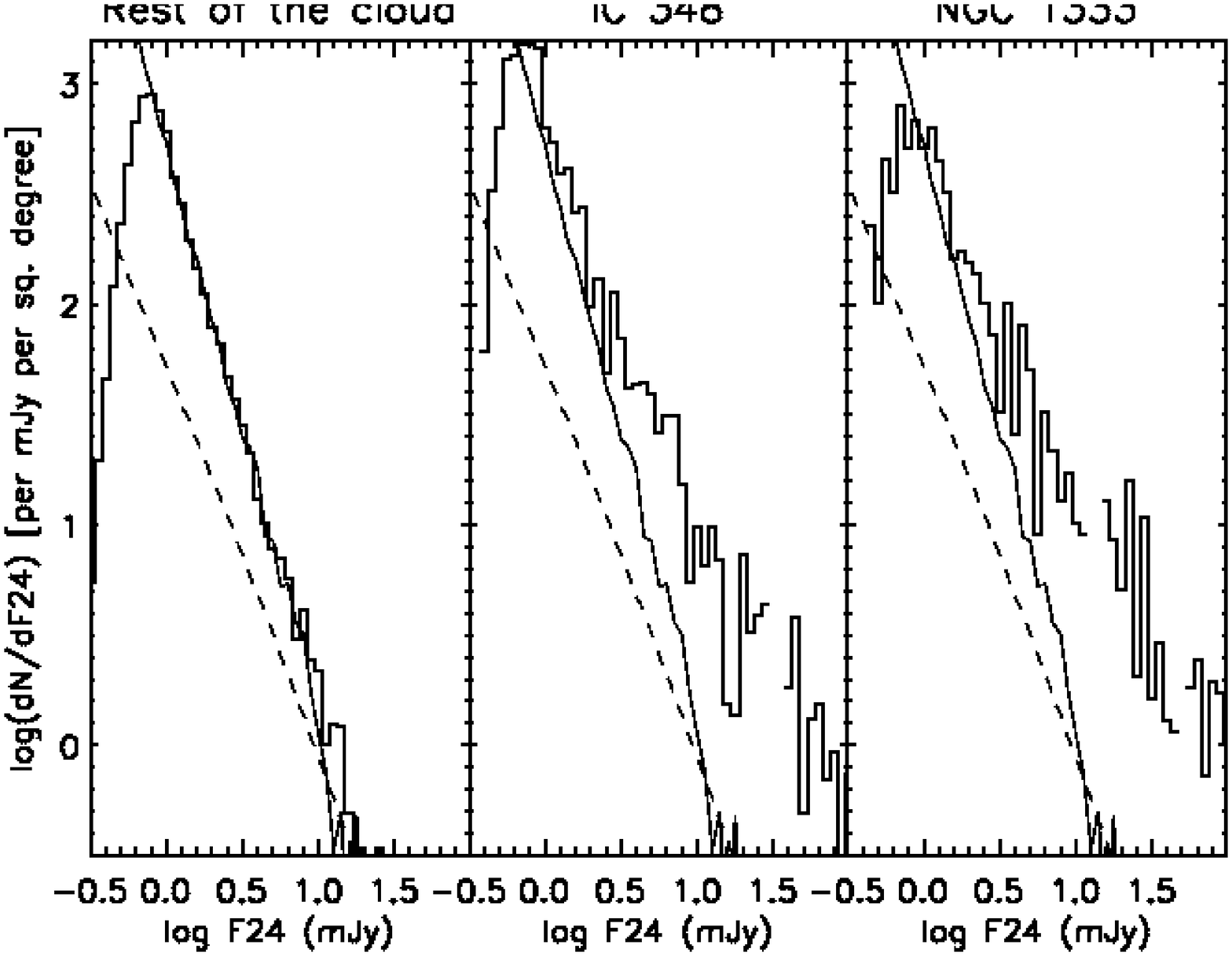}
\caption{Differential number counts at 24 \mum\ in Perseus. 
Extragalactic background source counts from the 6.1 square
degree SWIRE ELAIS N1 field are shown for comparison (thin
line), as well as predicted background star counts from the
models of Wainscoat \etal\ (1992; dashed line).  There is a
clear excess of sources in the clusters that are likely cluster
members.  In the ``rest of the cloud,'' an excess of on-cloud
sources is seen only for flux densities $>$10 mJy; below this
level, the Perseus source counts follow the extragalactic
background.}
\label{fig:diffnumbercounts24}
\end{figure*}

\begin{figure*}[tbp]
\epsscale{1}
\plotone{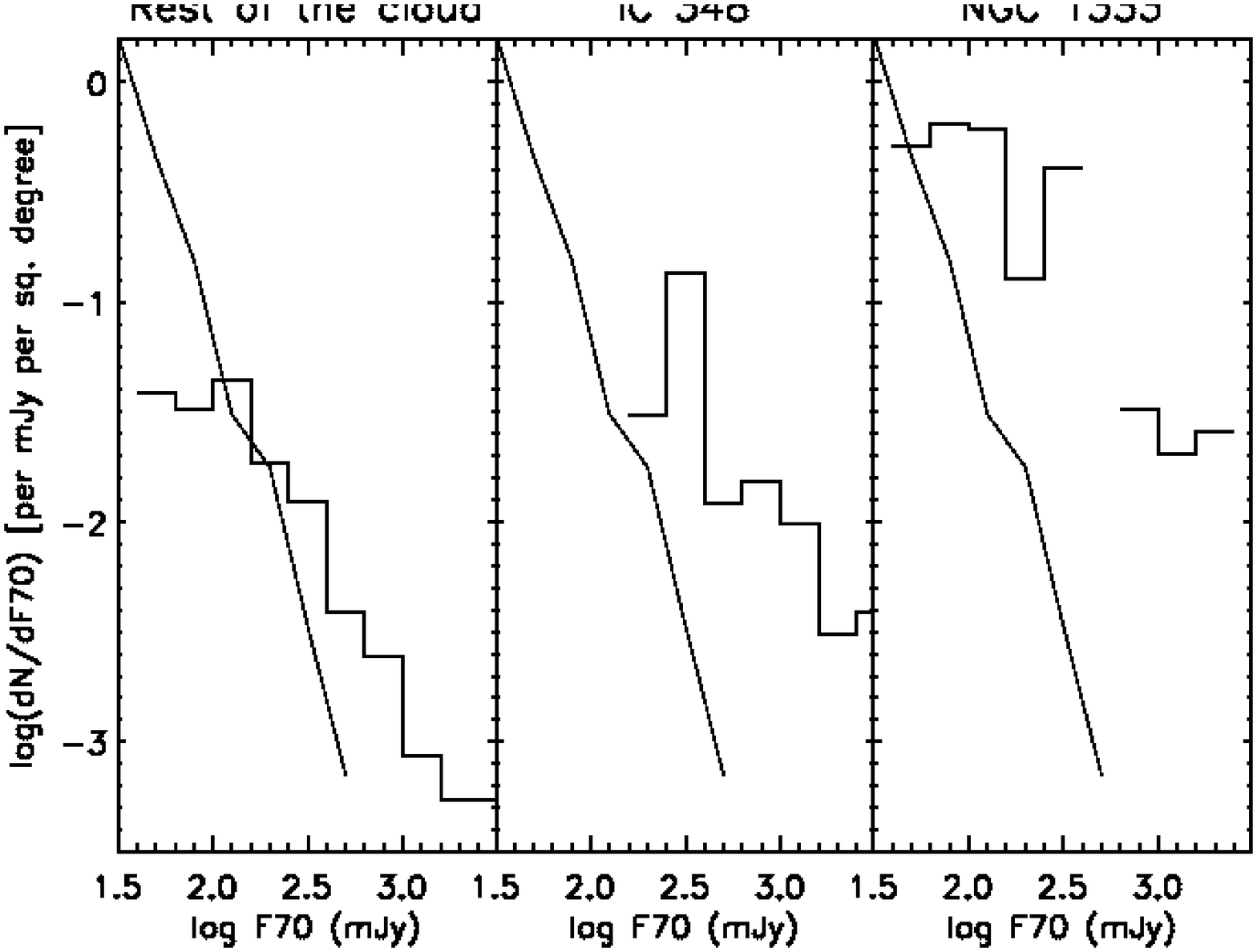}
\caption{Differential number counts at 70 \mum\ in Perseus. 
Extragalactic background source counts from the 6.1 square
degree SWIRE ELAIS N1 field are shown for comparison (thin
line).  There is a clear excess of Perseus cloud sources  above
the extragalactic background, for flux densities $>$ 200 mJy, in
all three regions.}
\label{fig:diffnumbercounts70}
\end{figure*}

MIPS sources in Perseus are a combination of cloud
members, foreground/background stars in the galaxy, and the
extragalactic background. Figure~\ref{fig:diffnumbercounts24}
shows the observed Perseus 24 \mum\ differential source counts
in comparison to observed source counts from the SWIRE ELAIS N1
extragalactic field, and to the prediction for galactic star
counts in  the IRAS 25 \mum\ band from the Wainscoat \etal\
(1992) model provided by J.\ Carpenter (2001, private
communication). In the IC 348 and NGC 1333 clusters, there is a
clear excess of cloud member sources above both backgrounds, for
flux densities $>$2-3 mJy.  In IC 348, there is suggestion of an
excess even at the 1 mJy level.  In NGC 1333, the source counts
turn over at higher flux densities, presumably due to reduced
sensitivity  caused by source crowding and bright extended
emission.  Across the much larger ``rest of the cloud'' region,
an excess of on-cloud  sources is seen only for flux densities
$>$10 mJy; below this level,  the Perseus source counts follow
the extragalactic background.  Galactic star counts make only a
minor contribution to the MIPS source counts in Perseus, in
contrast to the four other clouds surveyed by c2d (Evans \etal\
2003).  This is a natural consequence of Perseus' location at
$l$=160$\arcdeg$, $b= -20\arcdeg$, whereas the four other clouds
are located toward the inner galaxy and closer to the galactic
plane.

Figure~\ref{fig:diffnumbercounts70} shows 70 \mum\ source counts
in Perseus, along with extragalactic background counts from the
SWIRE ELAIS N1 field.  In the extended cloud outside the
clusters, a clear excess in source counts is present above a
flux density of 200 mJy.  A similar result is found in IC 348. 
In NGC 1333, however, the flux density limit above which cluster
number counts exceed the extragalactic background is about 65
mJy.  The brightest measurable objects in Perseus are found in
NGC 1333, where there is a clear excess of these relative to
other parts of Perseus.  It should be noted (see discussion in
\S\ref{sec:mips70} above) that the brightest sources (flux
densities $>$20 Jy) are under-represented in these plots due to
saturation effects, that the faintest sources are both subject
to the completion limitations discussed above and more difficult
to detect in regions of high surface brightness, and that only
point sources are included.

\subsection{Recovery of IRAS Sources}
\label{sec:iras}

\begin{deluxetable}{lll}
\tablecaption{IRAS Results in MIPS Perseus map region}
\label{tab:iras}
\tablewidth{0pt}
\tablehead{
\colhead{item} & \colhead{\ldots\ at 24  } & \colhead{\ldots\
at 70 \mum} }
\startdata
12 \mum\ PSC real (IRAS qual=3) detections &         51  &  51           \\
\hskip 36pt cleanly retrieved  &            34 (67\%)    &  19 (37\%)     \\
\hskip 36pt completely missing  &            1 ( 2\%)    &  16 (31\%)    \\
\hskip 36pt confused by nebulosity  &       11 (21\%)    &  15
(29\%)\\
\hskip 36pt resolved as multiple  &          5 (10\%)    &   1 ( 2\%)    \\
25 \mum\ PSC real (IRAS qual=3) detections &           56  &  56            \\
\hskip 36pt cleanly retrieved  &           34 (61\%)     &   31 (55\%)     \\
\hskip 36pt completely missing  &            0 ( 0\%)    &     6 (11\%)    \\
\hskip 36pt confused by nebulosity  &           13 (23\%)&        17 (30\%)\\
\hskip 36pt resolved as multiple  &            9 (16\%)    &     2 ( 4\%)    \\
60 \mum\ PSC real (IRAS qual=3) detections &       88  &  80           \\
\hskip 36pt cleanly retrieved  &           32 (36\%)     &   35 (44\%)     \\
\hskip 36pt completely missing  &            9 (10\%)    &     8 (10\%)    \\
\hskip 36pt confused by nebulosity  &           42 (48\%)&  36 (45\%)\\
\hskip 36pt resolved as multiple  &            5 (6\%)    &     1 ( 1\%)    \\
100 \mum\ PSC real (IRAS qual=3) detections &       78 &   75          \\
\hskip 36pt cleanly retrieved  &           10 (13\%)     &    9 (12\%)     \\
\hskip 36pt completely missing  &           41 (52\%)    &    40 (53\%)    \\
\hskip 36pt confused by nebulosity  &           25 (32\%)&    25 (33\%)\\
\hskip 36pt resolved as multiple &            2 (3\%)    &     1 ( 1\%)    \\
\hline
12 \mum\ FSC real (IRAS qual=3) detections &        56  &  57           \\
\hskip 36pt cleanly retrieved  &           48 (86\%)     &   24 (42\%)     \\
\hskip 36pt completely missing  &            4 (7\%)    &    27 (47\%)    \\
\hskip 36pt confused by nebulosity  &            2 (3.5\%)&  5 (9\%)\\
\hskip 36pt resolved as multiple  &            2 (3.5\%)    &     1 (1\%)    \\
25 \mum\ FSC real (IRAS qual=3) detections &           52  &  50            \\
\hskip 36pt cleanly retrieved  &           38 (73\%)     &   26 (52\%)     \\
\hskip 36pt completely missing  &            1 ( 2\%)    &    10 (20\%)    \\
\hskip 36pt confused by nebulosity  &            8 (15\%)&        11 (22\%)\\
\hskip 36pt resolved as multiple &            5 (10\%)    &     3 ( 6\%)    \\
60 \mum\ FSC real (IRAS qual=3) detections &            0  &   0            \\
100 \mum\ FSC real (IRAS qual=3) detections &            0& 0\\

\enddata
\end{deluxetable}

The MIPS data for the Perseus molecular cloud offer an
opporunity to assess how successful the IRAS survey was in
identifying point sources in a complex region.  We now
explicitly compare the IRAS PSC (Beichman \etal\ 1988) and IRAS
FSC (Moshir \etal\ 1992) results in this region to the Spitzer
c2d images and catalogs described in this paper.  Complex
extended emission is present in all three MIPS bands, posing a
significant source of confusion to the large IRAS aperture
measurements.  Many of the IRAS PSC objects were detected at 60
or 100 \mum\ only, with only upper limits at 12 and 25 \mum; so
even without Spitzer data, one might suspect that such sources
might correspond to texture in the extended emission.  As we
will show below, the Spitzer data comparison clearly shows that
significantly fewer spurious sources appear in the FSC than in
the PSC.

The statistics are presented in Table~3.  An IRAS 
data quality flag of 3 means a solid detection, whereas data
quality flags of 1 indicate upper limits.  ``Cleanly
retrieved'' means that IRAS reported a single source at a given
location, and the corresponding MIPS image shows a single
object which in several cases was saturated.  ``Completely
missing'' means that no source appears at that location in the
corresponding MIPS image, and there is no immediately apparent
reason for the discrepancy.  In contrast, ``confused by
nebulosity'' means that the corresponding MIPS image shows
structured nebulosity that could be confused for a point source
in the large IRAS beams.  Finally, ``resolved as multiple''
means that MIPS finds multiple far-infrared sources
where IRAS reported only one.

As can be seen in Table~3, most of the 12 and 25
\mum\ highest-quality PSC point sources are either retrieved or
it is clear why a point source was reported for that location. 
Overall, 20-30\% of the ``point  sources'' at 12 and 25 \mum\
resolve into knots of nebulosity.  Essentially none of these 12
and 25 \mum\ point sources are undetected at 24 \mum,
but a much larger fraction are completely missing at 70 \mum\ --
this is likely to be a result of the lower sensitivity of 70
\mum\ compared to 24 \mum\ combined with the lower expected flux
density from stars at 70 \mum\ compared with the shorter wavelengths.  

There are many more ``point sources'' reported at 60 and 100
\mum\ (than at 12 and 25 \mum) in the PSC.  A much lower
fraction of these objects are recovered; many more of these are
completely missing or fall apart into nebulosity when viewed
with MIPS.  More than half of the 100 \mum\ point sources are
completely missing, and 25-40\% of the 100 and 60 \mum\ point
sources are clearly confused by nebulosity.

Surprisingly few IRAS sources are resolved into multiple objects
by Spitzer.  However, our results may be a lower limit on the
true number of such objects, as we did not consider the
individual error ellipses for each IRAS source.  Two IRAS
sources clearly resolved into stellar aggregates are discussed
below in \S\ref{sec:iras1} and \ref{sec:iras2}.

Overall, a much larger fraction of FSC objects are recovered by
MIPS than PSC objects.  This is largely due to the FSC
being far more cautious about claiming to detect point sources
in the presence of cirrus confusion: no 60 or 100 \mum\
detections are listed in this region in the FSC, whereas the PSC
lists numerous detections that appear to be spurious.  At 12 and
25 \mum, unrecognized asteroids may be responsible for the lower
recovery rate of PSC sources relative to the FSC.  We conclude
that the IRAS FSC is much more robust for studies of stellar
sources in regions with complex extended emission, and should be
used in preference to the IRAS PSC unless Spitzer/MIPS data are
available.  A list of high-quality IRAS detections that are
missing or fall apart into nebulosity in our Perseus MIPS
dataset appears in Appendix A.

We note for completeness that direct comparison of the measured
MIPS and IRAS flux densities for the same sources are greatly
complicated by the different calibration approaches of the two
instruments, requiring detailed knowledge of the underlying SED
of the objects in question and large color corrections.  This is
beyond the scope of this paper.

\subsection{Color-Magnitude and Color-Color Plots}
\label{sec:cmdccd}

\begin{deluxetable}{llll}
\tablecaption{Classification based on the \ks vs. \ks$-$[24] diagram}
\label{tab:k24class}
\tablewidth{0pt}
\tablehead{
\colhead{item} &  \colhead{rest of cloud} & \colhead{IC 348} 
& \colhead{NGC 1333}}
\startdata
number objects with \ks\ and 24 \mum\ & 917 & 146 & 78 \\
number with \ks$-$[24]$>$2, \ks$<$14 & 138 & 107 & 52 \\
number with \ks$-$[24]$>$2, \ks$<$14, and Class I \ks$-$[24] color
& 9  (6\%) & 7 (6\%) & 4 (7\%) \\
number with \ks$-$[24]$>$2, \ks$<$14, and ``flat'' \ks$-$[24]
color & 12 (8\%) & 5 (4\%) & 11 (21\%) \\
number with \ks$-$[24]$>$2, \ks$<$14, and Class II \ks$-$[24]
color & 105 (76\%) & 92 (85\%) & 35 (67\%) \\
number with \ks$-$[24]$>$2, \ks$<$14, and Class III \ks$-$[24]
color & 12 (8\%) & 3 (2\%) & 2 (3\%)\\
\enddata
\end{deluxetable}

\begin{deluxetable}{lllllll}
\tablecaption{Very red objects selected by a variety of means}
\label{tab:redobj}
\tablewidth{0pt}
\rotate
\tabletypesize{\scriptsize}
\tablehead{
\colhead{SSTc2d name} & \colhead{\ks$-$[24]\tablenotemark{a}} &
\colhead{\ks$-$[70]\tablenotemark{b}}&
\colhead{[24]$-$[70]\tablenotemark{c}} &  
\colhead{missed match\tablenotemark{d}} &
\colhead{other name} &\colhead{notes} }
\startdata
032637.4+301528 & yes  &  yes &\ldots&\ldots         &IRAS 03235+3004 &see \S\ref{sec:all3} and J06                    \\
032800.4+300801 & yes  &\ldots&\ldots&\ldots         &IRAS 03249+2957  &see J06                                  \\
033015.1+302349 & yes  & yes  &\ldots&\ldots         &IRAS 03271+3013  &see \S\ref{sec:per6agg}                  \\
034202.1+314801 & yes  &\ldots&\ldots&\ldots         & \ldots          &see \S\ref{sec:iras1}                    \\
032845.3+310541 & yes  & yes  &\ldots&\ldots         &HH 340           &see \S\ref{sec:all3}                     \\
033925.5+321707 & yes  &\ldots&\ldots&\ldots         & \ldots          &see \S\ref{sec:all3}                     \\
032839.6+311731 & yes  &\ldots&\ldots&\ldots         &LAL 68           &in NGC 1333                              \\
032858.4+312217 & yes  &\ldots&\ldots&\ldots         &LAL 166          &in NGC 1333                              \\
032903.3+312314 & yes  &\ldots&\ldots&\ldots         &LAL 191          &in NGC 1333                              \\
032912.9+311814 & yes  &\ldots&\ldots&\ldots         &ASR 30           &in NGC 1333                              \\
033309.5+310531 & yes  &\ldots&\ldots&\ldots         &  \ldots         &in B1 region                             \\
033316.6+310755 & yes  &  yes &\ldots&\ldots         &IRAS 03301+3057  &in B1 region; see J06                    \\
033925.5+321707 &\ldots&  yes &\ldots&\ldots         &IRAS 03363+3207  &see \S\ref{sec:all3}                     \\
032548.2+305537 &\ldots&  yes &\ldots&yes, but ext@24&BD+30$\arcdeg$540&see \S\ref{sec:bd+30d540}              \\
032904.2+311608 &\ldots&  yes &\ldots&yes, but sat@24&  \ldots         &in NGC 1333, see \S\ref{sec:kk70}  \\
033554.4+304500 &\ldots&  yes &\ldots&\ldots         &2MASS 03355439+3045011&  see \S\ref{sec:kk70}              \\
034443.9+320136 &\ldots&  yes &\ldots&\ldots         &IRAS 03415+3152  &   on edge of IC348; see \S\ref{sec:kk70}\\
032910.5+311330 &\ldots&\ldots& yes  &\ldots         &IRAS 4 A1/A2     & in NGC 1333                             \\
033121.0+304530 &\ldots&\ldots& yes  &\ldots         &IRAS 03282+3035  & see J06                                 \\
033316.5+310652 &\ldots&\ldots& yes  &\ldots         &B1-d             & see J06                                 \\
033218.0+304946 &\ldots&\ldots& yes  &\ldots         &IRAS 03292+3039  & see \S\ref{sec:all3} and J06            \\
034356.9+320304 &\ldots&\ldots& yes  &\ldots         &IC 348 MMS       & see \S\ref{sec:hh211}                   \\
034741.6+325143 &\ldots&\ldots&\ldots& yes, no 24     &B5 IRS 1         &see J06                                  \\
032851.6+304502 &\ldots&\ldots&\ldots& yes, no 24     &LkHa 325         &see \S\ref{sec:2470160}                  \\
033227.5+310236 &\ldots&\ldots&\ldots& yes, no 24,70  &  \ldots         &in B1 region, shorter saturated         \\
033310.3+312108 &\ldots&\ldots&\ldots& yes, no 24,70  &  \ldots         &in B1 region, shorter saturated         \\
033327.0+310647 &\ldots&\ldots&\ldots& yes, no 24,70  &  \ldots         &in B1 region, shorter saturated         \\
032842.8+311744 &\ldots&\ldots&\ldots& yes, no 24,70  &  \ldots         &shorter saturated                        \\
034107.8+314411 &\ldots&\ldots&\ldots& yes, no 24,70  &  \ldots         &shorter saturated                        \\
033511.5+312026 &\ldots&\ldots&\ldots& yes, no 24,70  &  \ldots         &very weak @160, real?                    \\
032923.9+313320 &\ldots&\ldots&\ldots& yes, no 24,70  &  \ldots        &diffuse@160, offset from nearby bright obj  \\
033046.4+303242 &\ldots&\ldots&\ldots& yes, no 24,70  &  \ldots        &diffuse@160, offset from nearby bright obj  \\
\enddata
\tablenotetext{a}{Objects where $K_s-[24]>$9.7 mag.}
\tablenotetext{b}{Objects where \ks$-$[70]$>$ 15 mag.}
\tablenotetext{c}{Objects where [24]$-$[70]$>$7 mag.}
\tablenotetext{d}{Objects which were detected at 70 without a 24
counterpart, or detected at 160 without 24 or 70 counterparts.}
\end{deluxetable}

\subsubsection{The $K_s$ vs.\ $K_s-[24]$ Diagram}
\label{sec:kk24}

\begin{figure*}[tbp]
\epsscale{1}
\plotone{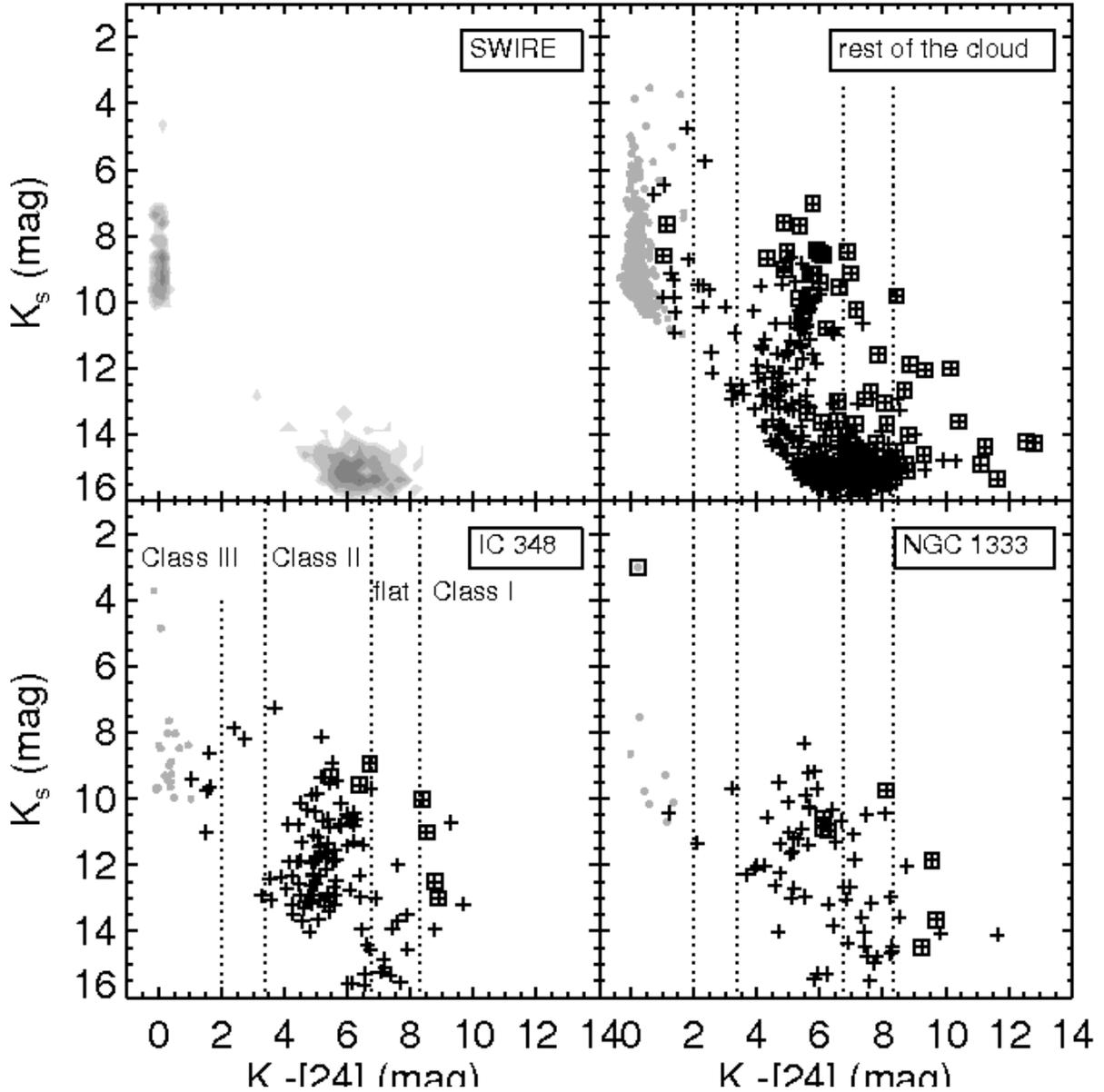}
\caption{$K_s$ vs.\ $K_s-[24]$ for objects in SWIRE (contour
plot, upper left), the ``rest of the cloud'' (upper right), IC
348 (lower left), and NGC 1333 (lower right).  Objects in SWIRE
are expected to be mostly galaxies (objects with $K_s\gtrsim$14)
or stellar photospheres (objects with $K_s-[24]\lesssim$1).  For
the remaining plots, filled gray circles are objects with SEDs
resembling photospheres, and plus signs are the remaining
objects.  An additional box around a point denotes that it was
also detected in 70 \mum.  Objects that are candidate young
objects have colors unlike those objects found in SWIRE, e.g.,
$K_s\lesssim$14 and $K_s-[24]\gtrsim$1. Dashed lines denote the
divisions between Class I, flat, Class II, and Class III
objects; to omit foreground and background stars, we have
further imposed a $K_s-[24]>$2 requirement on our Class III
objects (see text).  Based on the fraction of stars in the
various classes, NGC 1333 is younger than IC 348; there are
substantial numbers of young stars in the rest of the cloud.}
\label{fig:k_k24}
\end{figure*}

\begin{figure*}[tbp]
\epsscale{1}
\plotone{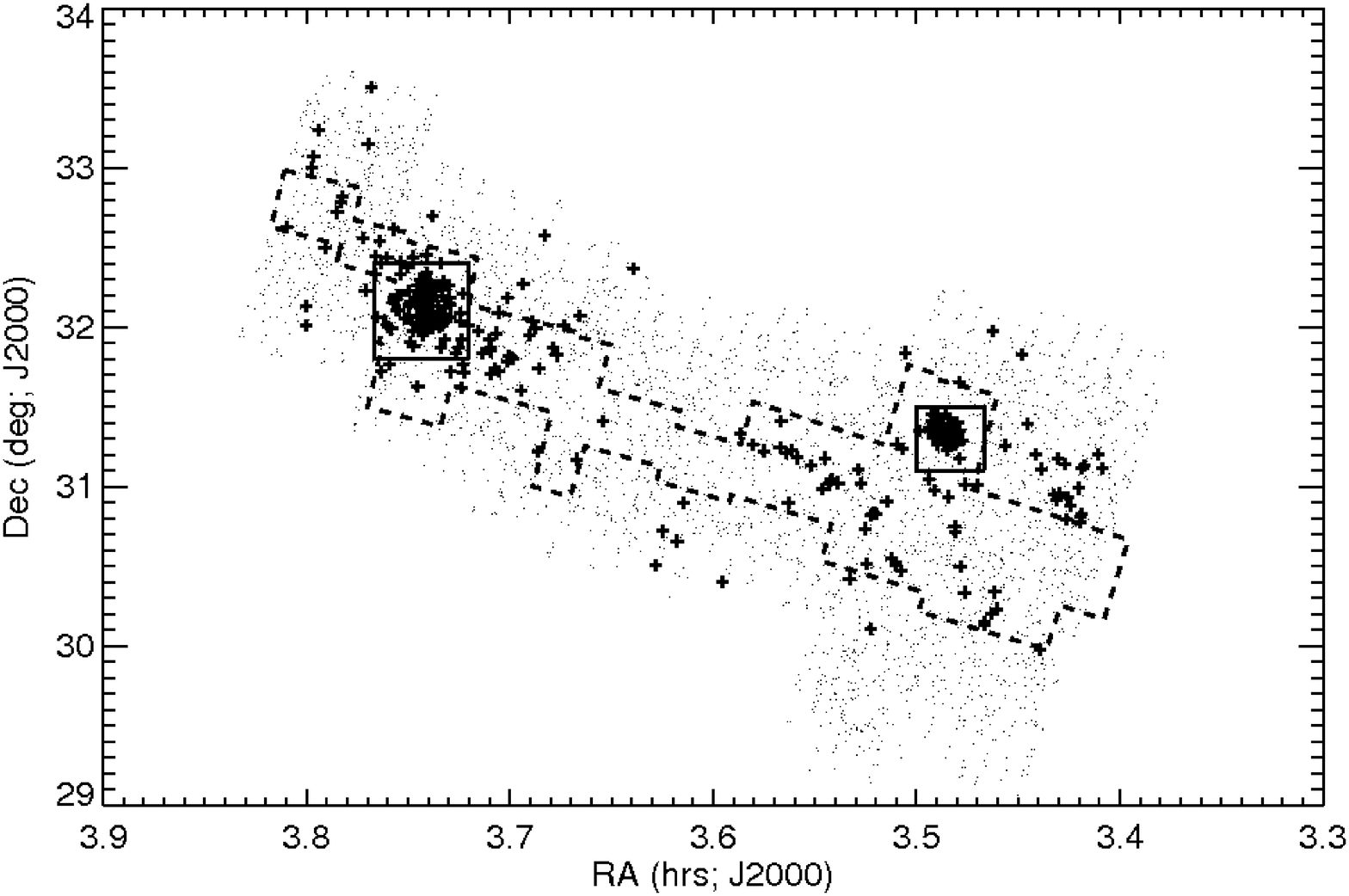}
\caption{Locations of all objects detected at MIPS-24 (small
dots) and the MIPS YSO candidates (plus signs), as defined by
$K_s-[24]>2$ and  $K_s<14$.  The dashed line denotes the region
of A$_V>3$ covered in the Spitzer/IRAC images of J06.  Solid
boxes outline the regions defined for the IC 348 and NGC 1333
clusters, as in Fig.~\ref{fig:where}.  There are 36 objects with
excesses outside the high-extinction region. Some of these are
located near the ends of scan legs comprising the mosaics, so
their photometry is less reliable.  A few may be background
galaxies.  Some are likely to be more distributed cloud members,
especially those clustered just north of L1448 on the western
side of the mosaic. }
\label{fig:wherek24x}
\end{figure*}

Figure~\ref{fig:k_k24} shows the $K_s$ vs.\ $K_s-[24]$
color-magnitude diagram (CMD) for Perseus, combining our results
with 2MASS survey data.  Ordinary stellar photospheres (likely
foreground or background stars) have $K_s-[24]\sim$0.  The lack
of sources in the lower left of the $K_s$ vs.\ $K_s-[24]$ plot
is entirely a sensitivity effect, explained as follows.  Our
\ks\ magnitudes are limited to those found in 2MASS, \eg, to
brighter than $\sim$16th magnitude in this region.  Our 24 \mum\
survey goes down to about 1 mJy (see
Fig.~\ref{fig:diffnumbercounts24}), which, for a stellar
photosphere, corresponds to about \ks$\sim$10, so we do not see
any stellar photospheres with \ks\ fainter than $\sim$10. As we
move to redder objects, source matches to fainter \ks\ are
obtainable down to the 2MASS sensitivity limit. Background
galaxies become numerous at \ks$>$14. 

The top left panel of Figure~\ref{fig:k_k24} shows data from the SWIRE
extragalactic survey, as discussed above, which have been flux-trimmed
to match the c2d survey sensitivity.  Since all of  the objects in
SWIRE are likely foreground stars or background galaxies, this panel
shows which regions of the CMD are least likely to provide a clean
sample of likely YSOs: objects with $K_s-[24]>$2 and \ks$<$14 are
least likely to be part of the galactic or extragalactic backgrounds,
and thus likely to be young stellar objects with a 24 \mum\ excess. 
There are many such stars with $K_s-[24]$ excesses in the ``rest of
the cloud'' despite some expectations that there would be little star
formation outside of the canonical clusters.  Looking at the IC 348
and NGC 1333 clusters, there are comparatively not many stars nor
background galaxies; most of the objects are likely young cluster
members.  

There remains a possibility that some of the objects seen at 24 \mum\
and having a \ks$-$[24] color suggestive of disks may be background
AGB stars.  Harvey \etal\ (2005,2006) find that any AGB star in our
Galaxy is likely to be saturated in the c2d IRAC observations, but the
MIPS survey effectively reaches a different depth.  The galactic
coordinates of Perseus are (160$\arcdeg$, $-$19.5$\arcdeg$), toward
the galactic anticenter and a significant angle out of the plane; the
expected background star counts are relatively small.  Assuming as a
worst-case scenario  that the Milky Way disk is 15 kpc radius and 1.5
kpc thick, the furthest  reaches of our Galaxy that could be included
in these observations of  Perseus are $\sim$1-4 kpc away.  We compared
Fig.~\ref{fig:k_k24} with  data from Blum \etal\ (2006) obtained in
the LMC.  Assuming the LMC is at 44 kpc, and correcting for the
relative distances along the line of sight to Perseus, a typical ABG
star would appear so bright as to be  saturated in the 2MASS survey. 
Only 0.06\% of the distance-adjusted LMC  sample would be unsaturated
and appear in Figure~\ref{fig:k_k24} -- equivalent to just one of the
1141 2MASS sources within the Perseus map area. Larger AGB counts
would be expected if the background population was highly reddened,
but this is unlikely given that only 2\% of the MIPS map  area lies
above the A$_v=$ 10 contour.  We therefore conclude that vast majority
of 297 objects with $K_s-[24]>$2 and \ks$<$14 are likely to be young
stars associated with the Perseus clouds.

As a simple first attempt at classifying the objects found
throughout our map into Class I, flat, Class II, and Class III
objects, we can use the observed \ks$-$[24] colors and assign an
$\alpha$ index following Greene \etal\ (1994), where \ks$-$[24]
values $>$8.31 are Class I objects, those between 6.75 and 8.31
are flat spectrum objects, those between 3.37 and 6.75 are Class
II objects, and those with $<$3.37 are Class III.  Note that the
formal Greene \etal\ classification puts no lower limit on the
colors of Class III objects (thereby including those with SEDs
resembling bare stellar photospheres, and allowing for other
criteria to define youth).  In our case, since we know little
about many of these new objects, we have minimized the contamination
from foreground/background stars by requiring \ks$-$[24]$>$2, and
limited the contamination from background galaxies by requiring
\ks$<$14.  Note that there are likely to be true Perseus members
that are excluded by these (conservative) criteria, but identifying 
them is beyond the scope of this paper.  The number of reported 
Class III objects we derive are certainly lower limits to the true 
values.  Table~4 summarizes the numbers of
the objects that meet these criteria found in each region. 
Comparable fractions ($\sim$7\%) in both of the clusters and the
rest of the cloud have colors consistent with Class I objects. 
A much larger fraction of objects in NGC 1333 (21\%) than IC 348
(only 4\%) have colors consistent with the ``flat''
classification, while only 8\% of the sources in the rest of the
cloud have flat SEDs.  Our results for the frequency of Class I
and flat spectrum SEDs differs from J06, who reports that 47\%
of the Perseus YSO population outside the two clusters fall into
these SED categories.  The different results can be traced to
how YSO candidates are selected.  The MIPS ``rest of cloud" area
covers $\sim$10 square degrees, whereas the IRAC "rest of cloud"
area of J06 is a factor of 3 smaller, and covers a region of
higher extinction.  The $\sim$140 MIPS YSO candidates outside the
clusters are selected from their postion in the $K_s$ vs.
$K_s-[24]$ diagram, implicitly limiting this selection to
objects bright enough to be detected at \ks\ in 2MASS; J06
considers 144 YSO candidates selected using the 2005 c2d SED
classification scheme combining IRAC and MIPS measurements, with
no flux cutoff.  The J06 results therefore extend to less
luminous YSOs than we have considered here.
The MIPS data show that the abundance of {\it bright} Class I and
flat spectrum sources is not higher in the rest of the cloud
versus the NGC 1333 and IC 348 clusters.  The results of J06
suggest that {\it faint} objects of these types are more
abundant in the intercluster region of Perseus.

Just from Figure~\ref{fig:k_k24}, it can be seen that there are
many more Class II objects in IC 348 than in NGC 1333; 85\% of
the candidate young objects have colors consistent with Class II
objects in IC 348, whereas 67\% of the objects in NGC 1333 have
similar Class II colors.  These numbers suggest that NGC 1333 is
younger than IC 348, consistent with expectations that IC 348
should be 1-2 Myr (\eg, Lada \etal\ 2006 and references therein)
and NGC 1333 should be younger at $<$1 Myr (\eg, Wilking \etal\
2004 and references therein).  From Fig.~\ref{fig:k_k24}, in the
``rest of the cloud,'' there are many more stars and
background galaxies than in the clusters.  Somewhat
surprisingly, as was found in J06, there are many YSO candidates
in the rest of the cloud as well; in sheer numbers, there are as
many candidate young objects outside the clusters as in the clusters. 
Using the relative fractions of ``flat'' and Class II object
colors as a proxy for age, objects in the rest of the cloud are
intermediate in age, on average, between IC 348 and NGC 1333.
However, there is also a larger fraction of Class III objects in
the ``rest of the cloud'' (8\%) than there is in either of the
clusters (2-3\%).  These Class III objects are located
essentially randomly across the cloud.    

Of course, for much of this region, we have more information than just
\ks\ and 24 \mum\ measurements, and a more sophisticated
classification scheme can be implemented.  In the clusters, because
these regions have IRAC coverage, essentially all of the YSO
candidates have already been identified and classified (using the 2005
c2d classification scheme as described there) as part of J06;
additional discussion will be provided by S.-P.\ Lai, in preparation. 
In the ``rest of the cloud,'' there are vast regions without 4-band
IRAC data; these include 36 objects with $K_s-[24]$ excesses (e.g.,
with $K_s<$14 and $K_s-[24]>$2) that are additional YSO candidates.  A
few of these are found near the ends of MIPS scan legs where their
photometry is less reliable, and thus could be spurious.  But most are
likely to be true cloud members, especially those clustered  near
L1448 on the far northwest side of the MIPS mosaic.
Figure~\ref{fig:wherek24x} shows where all $\sim$300 sources with
$K_s-[24]$ excesses are located in our mosaic.  Most of these are
projected against the regions of highest extinction covered by the
IRAC data of J06.  

The objects in Fig.~\ref{fig:k_k24} that also have 70 \mum\
detections are generally the brightest objects.  This is
effectively an instrumental artifact in the following sense. 
The intrinsic instrumental point source sensitivity is worse at
70 than at 24 \mum; for our observing strategy, the expected
1-$\sigma$ point source sensitivity is 0.25 mJy at 24 \mum\
and 14.5 mJy at 70 \mum. The faintest object we actually
see in this cloud at 24 \mum\ is 100 times fainter than the
faintest object seen at 70 \mum.  Based on very simple SEDs, we
expect that Class 0 objects through Class III objects could have
a 24:70 \mum\ flux ratio of $\sim$10 to $\sim$0.1.  Thus, only
the brightest cloud member objects will be detected at 70 as
well as 24 \mum.  Conversely, most of the objects detected at 70
\mum\ are bright enough that they are likely cloud members. 
There are three objects detected at 70 \mum\ with $K_s-[24]<$3
whose SEDs through 24 \mum\ resemble photospheres.   They are 
SSTc2d 032740.5+311539 (also known as VSS IX-12 [Vrba \etal\
1976]),  SSTc2d 032807.6+311040 (also known as VSS IX-11 and
IRAS 03250+3100), and SSTc2d 032820.9+294757 (also known as SAO
75942).  VSS IX-11 is in NGC 1333 (a portion with only 2-band
IRAC coverage), and it appears to retain a stellar
(Rayleigh-Jeans) slope through 70 \mum.   The other two objects
(VSS IX-12 and SAO 75942) are from outside the clusters, in
regions with no IRAC coverage, and they both appear to have a
clear excess at 70 \mum, suggesting Class II SEDs.  VSS IX-12 is
extremely faint at 70 \mum; SAO 75942 is a clear detection. 
Additional followup observations (such as MIR spectroscopy) will
be needed required to determine if these are true debris disks
objects.  

Low luminosity cloud members are particularly interesting
objects for study.  Unfortunately the SWIRE observations show
that there is a large population of red background galaxies at
$K_s>$ 14,  and so it is not generally possible to distinguish
faint YSOs from background galaxies using broad-band photometry
alone.  An important exception is the case of the reddest
sources; at $K_s-[24]>$9.7, there are no background galaxies
with 2MASS counterparts. There are 12 such objects
in Perseus (see Table~5); these are potentially the most
embedded and therefore youngest cloud members.  Four are located
in NGC 1333, and two more are in the B1 region.  IRAS 03301+3057
is one of the reddest objects in the entire cloud, turning up as
red by using more than one criterion (see Table 5).  

\subsubsection{The $K_s$ vs.\ $K_s-[70]$ Diagram}
\label{sec:kk70}

\begin{figure*}[tbp]
\epsscale{1}
\plotone{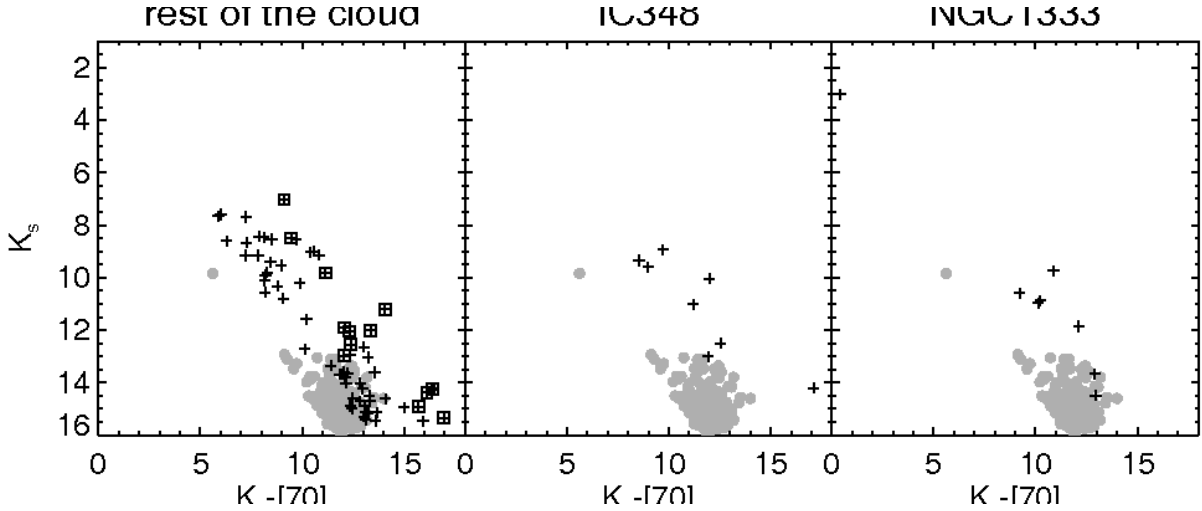}
\caption{$K_s$ vs.\ $K_s-[70]$ color-magnitude diagram for
Perseus (plus signs), with data for the full SWIRE survey (grey
dots) included for comparison.  An additional box around an
object denotes that it was also detected at 160 \mum.  The
fainter clump of objects near $K_s\sim15$ is extragalactic and
the brighter clump ($K_s<12$) is YSO candidates. The reddest
objects in SWIRE are $K_s-[70]\sim14$, all with $K_s>13$; in
Perseus, objects redder than $K_s-[70]\sim14$ are the most
embedded objects;  there are several such candidates for very
embedded objects in the ``rest of the cloud,'' and one in IC
348.   There are YSO candidates in all three Perseus regions.
There is one very bright object at 70 in Perseus that is a
likely a stellar photosphere -- it is in NGC 1333, near
\ks$\sim$3 and \ks$-$[70]$\sim$0.}
\label{fig:k_k70}
\end{figure*}

Figure~\ref{fig:k_k70} shows the $K_s$ vs.\ $K_s-[70]$
color-magnitude diagram for Perseus, with data for the full
SWIRE survey included for comparison.  The morphology of this
parameter space is similar to that of Figure~\ref{fig:k_k24} in
that the survey sensitivity limits create the absence of points
in the lower left, stellar photospheres are in the upper left
(only one very bright Perseus object, in NGC 1333, appears in
this region), and likely galaxies are fainter in \ks.  There are
roughly two clumps of objects in Perseus; the fainter clump is
reasonably well-matched in color and \ks\ to the objects found
in SWIRE.  There are very few such candidate galaxies
in the two clusters because the bright ISM cuts off the
sensitivity at a shallower level than elsewhere in the cloud. 
The SWIRE survey, because it is deeper than the c2d surveys, has
objects that extend bluer in \ks$-$[70] at the faintest \ks\
values than are found in Perseus. However, the reddest objects
in SWIRE at these \ks\ magnitudes levels are $K_s-[70]\sim14$,
all with $K_s>13$.   The reddest objects (redder than
$K_s-[70]\sim14$) are therefore likely to be Perseus cloud
members, and correspond to the most highly embedded  objects. 
Most of these very red objects are distributed outside the two
clusters, with only one in IC 348.  

The brighter clump of objects (with \ks$\sim$10) are also likely
YSO candidates, and there are such candidates in all three
Perseus regions; there is even one in SWIRE, suggesting that
perhaps not all of the brighter objects are guaranteed to be
YSOs.  There is just one very bright object at 70 that is a
likely stellar photosphere -- it is SSTc2d 032807.6+311040 (also
known as VSS IX-11 and IRAS 03250+3100), in NGC 1333.  There is
only one photosphere because of the shallow limits of our
survey; this object, as bright as it is in \ks, is near our
detection limit in 70 \mum.  There are not many objects in
Perseus that are bare photospheres and also bright enough at 70
\mum\ to be detected by our survey.  

We attempted to compare these observed colors with IRAS
measurements of two famous well-known young objects in Taurus
from Kenyon and Hartmann (1995), even though the IRAS and MIPS
bandpasses are not well-matched.  DO Tau is a CTTS with a
$K-[60]$ color of 9.4 mag, and IRAS 04016+2610 is a more
embedded object with a $K-[60]$ color of 13.3 mag.  The brighter
CTTS would be found in the brighter clump of young candidate
objects, were it at the Perseus distance.  The fainter, more
embedded object would appear on the bright side of the clump
that is more consistent with SWIRE galaxy colors.  Therefore,
some of the objects with colors consistent with SWIRE galaxies
could also be candidate YSOs; one cannot simply make a
brightness cut to cleanly distinguish between YSOs and galaxies.

As with the 70 \mum\ detections in Fig.~\ref{fig:k_k24}, here
only the brighter objects are also detected at 160 \mum.  
The effects of the bright nebulosity in the clusters is vividly
apparent, as no objects in this plot in the clusters are also
detected at 160 \mum.

There are nine very red objects with \ks$-$[70]$>$ 15 mag, a
region of color space in which no background SWIRE galaxies are
seen.  They are listed in Table~5.  For some of
these sources, 24 micron fluxes were not available due to
saturation or extended emission.

\subsubsection{The [24] vs.\ [24]$-$[70] Diagram}
\label{sec:242470}

\begin{figure*}[tbp]
\epsscale{1}
\plotone{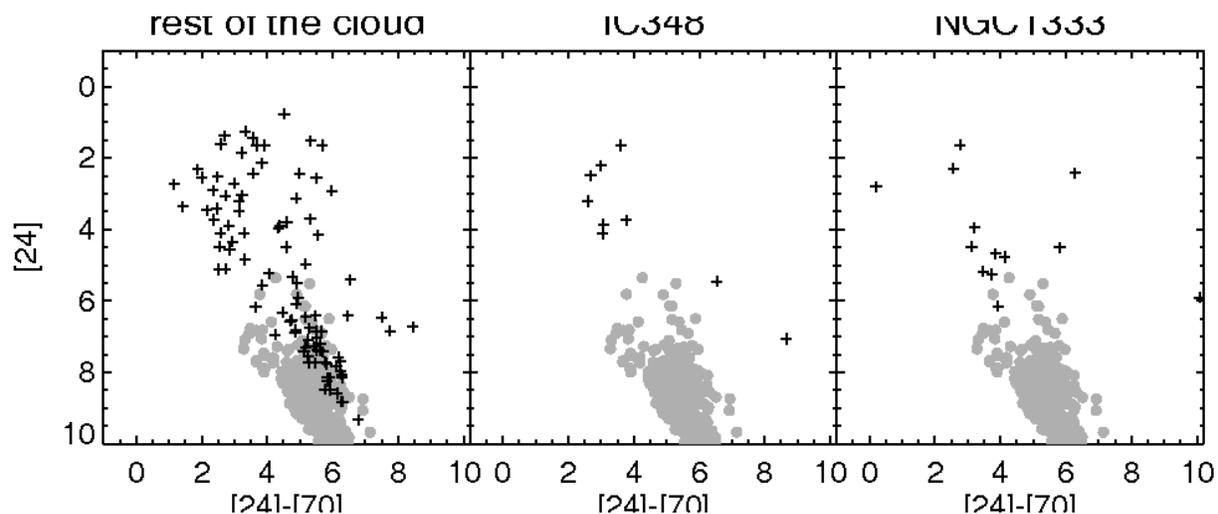}
\caption{[24] vs.\ [24]$-$[70] color-magnitude diagram for
Perseus (plus signs), with data for SWIRE (grey dots) included for
comparison.  As in Figure~\ref{fig:k_k70}, the fainter objects
like those found in SWIRE correspond to extragalactic
([24]$\sim$7-10), and the brighter objects are YSO candidates
([24]$\lesssim$5); there are few faint objects in the clusters 
because the bright nebulosity there limits the survey sensitivity.
Very red sources ([24]$-$[70]$>$6) are the most
embedded objects; note the NGC 1333 object near
[24]$-$[70]=10.1.  The Perseus YSO population in the ``rest of
the cluster'' is more dispersed (towards the upper right) than
in the clusters or lower on the diagram; this suggests young
and/or embedded objects, or there could be a significant
dispersion in distance to the objects.  }
\label{fig:24_70}
\end{figure*}

Figure~\ref{fig:24_70} is an all-MIPS color-magnitude diagram,
presenting [24] vs.\ [70] for objects in Perseus, and again, for
comparison, objects in the full SWIRE survey.   As in
Figure~\ref{fig:k_k70}, (a) the deeper SWIRE survey extends to
bluer objects for the faintest 24 \mum\ values in the Perseus
survey; (b) objects found in Perseus with 24 \mum\ and colors
consistent with those found in SWIRE are potentially galaxies,
though there are clearly objects in Perseus with redder colors
for a given 24 \mum\ flux density than  are found in SWIRE; and
(c) the brighter objects are likely to be  cluster members.  

The Perseus population from the ``rest of the cloud'' in
Fig.~\ref{fig:24_70} does not have two clumps that are as
distinct as those found in Fig.~\ref{fig:k_k70}; it is much more
dispersed, both in the [24] and in the [24]$-$[70] directions. 
It is difficult to make comparisons between the clusters and the
rest of the cloud because there are so few objects available in
the clusters; as before, for IC 348 and NGC 1333, the brightness
of the nebulosity enforces a brightness cutoff such that there
are few faint and red objects detected.  The nebula is bright
enough that it also limits the number of brighter legitimate
cluster members that can be detected in this parameter space. 
Certainly a young and/or embedded population is present in the
rest of the cloud.  A significant distance dispersion could
broaden the distribution of objects seen in the ``rest of the
cloud'' in comparison to the clusters, or the increased number
of objects could simply better represent the true distribution
of young objects.  At the faintest levels, the reddest SWIRE
objects are found near $[24]-[70]\sim$7, so the handful of
objects found to be redder than that are likely to be cloud
members and highly embedded.  

There are five very red objects with [24]$-$[70]$>$7 mag, and
they appear in Table~5.  They are all young, very embedded
objects. SSTc2d 032910.5+311330 is by far the reddest object,
with [24]$-$[70]=10.1; its position is consistent with IRAS 4
A1/A2 in NGC 1333, which is indeed a well-known embedded object
(see, \eg, Reipurth \etal\ 2002, Choi 2005, Rebull \etal\ 2003,
or Mott \& Andre 2001, and references therein).  Although this
object is clearly detected in 70 \mum, the IRAS 4B and 4C
components are nearby and also clearly detected, so the 70 \mum\
flux as measured for IRAS 4A may have imperfectly accounted for
contributions from IRAS 4B and C, affecting the [24]$-$[70] color
as measured.

\subsubsection{The [70]$-$[160] vs.\ [24]$-$[70] Diagram}
\label{sec:2470160}

\begin{figure*}[tbp]
\epsscale{1}
\plotone{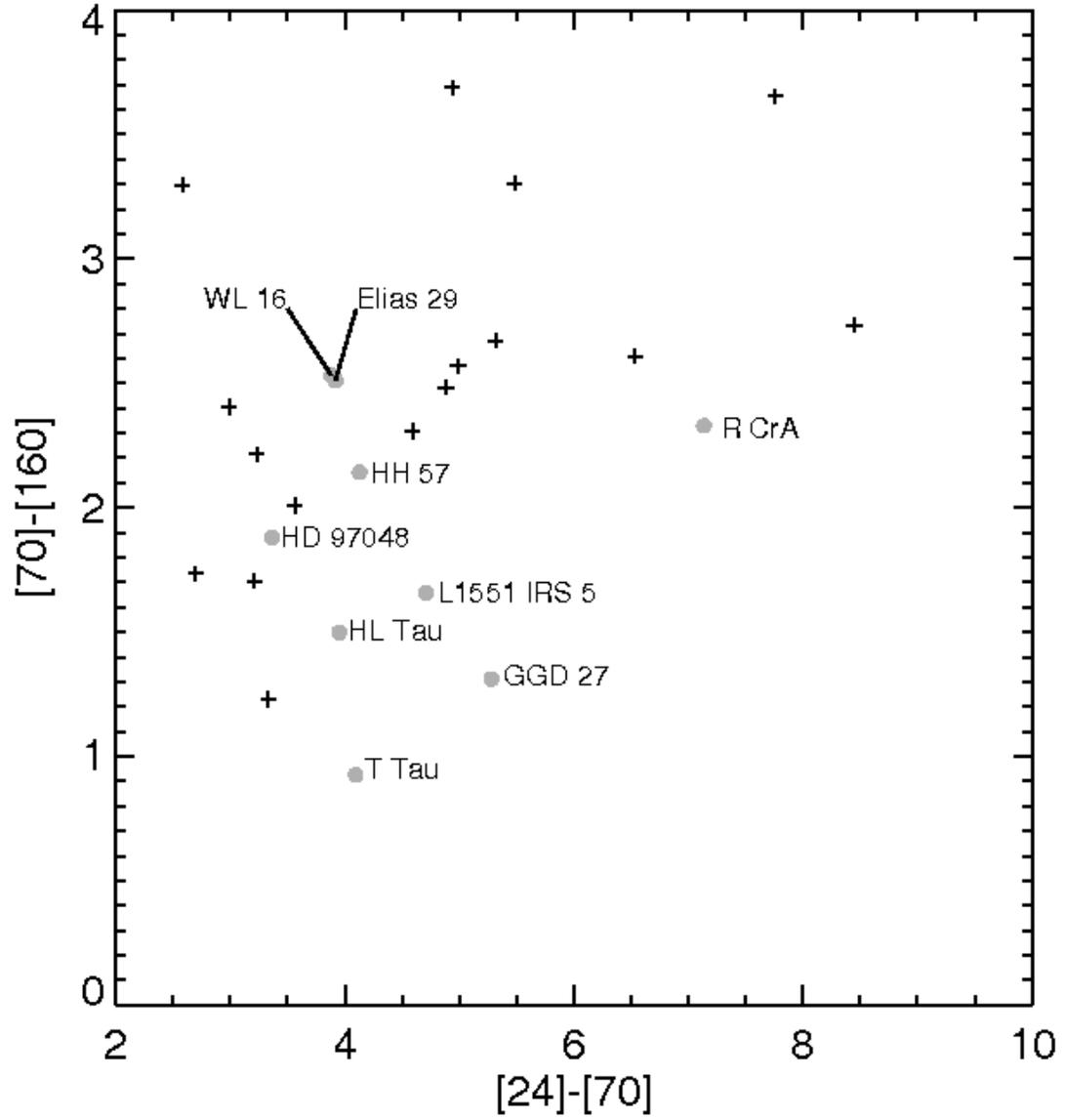}
\caption{MIPS color-color diagram for objects in Perseus
(plus signs) and for a variety of well-studied objects (grey dots,
A.\ Noreiga-Crespo, private communication). The colors of these
objects in Perseus are consistent with young objects. }
\label{fig:2470160}
\end{figure*}

\begin{figure*}[tbp]
\epsscale{1}
\plotone{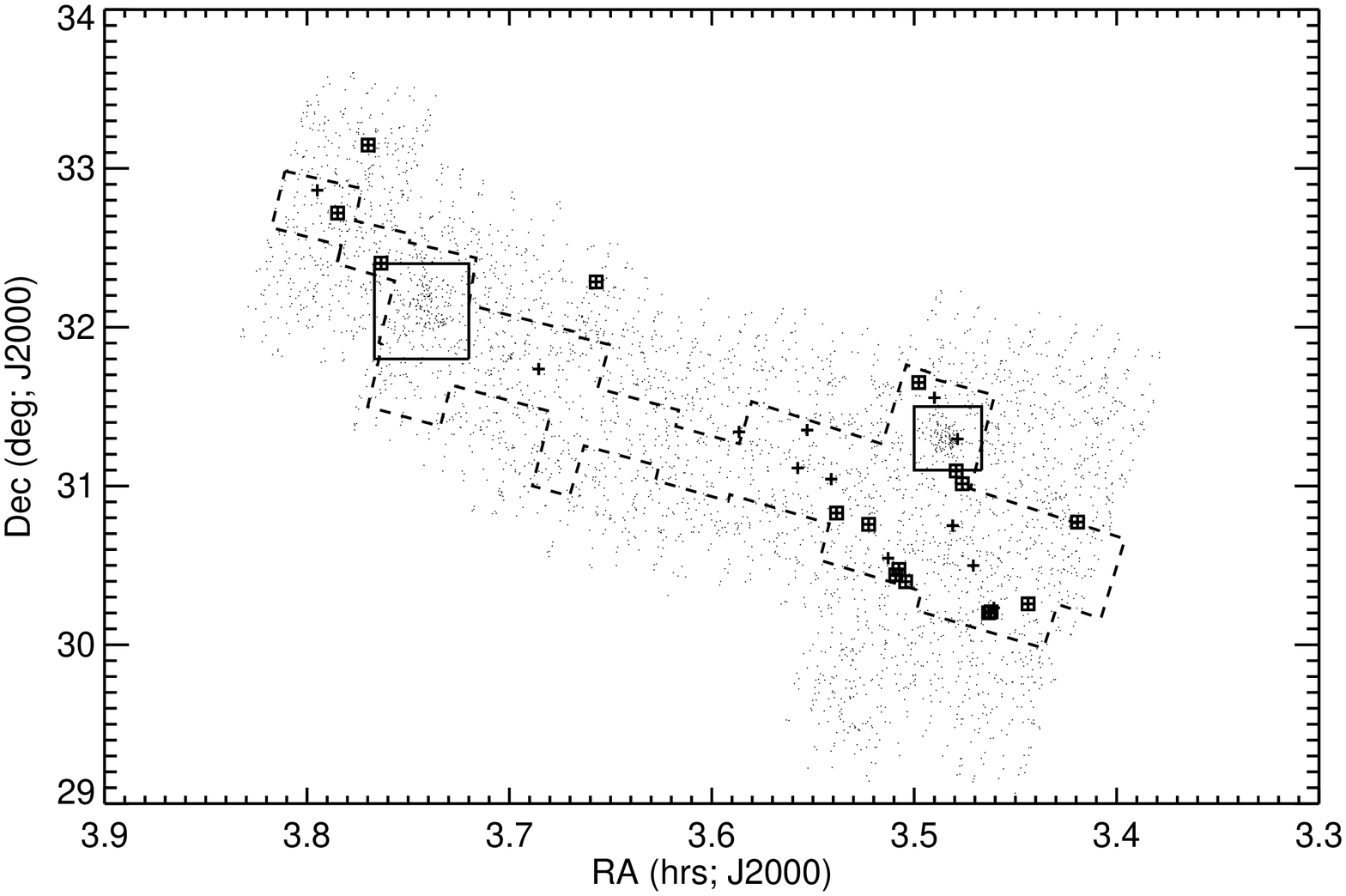}
\caption{Locations of all sources in the 24 \mum\ catalog
(small dots), sources detected at 160 \mum\ (plus signs), and those
detected at all three MIPS bands (additional squares). The rest
of the notation is as in Fig.~\ref{fig:where} and
Fig.~\ref{fig:wherek24x}. Most of the objects are within the
IRAC map, e.g., within the region of highest extinction. Objects
detected at all three MIPS bands are discussed individually in
\S\ref{sec:all3}. }
\label{fig:whereall3}
\end{figure*}

Figure~\ref{fig:2470160} shows the MIPS color-color diagram for
Perseus.  Included for comparison on this plot are colors
calculated for MIPS bands from ISO SWS data for a variety of
very famous well-studied embedded objects from A.\
Noriega-Crespo (private communication).  This parameter space
has not been commonly explored until very recently, in part
because the ``famous'' regions of most clusters have bright
enough nebulosity (as we do in our clusters here) so as to
preclude detection at at least one of the three MIPS bands. In
the case of Perseus, the large area we have covered with our
MIPS map includes many interesting objects in regions beyond the
famous clusters.  All of the Perseus objects in
Figure~\ref{fig:2470160} are from the ``rest of the cloud''
where the nebulosity is low enough to enable measurements at all
three MIPS wavelengths; the colors of these objects, despite
being from the ``rest of the cloud,'' are clearly consistent
with being young stars.  The locations of these objects are
shown in Figure~\ref{fig:whereall3}.  Only one object is
detected at 160 \mum\ within our defined cluster regions (it is
in NGC 1333; see Fig.~\ref{fig:whereall3}), and that object has
no measured (\eg, unsaturated) counterpart in the catalog at
24 and 70 \mum.  We expected that none of these objects are
likely to be extragalactic in that the sensitivity of our 160
\mum\ survey is such that it is really only likely to probe
galactic objects, e.g., cluster members.  However, one object may
be extragalactic; see the discussion in Section~\ref{sec:all3}
below which discusses each of these objects detected at all
three MIPS bands.

There are several objects detected at the longer MIPS bands but
lacking counterparts at the shorter MIPS bands.  These could be
indicative of very embedded objects, but, because our survey
becomes shallower and shallower at longer wavelengths, they are
also often objects where the shorter-wavelength counterpart is
simply saturated.  There are 18 objects with 70 \mum\
measurements but no 24 \mum\ counterparts; four of note appear
in Table~5.  As can be seen in the table, most of these objects
fall into one of two categories: objects that are saturated at
24 \mum\ (10 of these, most of which are in NGC 1333), and
objects that are faint and/or extended at 24 \mum\ (4 objects). 
The last of these objects, LkHa 325, is relatively isolated and
seems to have a 70 \mum\ PSF peak slightly offset from the 24
\mum\ peak; this could be an instrumental effect in that the
detection is on the edge of a scan leg, which could affect the
centering of the detection.

Finally, we looked for objects with 160 \mum\ measurements but no 24
or 70 \mum\ counterparts; there are 8 such objects, and they appear in
Table~5.  Five of these objects are saturated at the shorter
wavelengths.  Two objects are diffuse and the center of their 160
\mum\ emission is clearly offset from the center of a nearby star seen
at the shorter wavelengths.   We suspect that the emission in these
cases arises in a clump of ISM material heated by an adjacent stellar
source.

\subsection{24 $\mu$m Variability}
\label{sec:variability}

\begin{figure*}[tbp]
\epsscale{0.75}
\plotone{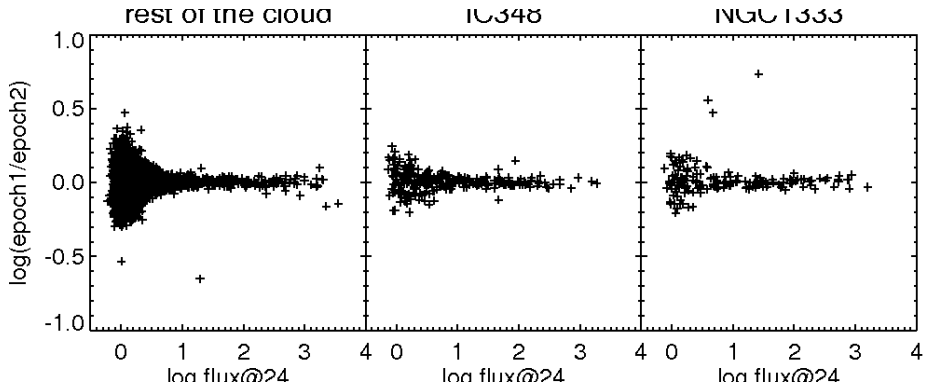}
\caption{The search for time variability at 24 \mum. All of the
outliers are explicable in terms of image defects or extended
sources; see text.  There is no real variability of any of the
24 \mum\ sources in our field on timescales of $\sim$6 hours to
$\sim$10\%.}
\label{fig:timevar}
\end{figure*}

To reject asteroids, two 24 \mum\ observations separated by 3-6
hrs were made for every position in the MIPS Perseus map.  This
dataset provides an opportunity to search for time variability
of cloud members and extragalactic background sources on this
timescale.   Figure~\ref{fig:timevar} shows the ratio of flux
densities measured at the two epochs, as a function of the
combined 24 \mum\ flux density.   The results show an RMS
difference of $\sim$10\% between epochs, consistent with our
expected measurement uncertainties.  Below flux densities of 10
mJy, the detected signal/noise ratio falls below 10, and the
dispersion between the measurements increases accordingly. 
Several objects appear as outliers from this general
distribution.  Investigation of these individually shows that
they suffer from one of several defects: they fall on the edges
of the map (where photometry is less reliable), are contaminated
by artifacts from nearby bright objects (as in the case of NGC
1333), are confused by bright extended emission, or are extended
and thus not measured accurately by our point-source-fitting
photometry.  The apparent variability in all these cases is
therefore of instrumental origin.  We therefore conclude that
none of the $\sim$ 4,000 24 \mum\ sources in Perseus shows real
variability above the 10\% level on timescales of 3-6 hours.

\subsection{Large-Scale Extended Emission}
\label{sec:extemiss}

\begin{figure*}[tbp]
\plotone{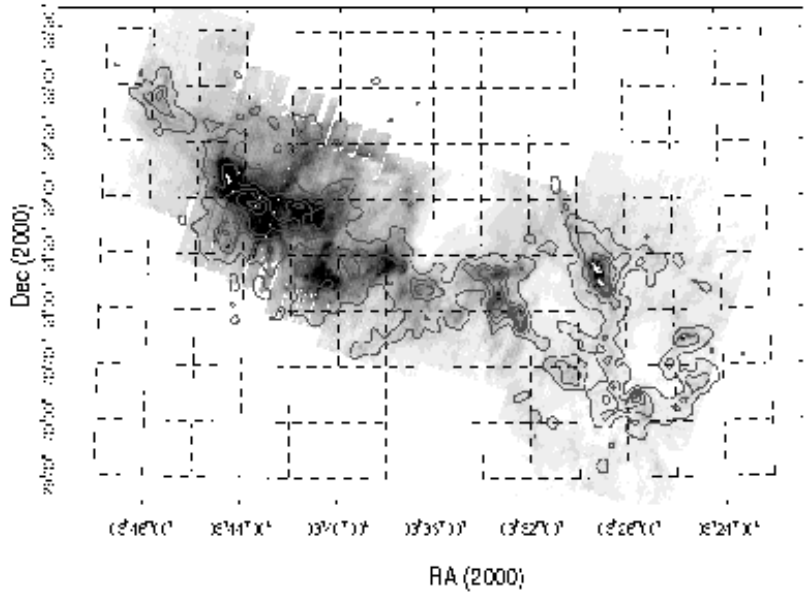}
\caption{Mosaic of MIPS data at 160 \mum\ (greyscale) with \av\
contours overlaid (\av\ levels are 2,4,6,8,10,15,20,25,30,35,40
magnitudes; see J06 and Enoch \etal\ 2006 for more discussion
on \av).  The 160 \mum\ emission closely follows the \av, except
in the large ring illuminated by HD 278942; see text. }
\label{fig:160ext}
\end{figure*}

Bright extended emission is present at all three wavelengths
throughout the MIPS Perseus mosaics
(Figs.~\ref{fig:24mosaic}-\ref{fig:3color}).  While MIPS has much more
sensitive detectors than IRAS had, it also samples much smaller
pixel/beam sizes.  The result is that in our MIPS fast scan
observations, the achieved surface brightness sensitivity is only
incrementally better than that of IRAS, and the majority of the
extended structures we see have direct counterparts in the IRAS maps. 
The major value added by the MIPS observations is a factor of 3-5
better spatial resolution, enabling studies of structural detail in 
the extended emission.  A drawback is that MIPS saturates on the
extended emission in parts of IC 348 and NGC 1333 at 70 and 160 \mum.

At 24 \mum\ (Fig.~\ref{fig:24mosaic}), the brightest extended
emission is seen on the East side of the cloud complex where
several B stars act as illuminating sources.  The largest single
feature is the 1.4$\arcdeg$ diameter IRAS ring illuminated by HD
278942 (Andersson \etal\ 2000; Ridge \etal\ 2006).  Even with
the added resolution supplied by MIPS, the ring edges remain
diffuse -- unlike a swept-up shell.  A prominent 24 \mum\
nebulosity surrounds the central star, and is discussed further
in Sect.~\ref{sec:hd278942}.  A second, much smaller ring
appears within  the larger one.  Its center is offset to the
East by 7$\arcmin$, and it appears to be illuminated by the IRAS
source 03382+3145.  Bright emission in IC 348 takes the form of
a large cavity open to the northwest, again with diffuse
edges.  The peculiar structure in the 24 \mum\ nebulosity at the
center of IC 348 is discussed below (\S\ref{sec:bd+31}).  With
the exception of local nebulosities illuminated by
BD+30$\arcdeg$549 (in NGC 1333) and BD+30$\arcdeg$540 (see
\S\ref{sec:bd+30d540}), there is very little 24 \mum\ extended
emission in the West side of the cloud complex.

At 70 \mum\ (Fig.~\ref{fig:70mosaic}), the extended emission
structure is similar to that seen at 24 \mum.  One important
difference is that the nebulosity around HD 278942 fades
considerably.  Streaking artifacts appear along the scan
direction after a scan crosses over a bright source.

At 160 \mum\ (Fig.~\ref{fig:160mosaic}), however, a wealth of
new structure appears in the extended emission.  On the West
side of the cloud complex, a bright filament appears along the
B1 submm ridge (Enoch \etal\ 2006), and extends southwest to the
Per 6 stellar aggregate (see \S\ref{sec:per6agg}.  Another large
filament of 160 \mum\ emission extends northeast to southwest
across NGC 1333, and continues down towards L1455. There is
extended emission around L1448.  Just South of IC 348, a
East-West region of 160 \mum\ emission becomes prominent,
offset from the 24 \mum\ emission in the direction away from the
illuminating stars.  To appear so prominently at 160 \mum\ while
going largely unseen at 24 and 70 \mum, these regions must trace
cold cloud material.  This is confirmed by a comparison of the
distribution of 160 \mum\ emission to the contours of visual
extinction (Figure~\ref{fig:160ext}; J06 and Enoch \etal\
2006).  The extinction follows the 160 \mum\ emission closely,
except for the large ring illuminated by HD 278942.  The
extended emission around this star, which was so bright at 24
\mum, is not detected at 160 \mum.

Finally, the 160 \mum\ extended emission also highlights the presence
of three large voids on the West side of the Perseus cloud  complex. 
The darkest of these is centered at $\alpha$=03$^h$35$^m$40$^s$,
$\delta$=+31$\arcdeg$50$\arcmin$; two others are present at
$\alpha$=03$^h$31$^m$, $\delta$=+31$\arcdeg$ and
$\alpha$=03$^h$27$^m$10$^s$, $\delta$=+30$\arcdeg$30$\arcmin$ (between
L1455 and L1448).  These voids have characteristic sizes of
$\sim0.7\arcdeg$ (3 pc at a distance of 250 pc).  Smaller voids are also 
present.  The absence of young stars and extended emission in these
voids, and the corresponding concentration of star formation along the
nearby extended emission/extinction filaments, is similar to  what is
seen in the Taurus molecular clouds (Hartmann 2002).

\begin{figure*}[tbp]
\epsscale{0.75}
\plotone{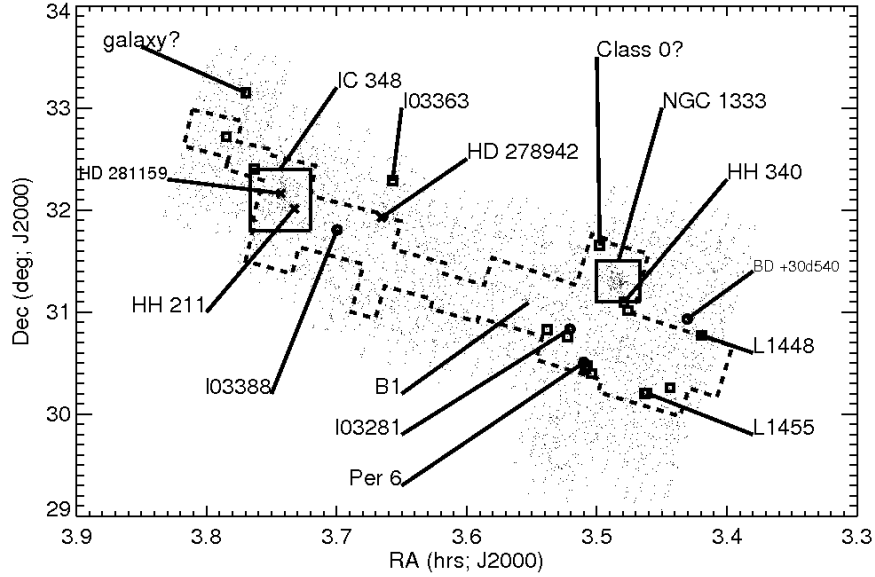}
\caption{Locations of all objects in MIPS-24 catalog (small
dots), and objects discussed in \S\ref{sec:indobj}.  Most of the
rest of the notation is as in previous figures (\ref{fig:where},
\ref{fig:wherek24x}, \ref{fig:whereall3}), but here boxes are
objects detected at all 3 MIPS bands, $\times$ symbols denote
specific individual objects, and circles denote small groups of
objects, all of which are discusssed in \S\ref{sec:indobj}.
(Note that Per 6, which has a circle because it is a small
group, also has several objects detected at all 3 bands and thus
plotted as squares.) Object names are given in most cases; to
conserve space, IRAS sources are abbreviated ``I'' followed by
the first 5 digits of the formal name.}
\label{fig:wherehilite}
\end{figure*}

\clearpage

\section{Results \& Discussion for Individual Regions \& Objects}
\label{sec:indobj}

This large map is incredibly rich in interesting objects and
features.  In this section, we select several objects or small groups
of objects to discuss in more detail.  Fig.~\ref{fig:wherehilite}
shows the locations of these specific objects within the overall
Perseus cloud. Table~6 lists the MIPS flux densities for the point
sources  discussed here (IRAC flux densities are available from the
online c2d delivered catalogs).  \av\ estimates for individual stars
were made  using the method described in Evans \etal\ (2005).  The
SEDs presented below  are all in $\lambda F_{\lambda}$ in cgs units
(erg s$^{-1}$ cm$^{-2}$), against $\lambda$ in microns.  In these
SEDs, we have included the 2MASS and IRAC data from J06, but have not
comprehensively included every other flux measurement from the
literature of these objects, if relevant; some objects presented here
are entirely new.  Some objects whose SEDs are presented here were
mentioned in J06,  but their SEDs were deferred for presentation until
this paper.

\subsection{Spectral Energy Distributions in Young Stellar Aggregates}
\label{sec:agg}

The term ``stellar aggregates'' was originally coined by Strom \etal\
(1993)  in their near-infrared study of the young stellar population
of the Orion  L1641 cloud.  They defined them as ``regions of enhanced
stellar surface  density'' --  ``sparse groups \ldots likely to be
gravitationally unbound,  transient structures.''  They applied the
term to isolated stellar groups  $<$1 pc in size (14$\arcmin$ at the
assumed distance of Perseus) and with as  many as several dozen
members.  Their underlying assumption was that the  aggregate members
were born contemporaneously (i.e., are coeval) from a  single parent
molecular cloud core.  

Throughout the Perseus map, and outside the two large clusters, there
are  several small groups of 24 \mum\ sources that fit the above
definition of a stellar aggregate.  In some of them, the stars are
distributed in small linear strings or arcs, similar to those seen in
NGC 2264 (Teixeira \etal\ 2006).  The distribution of stars often
follows structure  in the millimeter continuum and 160 \mum\ emission
(see Fig.~\ref{fig:3color}). In this section, we discuss three new
aggregates of stars, two of which  appeared to the IRAS survey as
single objects.

\subsubsection{Per 6: A New Aggregate Between L1455 and B1}
\label{sec:per6agg}

\begin{figure*}[tbp]
\plotone{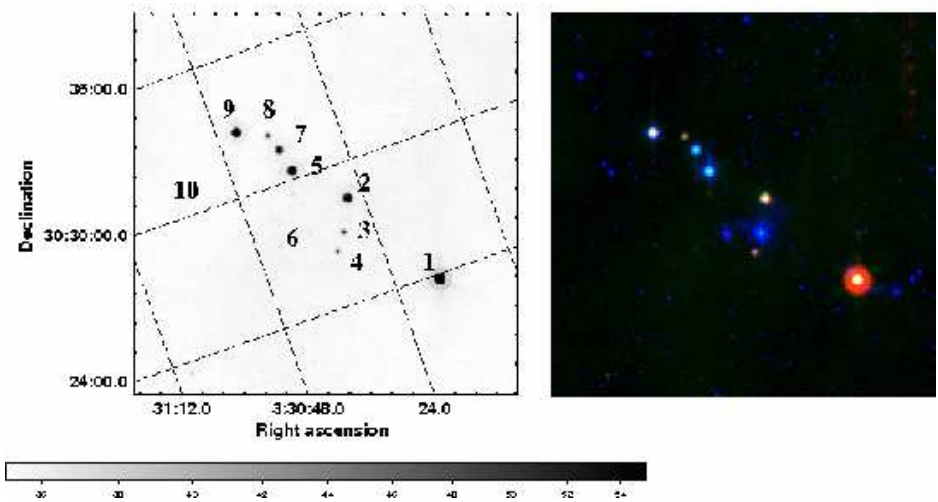}
\caption{Per 6, a new aggregate of objects bright at MIPS
wavelengths. All of the 
objects are within 11$\times$2.5$\arcmin$. LEFT: greyscale of 24
\mum\ image with sources numbered to correspond to discussion in
text.  RIGHT: three-color image
with IRAC 4.5 \mum\ (blue), MIPS 24 \mum\ (green), and MIPS 70
\mum\ (red), all log scale.  Note extended emission around source
\#1 in IRAC. }
\label{fig:wherenew}
\end{figure*}

\begin{figure*}[tbp]
\includegraphics[angle=90,scale=.8]{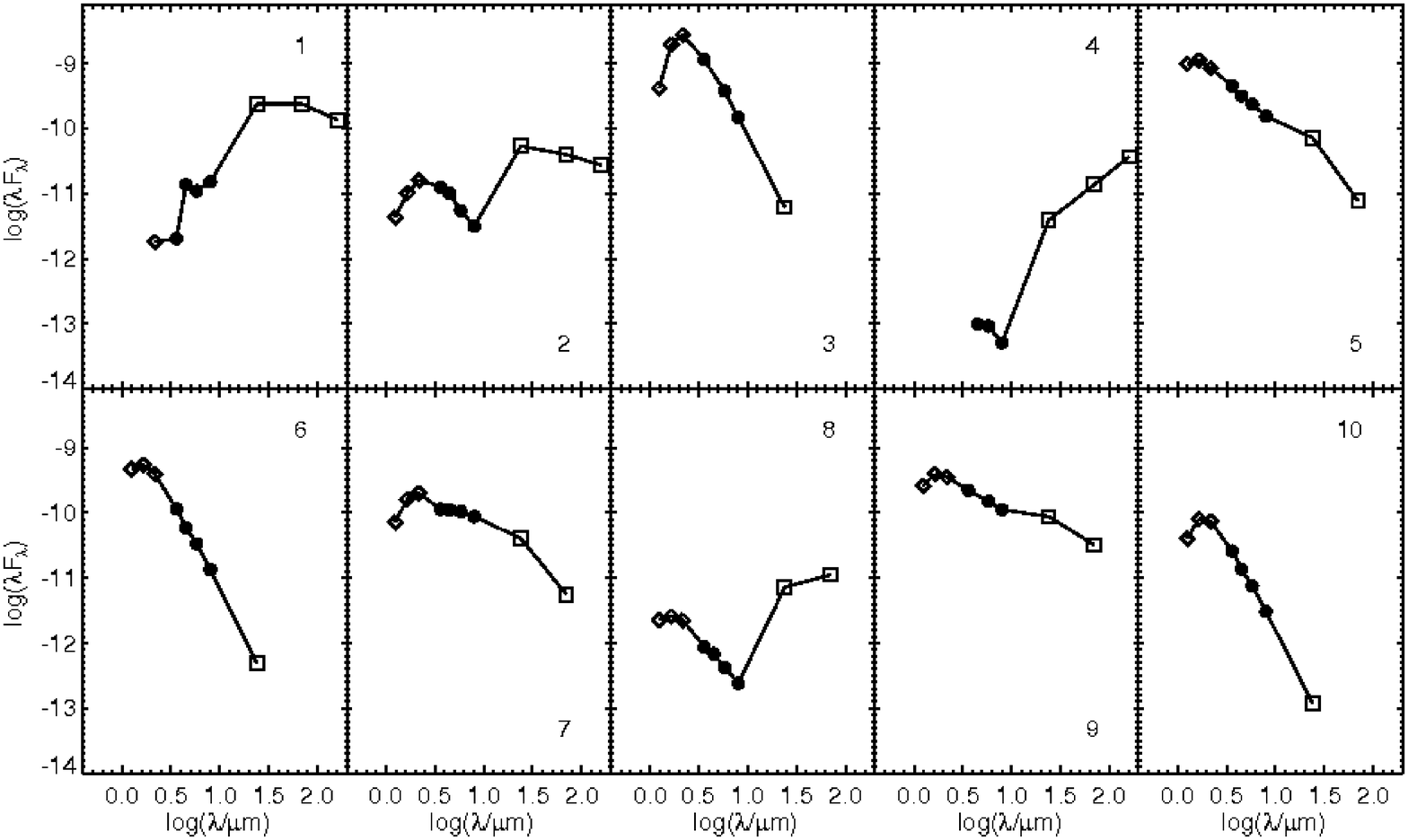}
\caption{Spectral energy
distributions (SEDs) for the 10 components of the new Per 6
aggregate.  Diamonds are 2MASS detections, circles are IRAC
detections, and squares are MIPS detections.  The units of 
$\lambda F_{\lambda}$ are cgs (erg s$^{-1}$ cm$^{-2}$); it is
plotted against $\lambda$ in microns.  The objects here have
diverse SEDs; some objects (like 6) are photospheres, some
objects (like 2) clearly have circumstellar disks, and some objects
(like 1) are deeply embedded. }
\label{fig:sedper6}
\end{figure*}

There is a ridge of molecular gas that extends southwest from B1
towards L1455.  Roughly in between B1 and L1455, Ladd \etal\
(1994) find an ammonia core they dubbed Per 6.  Olmi \etal\
(2005) find here two cores, an N$_2$H$^+$ core, and, slightly
offset, a CS core.  Hatchell \etal\ (2005) find an 850 \mum\
core here (\# 81); Enoch \etal\ (2006) find 3 millimeter
continuum cores in this region (Bolo 60, 61, 62).  Based on
sub-millimeter continuum and NIR extinction maps, in between
Enoch's Bolo 61 and 62, Kirk \etal\ (2006) find what they dub an
``extinction core'' (their \#25) and an ``extinction super
core'' (their \#6).  The Per 6 region also contains 4 IRAS
sources, IRAS 03271+3013, IRAS 03273+3018, IRAS 03275+3020, and
IRAS 03276+3022.

In this Per 6 region, there is a grouping of 10 objects detected
at MIPS wavelengths, 8 of which are bright at 24 \mum, 7 of
which are detected at 70 \mum, and 3 of which are detected at
160 \mum.  Figure~\ref{fig:wherenew} contains a 24 \mum\ image
and a key to object numbers for this discussion, as well as a
three-color image using 4.5, 24, and 70 \mum.  SEDs for all of
these Per 6 components can be found in
Figure~\ref{fig:sedper6}.  There is a diversity of SED types
present here, ranging from apparent photospheres, to Class 0
candidates.  Additionally, there are 8 objects in this region
with similar SEDs, which are very faint at all available bands
and undetected by 2MASS; these objects are most likely
background galaxies, so they are not included in the subsequent
discussion here.  

All of the bright objects are located within a region $11
\arcmin \times 2.5 \arcmin$ (0.8 $\times$ 0.2 pc at a distance
of 250 pc). The positions of the Ladd \etal,  Olmi \etal, and
Hatchell \etal\ cores are coincident with source \#1, the
brightest 24 \mum\ source in the grouping, so we will refer to
this new aggregate of young objects as ``near Per 6,'' or simply
``Per 6.''  

Enoch \etal\ (2006) found three cores here, one
of which, Bolo 60, is coincident with source \#1.  Core Bolo 62
is coincident with source \#4, and the core Bolo 61 is not
coincident with a 24 \mum\ source; it is about an arcminute
southwest of \#2 and about an arcminute west of \#3. Texture in
the mm map from Enoch \etal\ reveals additional structure in the
extended dust continuum emission that threads between sources 2
and 5, curving back around to \#9; a similar structure can be
seen in the 160 \mum\ emission.  

The brightest object at 24 \mum, object \#1, can be identified
with IRAS 03271+3013, which was seen in Aspin (1992) as a
bipolar outflow source, and listed in Ladd \etal\ (1993) as a
young object.  It is a point source in all three MIPS
wavelengths, but has ``cometary'' morphology in IRAC, such that
there is diffuse emission coming off of one side of the IRAC
source (see Fig.~\ref{fig:wherenew}).  J06 finds that deeply
embedded objects often have extended emission in 1 or more IRAC
bands, lending support to the idea that this object is deeply
embedded.  This object is bright enough to be detected at all 3
MIPS bands; see Figure~\ref{fig:sedper6}.  Most of the energy is
being emitted at the longest wavelengths, consistent with an
embedded very young object.  It is one of the reddest dozen
objects seen in Fig.~\ref{fig:k_k24} with \ks$-$[24]$>$9.7 mag,
and also one of the reddest objects seen in Fig.~\ref{fig:k_k70}
with \ks$-$[70]$>$15 mag.  It does not have a smooth SED at IRAC
wavelengths, possibly because the image is slightly extended,
which leads to errors in our PSF-fitting photometry.  As 
mentioned above, the position for this object is consistent with
cores measured at long wavelengths such as the millimeter
continuum observations in Enoch  \etal\ (2006). 

The position for IRAS 03273+3018 is consistent with both the
position given in the literature for HH 369 and with our Spitzer
detection \#2.  The position given for HH 369 (Alten \etal\
1997) is 4$\arcsec$ north of the MIPS position. The object found
at this location is a point source at MIPS wavelengths, but
slightly extended in IRAC-2, as can be seen in
Figure~\ref{fig:wherenew}.  The position given for HH 369 is in
the direction of a slight extension of the IRAC image, but still
within the extent of the object as seen in
Figure~\ref{fig:wherenew}.  The SED for this object, as can be
seen in Figure~\ref{fig:sedper6}, includes all three MIPS bands,
and has most of the energy being emitted at the longer
wavelengths.  The shape of the SED at the long wavelength end 
is consistent with being composed of photosphere plus strong
long wavelength excess, consistent with a YSO seen only via
scattered light at short wavelengths.  Source \#2 is thus
presumably the protostar driving the HH 369 outflow.  

Object \#4 has no known literature counterpart.  It is the third
(of 3) objects in this aggregate detected at all three MIPS
bands.  It has a very steep SED (Fig.~\ref{fig:sedper6}), with
only marginal detections at the three longest IRAC bands, and no
IRAC-1 or 2MASS counterparts at all.  This SED resembles that
found for IRAS 03282+3035, discussed below in \S\ref{sec:all3}
and in J06 as an outflow-driving source, except that this SED
does not level off by 160 \mum.  This is evidently a deeply
embedded object.  The similarities between the morphology at
IRAC bands and the SED of this object and IRAS 03282+3035
suggests that this new object may also be a Class 0 object.  

Object \#5 can be identified with IRAS 03275+3020, which is
optically visible as GSC 02342-00390 and the ASCA X-ray source
AX J0330.5+3030 (Yamauchi \etal\ 2001).  Yan \etal\ (1998)
identify this object as a Class I object; Yamauchi \etal\
identify their X-ray detection with the IRAS source and classify
it as a T~Tauri object.  The SED that can be seen in
Fig.~\ref{fig:sedper6} suggests that it is certainly a Class II
object.  

Object \#9 is consistent with the position for IRAS 03276+3022,
which appears in Ladd \etal\ (1993) as a candidate young
object.  This same object is identified with LkHa 326 and HBC 14
(Casali \& Eiroa 1996), an emission line star in Liu \etal\
(1980), and even a candidate AGN (de Grijp \etal\ 1987). Based
on the SED seen in Fig.~\ref{fig:sedper6}, we suggest it is an
embedded young object, perhaps a Class II. 

Objects \#7 and 8 also have no known literature counterparts.
Object 7 has an SED consistent with that for a young
and/or embedded Class II object.  Object 8 has an SED where
most of the energy is emitted at the longer wavelengths,
suggesting a classification similar to object \#2 in this same
aggregate.  

Objects \#3, 6, and 10 are all consistent with photospheres, with no
known counterparts in the literature.  They are distributed along the
ridge of 160 \mum\ emission, suggesting that they are  Perseus
members.  The Wainscoat (1992) galactic starcount models predict  a
70\% chance that a background source with a 24 \mum\ flux density as 
faint as object \#10 should randomly fall within the area of the Per
6  aggregate.  While this object is probably not a Perseus member, the
same  analysis gives probabilities of just 20\% and 3\% that objects
\#6 and 3  are part of the background galactic population.  Object \#3
has the highest  extinction of these three, with an estimated \av\ of
15 mag; object \#6 has  \av\ of about 4 mag; and object \#10, 9 mag.


There are at least two different possible methods of
classification of the objects in this aggregate. Based on the
\ks$-$[24] color, three objects (\#1,2,8) are Class I, three
objects (\#5,7,9) are Class II, and three objects (\#3,6,10) are
Class III.  The remaining object (\#4) is not detected at \ks. 
Alternatively, we can classify objects based on a value of
$\alpha$ fit to the SED between \ks\ and 24 \mum, following the
original Greene \etal\ (1994) classification scheme, where
$\alpha \geq$0.3 is Class I, $-0.3 \leq \alpha <0.3$ is ``flat
spectrum,''  $-1.6 < \alpha < -0.4$ is Class II, and $\alpha <
-1.6$ is Class III. Using this classification, four of the
objects (\#1,2,4,8) are Class I, none are ``flat spectrum''
objects, three objects (\#5,7,9) are Class II, and three objects
(\#3,6,10) are Class III.  The objects with available
classifications based on \ks$-$[24] color have the same
classification using an $\alpha$ fit to the entire available SED. 

This aggregate, while it spans $\sim$0.2 pc in the narrow
dimension, is $\sim$0.8 pc in the long direction, larger than
the average core size of $\sim$0.08 pc found by Enoch \etal\
(2006).  Objects 2, 3, 4, and 6 and a millimeter continuum core
without a 24 \mum\ counterpart from Enoch \etal\ (2006) are all
within $\sim2.3\arcmin$ ($\sim$0.2 pc); even this subgroup has a
diversity of SEDs.  An apparently starless millimeter continuum
core and two of the most embedded objects in this aggregate are
found in close proximity to two of the least embedded.

\subsubsection{IRAS 03388+3139 stellar aggregate}
\label{sec:iras1}

\begin{figure*}[tbp]
\plotone{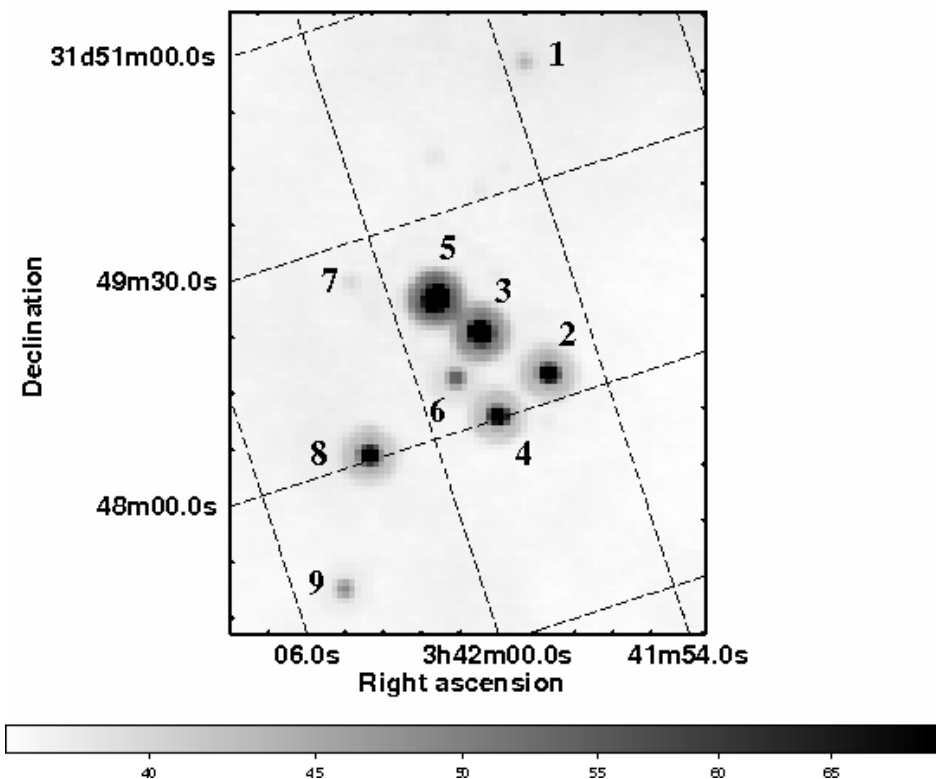}
\caption{IRAS 03388+3139 resolves into
several pieces within 90$\arcsec$ (diameter) when viewed by
MIPS-24 (greyscale).  The numbers correspond to the numbers used
in the next figure, which presents SEDs.}
\label{fig:where03388}
\end{figure*}

\begin{figure*}[tbp]
\plotone{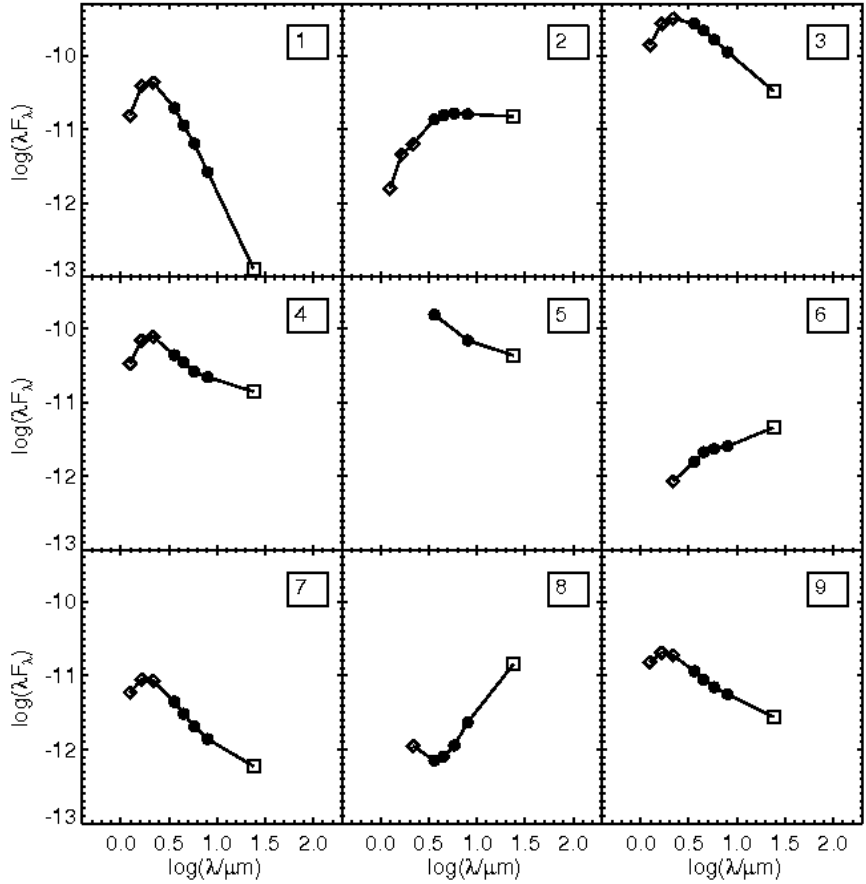}
\caption{SEDs for the components of IRAS 03388+3139 within 90
$\arcsec$ diameter. The numbers correspond to objects identified
in the previous Figure.  All other notation is as in previous SED
figures. There are a variety of SEDs found here, ranging from
photospheric (object 1) to more embedded (object 2). }
\label{fig:sed03388}
\end{figure*}


IRAS 03388+3139 was seen by IRAS as a single object, but MIPS
reveals that there are actually at least 9 objects seen at 24
\mum\ here within a region 90$\arcsec$ in diameter (0.1 pc at a 
distance of 250 pc); see Fig.~\ref{fig:where03388}. The location of the
original IRAS source is between sources \#3 and 5 in this finder
chart; the positional uncertainty encompasses both objects. This
object also appeared as seen by MSX in Kraemer \etal\ (2003; it
is listed as G160.2784-18.4216  and J034158.34+314852.6).  There
are no cores found by Enoch \etal\ (2006) or Hatchell \etal\
(2005) in this region.

As in Per 6, this aggregate of objects displays a striking variety of
SEDs within this small region; see Figure~\ref{fig:sed03388}.  Some
objects resemble photospheres, most with long-wavelength excesses (1,
3, 4, 7, 8, and perhaps 5, if we had unsaturated \jhk), and some (2,
6, 8) are clearly more deeply embedded.  None of the objects is
detected at 70 or 160 \mum.  Despite this, object 8 is one of the
reddest dozen objects in Fig.~\ref{fig:k_k24} with \ks$-$[24]$>$9.7
mag.  Object \#1 is the faintest of the group, with a nearly
photospheric SED  and an inferred \av\ $\sim$11 mag.  The Wainscoat
(1992) models assign a 16\% probability that a background source with
the 24 \mum\ flux density of Object \#1 would be found within the
$4\arcmin\times2\arcmin$ area of the aggregate.

As above, there are at least two different ways to classify
these objects.  A \ks$-$[24] color is available for all but one
of the objects; three objects (\#2,6,8) are Class I, four
objects (\#3,4,7,9) are Class II, and the last (\#1) is Class
III.  Based on a fit to the SED between \ks\ and 24 \mum,
similar classifications are obtained (only \#2 changes
classification) -- two Class Is (\#6,8), one flat spectrum
(\#2), five Class IIs (\#3,4,5,7,9), and one Class III (\#1).

\subsubsection{IRAS 03281+3039 stellar aggregate}
\label{sec:iras2}


\begin{figure*}[tbp]
\plotone{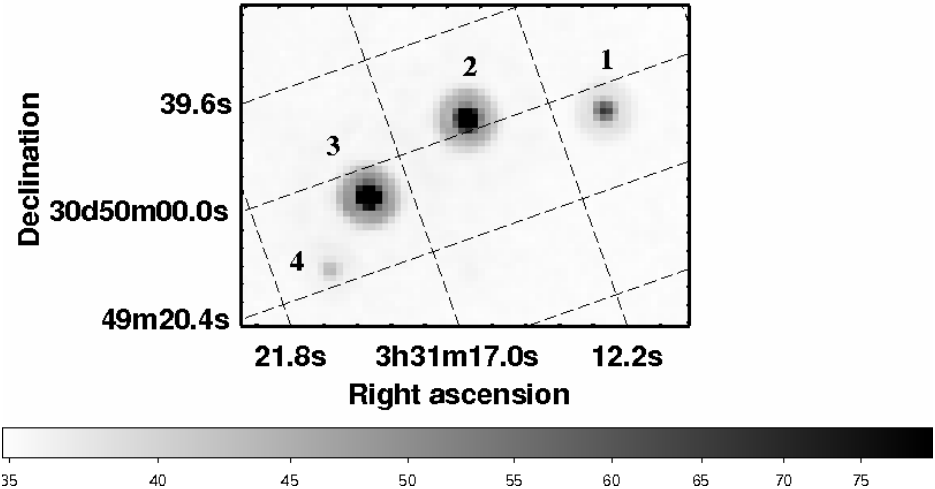}
\caption{IRAS 03281+3039 resolves into
several pieces within 120$\arcsec$ (diameter) when viewed by
MIPS-24 (greyscale).  The numbers correspond to the numbers used
in the next figure, which presents SEDs. }
\label{fig:where03281}
\end{figure*}

\begin{figure*}[tbp]
\plotone{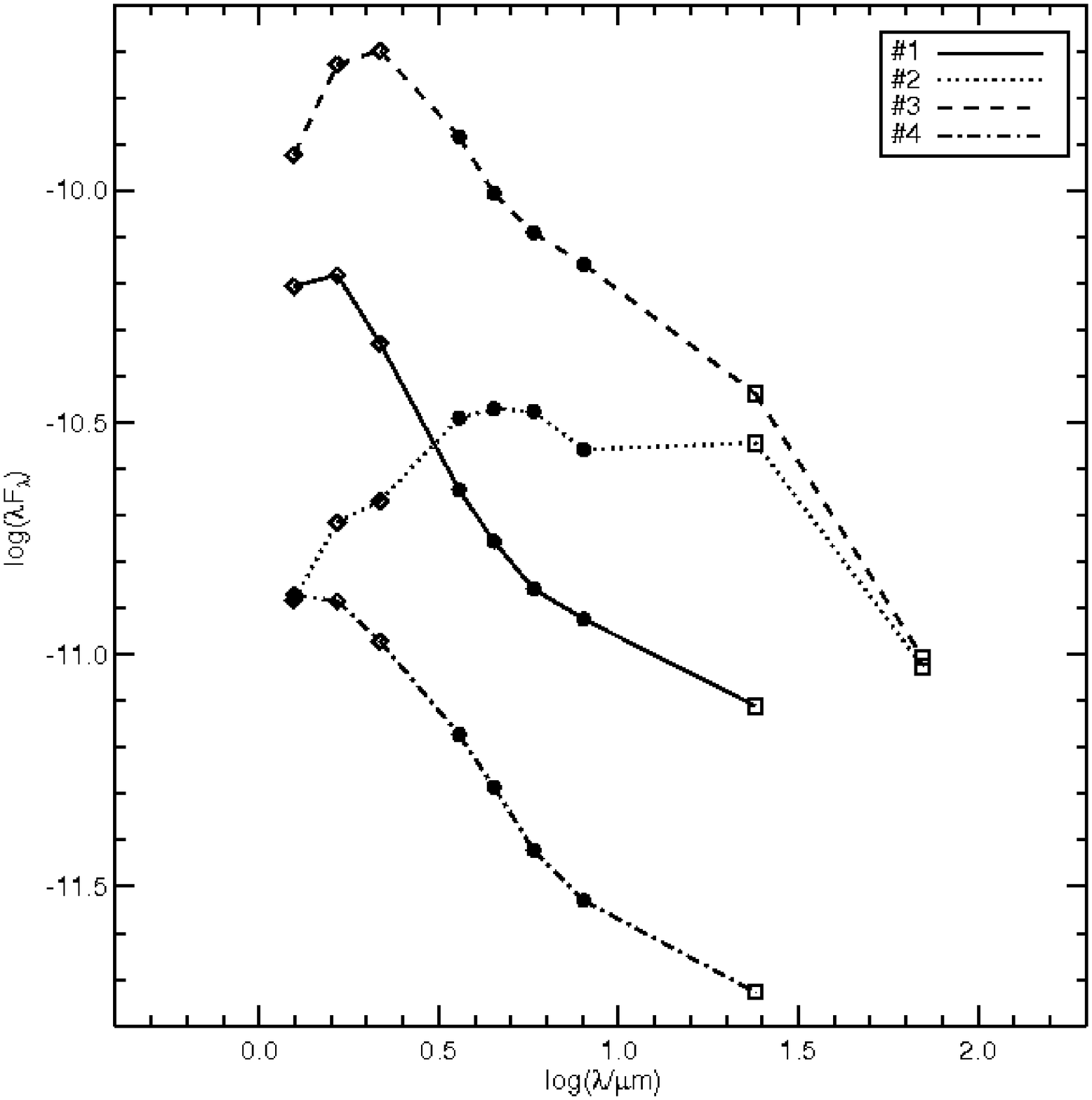}
\caption{SEDs for the components of IRAS 03281+3039 within
120$\arcsec$ diameter.  The numbers correspond to objects
identified in the previous Figure.  Notation is as in previous
SED plots. In this aggregate, 3 of the 4 objects are close to
photospheric; only one appears to have a substantial
long-wavelength excess.}
\label{fig:sed03281}
\end{figure*}

IRAS 03281+3039 is another object that, when viewed by MIPS,
is resolved into multiple sources; see Figure~\ref{fig:where03281}.  
(This object was also mentioned by J06; this aggregate of stars is
located along the ridge of molecular emission between B1 and the
Per 6 cluster discussed above.)  There are four components to
this object seen at MIPS-24 (2 of which are also seen at 70
\mum) within 120$\arcsec$ diameter, or 0.15 pc at a distance of
250 pc.  The IRAS position is located at a position equidistant
from the three brightest objects seen in
Figure~\ref{fig:where03281}, with a position uncertainty nearly
reaching across from source 1 to 3.  There are no cores found by
Enoch \etal\ (2006) or Hatchell \etal\ (2005) in this region.

As mentioned in J06, de Grijp \etal\ (1987) lists this as an AGN
candidate, but we suspect, based on the SEDs seen in
Figure~\ref{fig:sed03281}, that it is instead a small grouping
of young stars.  Objects 1, 3, and 4 look like photospheres plus
circumstellar dust; object 2 seems to be the most embedded.  As
above, there are several different ways to classify these
objects.  The classifications based on \ks$-$[24] and on a fit
to the SED between \ks\ and 24 \mum\  produces identical
classifications; all are Class II, except \#2, which is a flat
spectrum object. 


Objects 1 and 3 can be identified with emission line stars in
Liu \etal\ (1980). HH 770-772 can be seen in this vicinity in
the IRAC-2 image, suggesting recent jet activity in this area.
On the whole, these objects appear to be less embedded than the
objects from IRAS 03388+3139 or Per 6 above.  The inferred \av\
for these objects ranges from $\sim$6 to $\sim$12 magnitudes.

\subsubsection{Comments on the aggregates}

Three new aggregates of young stars have been discussed above.  In all
three cases, follow-up data will be required to better define their
SEDs, establish their multiplicity, and measure their spectral types
and presence of accretion diagnostics.  However, the facts that most
of the sources are bright at 24 \mum, that they have high inferred
extinctions, and that they have infrared excesses reminiscent  of
known YSOs suggest that they are indeed YSOs.  They are all in close
clumpings (within $\sim$0.1 pc\footnote{IRAS 03388+3139 is $\sim$0.1
pc, IRAS 03281+3039 is $\sim$0.15 pc, and Per 6 is $\sim0.2\times0.8$
pc, larger than $\sim$0.1 pc, but there is diversity of SEDs found
even within a subgroup of size $\sim$0.2 pc.  Therefore, we are taking
their typical size to be on the order of $\sim$0.1 pc, not $\sim$1
pc.}); these and other similar groupings can be seen in the full
Perseus map (Fig.~\ref{fig:24mosaic}), usually following the
filamentary structure found in the 160 \mum\ emission
(Fig.~\ref{fig:3color}).   A similar spatial relationshop between
strings of young stars and molecular cloud filaments has also been
seen in NGC 2264 (Teixeira \etal\ 2006).  The typical nearest-neighbor
distances within the aggregates are 0.08, 0.04, and 0.05 pc for Per 6,
IRAS 0338+3139, and IRAS 03281+3039 (respectively) - very similar to
the 0.08 pc spacing found in NGC 2264 ``spokes cluster'' which
Teixeira \etal\ (2006) identify with the Jeans length for thermal
fragmentation of the molecular cloud filaments.

Each of the Perseus aggregates contains objects in close physical
proximity  and with a wide diversity of SEDs.  In this, they are
reminiscent of the  L1228 South aggregate identified in {\it Spitzer}
data by Padgett \etal\ (2004), where highly embedded objects, objects
with substantial disks,  objects with debris disks, and photospheres
were all found in an aggregate  of nine bright 24 \mum\ sources within
a region of $d<$ 1 pc.  The potentially  youngest object found in the
Perseus aggregates is a cold millimeter continuum  core from Enoch
\etal\ (2006).  The earliest class of object seen at 24 \mum\ is Class
I.  The latest class found in these aggregates is Class III.  This
wide range of circumstellar environments present within a compact (and
presmably related) group of young stars raises interesting questions
about star formation and circumstellar disk evolution.

There are two possible formation scenarios that could account for the
diversity of SEDs in these aggregates.  The first is that  every
source spends a relatively fixed fraction of its life in each of the
class 0, I, II, and III phases of protostellar evolution.  In this
scenario, the observed diversity of SEDs would require a significant
age spread among the members of the aggregates.  The second
possibility is that the aggregate members formed at roughly the same
time (i.e., are coeval), and that the process of circumstellar
evolution proceeds at very different rates in different objects -- such
that some members to evolve  into class III sources over the same
timescale at which others remain class I objects.

In the first scenario, if the aggregate members objects all started
with  similar initial conditions of mass, angular momentum, and
initial cloud  temperature (which is not unreasonable to assume, given
their physical  proximity), then to have both Class I objects and
Class III objects in  the same region requires some stars to be
$\lesssim0.1$ Myr old and some to be 1-5 Myr or possibly as much as 10
Myr old, assuming the the canonical ages for objects in these SED
classes (\eg, Bachiller 1996).  If the Class III objects are not
truly associated, leaving us with the Class IIs, then the oldest ages 
of objects in these aggregates are $\gtrsim1-5$ Myr old, taking the
longest  disk lifetimes to be between 1 and 5 Myr (see, \eg, Cieza
\etal\ 2006, Rebull \etal\ 2004, and references therein).  

On the other hand, if all of the objects were formed at very
nearly the same time, which is also not unreasonable to assume
based on their physical proximity, then there is a real
diversity of evolutionary timescales for the various classes. A
Class I object is often assumed to be much younger than a Class
III object, and the timescale for the Class I to Class III
transition is usually taken to be long compared to the timescale
for the transition from Class 0 to Class I.  In order for these
clumpings of stars to be the same age, the evolution of
circumstellar matter around some objects must be fast, with the
timescale to evolve from Class 0 to Class I comparable to the
shortest possible time to evolve from Class 0 to Class III.
Influences such as the system inclination, local stellar density
and interactions between systems, initial disk mass (\eg,
Beckwith \etal\ 1990), mass accretion rate and history (\eg,
Calvet \etal\ 2005), initial core/clump and stellar
rotation rate, and/or close binaries could all affect the
disk/envelope dissipation timescale.  An obvious follow-on
investigation is an assessment of stellar multiplicity in the
aggregates -- if the class III objects are preferentially binary,
this could explain their more rapid circumstellar evolution.

Other authors are beginning to suggest that some stars can
lose their disks quickly (\eg, Silverstone \etal\ 2006). 
Recent work (\eg, McCabe \etal\ 2006, Prato \etal\ 2003 and
references therein) have found PMS binary components in
different states of disk evolution (such as a Class II paired
with a Class III), and in those cases, the evidence for the
stars being exactly the same age is much stronger than the
circumstantial evidence for the Perseus aggregates.  It is
therefore plausible that the aggregate members could be coeval 
and also display a wide range of disk and envelope properties.  

In any case, the observational evidence we present here is that
objects of at least Class I (if not younger) to at least Class
II (if not Class III) are located within close proximity
($\sim$0.1 pc) to each other.  Additional multi-wavelength
follow-up studies of these groupings of stars in Perseus (and
elsewhere) will shed light on the issues raised by this result.

\thispagestyle{empty}

\begin{deluxetable}{lllllll}
\rotate
\tabletypesize{\scriptsize}
\tablecaption{MIPS photometry for objects discussed in section 
\ref{sec:indobj}}
\label{tab:objects}
\tablewidth{0pt}
\tablehead{
\colhead{SSTc2d name} & \colhead{MIPS-24 (mJy)\tablenotemark{a}} & \colhead{MIPS-70 (mJy)\tablenotemark{a}} & 
\colhead{MIPS-160 (mJy)\tablenotemark{a}} & \colhead{other names} & 
\colhead{notes} }
\startdata
\cutinhead{new aggregate near Per 6}
033015.1+302349& 1900.0 & 5499& 7170  &IRAS 03271+3013&\#1 in \S\ref{sec:per6agg}\\
033027.1+302829&  431.0 &  920& 1450  &IRAS 03273+3018, HH 369&\#2 in \S\ref{sec:per6agg}\\
033030.2+302708&   48.9 &\ldots&\ldots&&\#3 in \S\ref{sec:per6agg}\\
033032.7+302626&   31.1 &  320& 1960  &&\#4 in \S\ref{sec:per6agg}\\
033035.9+303024&  572.0 &  180&\ldots &IRAS 03275+3020,GSC 02342-00390, AX J0330.5+3030&\#5 in \S\ref{sec:per6agg}\\
033036.7+302735&    3.9 &\ldots&\ldots&&\#6 in \S\ref{sec:per6agg}\\
033036.9+303127&  322.0 &  130&\ldots &&\#7 in \S\ref{sec:per6agg}\\
033038.2+303211&   57.5 &  260&\ldots &&\#8 in \S\ref{sec:per6agg}\\
033044.0+303246&  699.0 &  750&\ldots &LkHa 326=HBC 14 =IRAS 03276+3022&\#9 in \S\ref{sec:per6agg}\\
033054.5+303146&    1.0 &\ldots&\ldots&&\#10 in \S\ref{sec:per6agg}\\
\cutinhead{IRAS 03388+3139 stellar aggregate}
 034155.0+314939&    1.0 &  \ldots &  \ldots& & \#1 in \S\ref{sec:iras1}\\
 034155.7+314811&  120.0 &  \ldots &  \ldots& & \#2 in \S\ref{sec:iras1}\\
 034157.4+314836&  264.0 &  \ldots &  \ldots& & \#3 in \S\ref{sec:iras1}\\
 034157.7+314800&  112.0 &  \ldots &  \ldots& & \#4 in \S\ref{sec:iras1}\\
 034158.5+314855&  345.0 &  \ldots &  \ldots& & \#5 in \S\ref{sec:iras1}\\
 034158.6+314821&   36.5 &  \ldots &  \ldots& & \#6 in \S\ref{sec:iras1}\\
 034201.0+314913&    4.8 &  \ldots &  \ldots& & \#7 in \S\ref{sec:iras1}\\
 034202.1+314801&  115.0 &  \ldots &  \ldots& & \#8 in \S\ref{sec:iras1}\\
 034204.3+314711&   22.2 &  \ldots &  \ldots& & \#9 in \S\ref{sec:iras1}\\
\cutinhead{IRAS 03281+3039 stellar aggregate}
  033110.7+304940&   62.0 & \ldots & \ldots& LZK 19& \#1 in \S\ref{sec:iras2}; mentioned in J06\\
  033114.7+304955&  229.0 &  220   & \ldots& & \#2 in \S\ref{sec:iras2}; mentioned in J06\\
  033118.3+304939&  292.0 &  230   & \ldots& LZK 20& \#3 in \S\ref{sec:iras2}; mentioned in J06\\
  033120.1+304917&   15.0 & \ldots & \ldots& & \#4 in \S\ref{sec:iras2}; mentioned in J06\\
\cutinhead{objects detected at all 3 MIPS bands}
  032509.4+304622& 2000.0 & 2600& 2640&LDN 1448 IRS 1           & discussed briefly in J06; see \S\ref{sec:all3}\\
  032637.4+301528&  400.0 & 3900& 7880&IRAS 03235+3004          & discussed in J06\\
  032743.2+301228&  761.0 & 8190&17900&LDN 1455 IRS 4           & discussed in J06\\
  032747.6+301204& 1620.0 & 1900& 8130&IRAS 03247+3001, L1455 IRS 2& discussed in J06; see \S\ref{sec:all3}\\
  032834.5+310051& 1290.0 & 2700& 2660&IRAS 03254+3050          & see \S\ref{sec:all3}\\
  032845.3+310541&  215.0 & 1600& 2750&HH 340                   & see \S\ref{sec:all3}\\
  032951.8+313905&   49.3 & 2200& 4990&IRAS 03267+3128          & see \S\ref{sec:all3}\\
  033015.1+302349& 1900.0 & 5499& 7170&IRAS 03271+3013          & see \#2, \S\ref{sec:per6agg}\\
  033027.1+302829&  431.0 &  920& 1450&IRAS 03273+3018, HH 369  & see \#5, \S\ref{sec:per6agg}\\
  033032.7+302626&   31.1 &  320& 1960&\ldots                   & see \#7, \S\ref{sec:per6agg}\\
  033121.0+304529&   14.6 & 3800& 9660&IRAS 03282+3035          & discussed in J06\\
  033218.0+304946&   13.1 & 1800&10700&IRAS 03292+3039          & discussed in J06\\
  033925.5+321707&  234.0 & 3400& 8170&2MASS 03392549+3217070, IRAS 03363+3207   & see \S\ref{sec:all3}\\
  034548.2+322411& 2250.0 & 5265& 3350&LkHa 330                 & discussed in Brown (2007)\\
  034611.0+330848&   19.5 &  330& 1420&2MASS 03461106+3308488   &galaxy\\
  034705.4+324308&  587.0 & 1000& 1880&IRAS 03439+3233, B5 IRS 3 & discussed in J06; see \S\ref{sec:all3}\\
\cutinhead{Other individual point sources}
  032548.9+305725&  197.0 &  290.0 & \ldots& LZK 5 & \S\ref{sec:bd+30d540}\\
  032533.2+305544&   64.8 &   86.0 & \ldots& LZK 2 & \S\ref{sec:bd+30d540}\\
  032552.8+305449&   64.5 &  \ldots & \ldots& \ldots & \S\ref{sec:bd+30d540}\\
  032548.1+305537& $\sim$100 & 1200.0 &  \ldots & BD+30$\arcdeg$540, SAO 56444, IRAS 03227+3045\\
  033955.6+315533&   62.5 & \ldots & \ldots & HD 278942, IRAS 03367+3145 & \S\ref{sec:hd278942}\\
\enddata
\tablenotetext{a}{Absolute uncertainties on the 24 \mum\ data are
estimated to be 10-15\%; statistical uncertainties are much less
than this.  Uncertainties on 70 and 160 \mum\ flux densities are
estimated to be 20\%.}
\end{deluxetable}


\subsection{Objects detected at all 3 MIPS bands}
\label{sec:all3}

\begin{figure*}[tbp]
\plotone{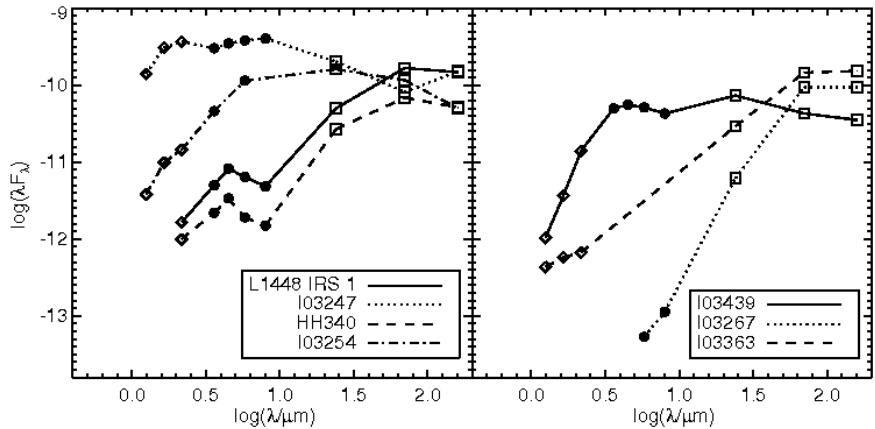}
\caption{SEDs for objects detected at all three MIPS bands and
not already discussed elsewhere.  Notation is as in
Fig.~\ref{fig:sedper6}. Discussion of each of these objects
appears in the text.}
\label{fig:seds-all3}
\end{figure*}

\begin{figure*}[tbp]
\plotone{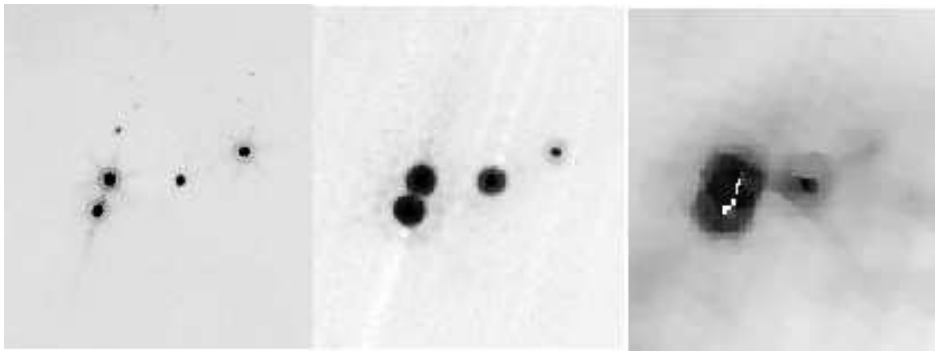}
\caption{The L1448 region in all 3 MIPS bands; 24, 70, and 160
\mum, from left to right.  See text for discussion.}
\label{fig:l1448}
\end{figure*}

\begin{figure*}[tbp]
\plotone{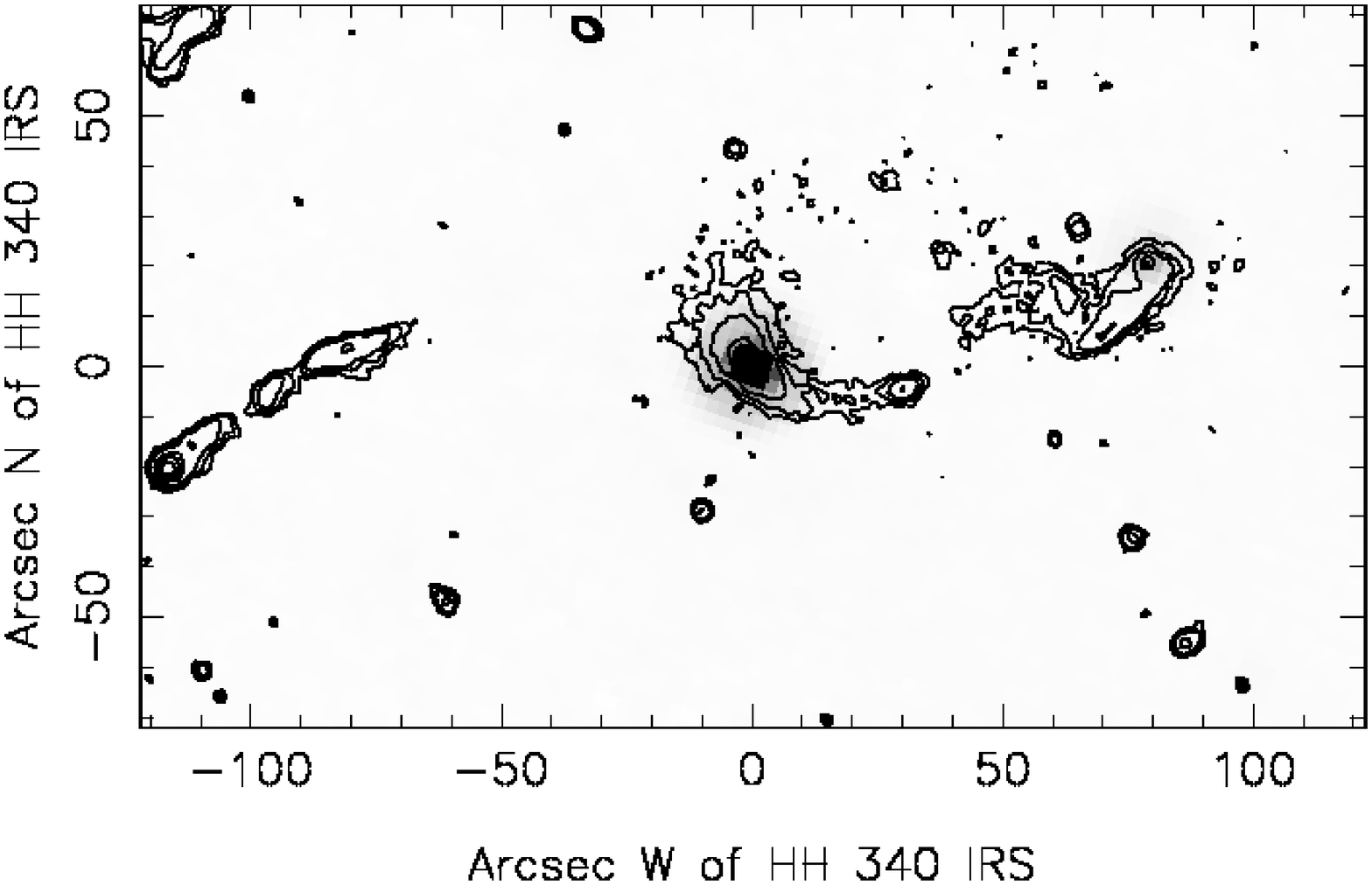}
\caption{Greyscale MIPS 24 \mum image of SSTc2d 032845.3+310541
(HH 340) overlaid by contours of 4.5 \mum\ emission.  The contour
spacing is a factor of two in surface brightness.  Scattered light
near the source is elongated at a different PA from the presumed line
emission at greater distances.}
\label{fig:hh340img}
\end{figure*}

While there are many objects detected at all four IRAC bands, or
even all four IRAC bands plus MIPS-24, there are few objects
detected at all three MIPS bands.  Objects detected at all three
MIPS bands are likely to be quite interesting because they are
likely to be the most embedded; moreover, if they are detected
at 70 and 160 \mum\ using our observing strategy, they are among
the brightest objects in the region, and therefore most likely
cloud members.  In this section, we discuss the 16 objects in
our catalog detected at all three MIPS bands.  Many of these
objects have already been discussed (and had SEDs presented) in
J06 or above in \S\ref{sec:per6agg}.  For completeness and ease
of reference, all of these objects appear in Table~6 under
``objects detected at all three MIPS bands'' even if they have
already appeared under the ``Per 6 aggregate'' heading or in
J06.  Because objects were detected in all 3 bands throughout
our large map, for lack of a better approach, we discuss these
objects in approximately RA-order (the same order as they appear
in Table~6), although we start with all of the objects discussed
by J06.  J06 considered the colors of deeply embedded sources,
and several Class I/II objects mentioned there did not have
SEDs, so they appear here.  The objects are portrayed by
position in Figure~\ref{fig:whereall3}.  We investigated the SED
and the morphology of each object individually in all our
available wavelength regimes (2MASS through 160 \mum).  

J06 has a discussion of the objects in the L1448 region,
including full SEDs of the embedded Class 0 objects (L1448-C and
L1448-N) and images of the outflows.  The complex morphology
here makes it occasionally difficult to extract the point
sources from their outflows.  L1448-C is clearly broken into two
pieces in the IRAC images.  L1448 IRS 1 is just off the edge of
the 4-channel IRAC map, and so was not included in detail.  The
full SED for L1448 IRS 1 (including the 2 IRAC bands which
observed it) is presented for the first time including Spitzer
points in Figure~\ref{fig:seds-all3}; indeed, it is a deeply
embedded object, as expected from earlier observations.  Images
of this region in all three MIPS bands are striking; see
Figure~\ref{fig:l1448}.  The southernmost bright 24 \mum\ source
in the image is L1448-C (L1448-mm).  Where IRAC sees two objects
here, only one of which is associated with the mm source (J06),
MIPS-24 clearly also sees two sources.  The object detected at
MIPS-70 also appears to be extended.  Moving north from this
object, we find L1448-N (L1448 IRS 3).  MIPS-24 also marginally
resolves L1448-N(A) from L1448-N(B); the ``B'' component is
substantially fainter than the ``A'' component, consistent with
results from Ciardi \etal\ (2003).   These objects are not
resolved at 70 \mum.  At 160 \mum, all of the IRS-3 components
are combined into one saturated blob.  Moving westward from
L1448 IRS 3, we encounter L1448 IRS 2 and then 1.  Both of these
objects appear point-like in MIPS-24 and -70.  There is a slight
extension in MIPS-160 towards IRS 3.  IRS 1 is faint at
MIPS-160, but detected.   Because IRS 2 and 3 are at least
slightly resolved in at least 1 band, they do not appear in
Table~6; IRS 1 is a point source in all three bands, and so it
does appear in the Table.  IRS 1 is one of the bluest two
objects in [24]$-$[70] colors in Fig.~\ref{fig:24_70}.

J06 also includes a discussion of objects from L1455, including
matching of outflows to driving sources.  J06 finds the three
previously-known YSOs (L1455 IRS 1,2, and 3), but also
identifies new infrared sources, including one named L1455 IRS
4, a candidate Class I object.  IRAS 03235+3004 is also
identified in this region as a candidate Class I object. In
IRAC, IRS 4 is undetected at the shorter two IRAC bands.  SEDs
appear for both of these objects in J06; both of them are
detected at all three MIPS bands and so appear in Table~6.  IRAS
03235+3004 is one of the reddest objects in Fig.~\ref{fig:k_k70}
with \ks$-$[70]$>$15 mag.  L1455 IRS 2 (IRAS 03247+3001) is also
seen at all three MIPS bands.  This less-embedded Class II did
not have an SED in J06, and so it appears in
Figure~\ref{fig:seds-all3} (abbreviated `I03247').  IRAS
03247+3001 is one of the bluest two objects in [24]$-$[70]
colors in Fig.~\ref{fig:24_70}.

The remaining objects detected at all three MIPS bands that are
discussed in J06 with SEDs are IRAS 03282+3035 and IRAS
03292+3039, both of which are outflow-driving sources on the B1
ridge.  The MIPS flux densities for these candidate Class 0
objects simply appear in  Table~6; for SEDs and
much more discussion, please see J06.  Both of these objects
have faint nebulosity around them in at least one of the IRAC
bands. For completeness, we note that IRAS 03282+3035 and IRAS
03292+3039 are two of the five reddest objects in
Fig.~\ref{fig:24_70} with [24]$-$[70]$>$7 mag.

Returning to our roughly RA-order discussion, the next object in
Table~6 is IRAS 03254+3050.  Another object
similar in morphology with detections in all three MIPS bands is
IRAS 03439+3233 (also known as B5 IRS 3).  Both of these objects
appear to be point sources at all available Spitzer bands, both
IRAC and MIPS.  SEDs for these objects, abbreviated ``I03254''
and I03439'' respectively, appear in
Figure~\ref{fig:seds-all3}.  These SEDs are consistent with
those for typical Class I sources, and indeed, both have already
been identified as such in, e.g., Ladd \etal\ (1993). 

Searching in the literature by position for our objects detected at
all three MIPS bands, we find one listed as HH 340 (Bally \etal\
1996). The SED for this object appears in Figure~\ref{fig:seds-all3};
a multi-wavelength image of HH 340 appears in
Figure~\ref{fig:hh340img}.  The SED for this object resembles that of
the object discussed in \S\ref{sec:per6agg} as being identified with
HH 369, with most of the energy being emitted at the longest
wavelengths.  The overall shape of the SED is also consistent with the
flux at the shortest wavelengths being scattered light (such as would
be found in edge-on disks).  Because we are not aware of jets or knots
of ISM alone possessing these properties, we suspected that this
object is not truly an HH object, but rather a protostar in its own
right, perhaps with shocked emission close to the central object. 
Indeed, others, including Hodapp \etal\ (2005), have already come to
this same conclusion, that HH 340 is actually a jet-driving source. 
Hodapp \etal\ identify the driving source with IRAS 03256+3055; they
find a bipolar jet of S-shaped morphology, very similar to that seen
in our IRAC images (Fig.~\ref{fig:hh340img}).  Young \etal\ (2003)
identify this object as a Class I.  This object, and HH 369, appear
point-like in MIPS bands (see Fig.~\ref{fig:hh340img}), but slightly
extended at IRAC bands\footnote{Note that flux densities in the IRAC
bands for this and other similar objects may not appear in the 2005
c2d catalog at IRAC wavelengths because the objects are extended; flux
densities from the 2005 c2d catalog were compared with those obtained
using extremely simple aperture photometry to approximate the flux
density, and the best apparent value is plotted here and for other
similar objects in this discussion.}, consistent with this
interpretation.  Moreover, this object is located in a region rich
with outflows and therefore young, embedded objects, and as such is it
reasonable to assume that it could also be a young, embedded object. 
This object is also one of the reddest dozen objects in
Fig.~\ref{fig:k_k24} with \ks$-$[24]$>$9.7 mags, and also one of the
reddest objects in Fig.~\ref{fig:k_k70} with \ks$-$[70]$>$15 mag. 

One object, IRAS 03267+3128 (abbreviated I03267 in
Figure~\ref{fig:seds-all3}), has no 2MASS counterpart at all and
is only marginally detected, if that, at IRAC-1 and IRAC-2; by
eye, it is difficult to see the object in the IRAC images.  An
SED for this object can be found in Figure~\ref{fig:seds-all3}. 
It resembles the SED found for IRAS 03282+3035, discussed as an
outflow-driving source in J06, and also that for one of the new
objects discussed above in \S\ref{sec:per6agg}.  All of these
SEDs steeply rise at the longer wavelengths; the two previously
identified in IRAS level off between 70 and 160 \mum. Because
all are faint between 3.5-8 \mum\ and undetected in 2MASS, all
three are evidently deeply embedded objects.  There are clearly
outflows in IRAC-2 associated with IRAS 03282+3035, as discussed
in J06, consistent with it being a young, embedded object; it is
listed as a Class 0 object in Froebrich (2005) (see also, \eg,
Barsony \etal\ 1998 and references therein).  The similarities
between the morphology at IRAC bands and the SED of this object
and IRAS 03267+3128 suggests that IRAS 03267+3128 may also be a
Class 0 object. 

Returning to our roughly RA-order discussion, the next three
objects in Table~6 are  IRAS 03271+3013, IRAS
03273+3018, and a new object (SSTc2d 033032.7+302626).  These
objects are presented as part of the new Per 6 aggregate
discussed above in \S\ref{sec:per6agg}.

The next object in the table, SSTc2d 033925.5+321707, with an
SED also portrayed in Figure~\ref{fig:seds-all3} (abbreviated
``I03363''), can be identified with 2MASS 03392549+3217070.  
The source IRAS 03363+3207 is $30\arcsec$ from this position,
with an error ellipse that overlaps the position of our MIPS
detection. We therefore conclude that IRAS 03363+3207 probably
should be identified  with our detection.  The shape of its SED
is similar to the other  young objects discussed here,
particularly IRAS 03235+3004, except that  it has a faint NIR
counterpart where IRAS 03235+3004 does not.  This object is is
located near the north edge of the large emission ring seen in
the extended emission, in a region outside the coverage of our
IRAC data, but also in a region of relatively low
nebulosity in MIPS-24.  It is one of the dozen reddest objects
in Fig.~\ref{fig:k_k24} with \ks$-$[24]$>$9.7 mag, and also one
of the reddest objects in Fig.~\ref{fig:k_k70} with
\ks$-$[70]$>$ 15 mag.  This source lies within a small dark
cloud about $1.5\arcmin\times2\arcmin$ in size, as seen on POSS
red plates. It appears to be a single YSO in a relatively
isolated cloud core.  Follow-up observations are warranted to
determine if it is in fact an isolated core. 

An object that is bright in all 7 Spitzer bands is LkHa 330;
this object is discussed further in Brown \etal\ (2007), but it
is also included in Table~6 for completeness.  It is the bluest
object in [70]$-$[160] colors in Fig.~\ref{fig:24_70}. 

We expected that none of the objects detected in 160 would be
extragalactic; the sensitivity of our 160 \mum\ survey is such
that it is really only likely to probe galactic objects, e.g.,
cluster members.  The last of the objects detected at all three
bands appears to be new (e.g., nothing appears in SIMBAD at that
position).  It is near the edge of the large MIPS map and has no
IRAC coverage; it is identified with 2MASS source
03461106+3308488. It is one of the reddest four sources in
[70]$-$[160] colors in Fig.~\ref{fig:24_70}.  By inspection of
the 24 \mum\ image, it is elliptical in MIPS-24, and appears by
eye to be a very marginal detection at MIPS-160.  Looking at
this position in the POSS and 2MASS reveals an extended object
that seems likely to be a galaxy.  We conclude that {\em most}
of the objects we see at 160 \mum, those that are bright and not
near the far reaches of the map, are likely to be cloud members.

\subsection{HD 281159 (BD+31$\arcdeg$643)}
\label{sec:bd+31}

\begin{figure*}[tbp]
\plotone{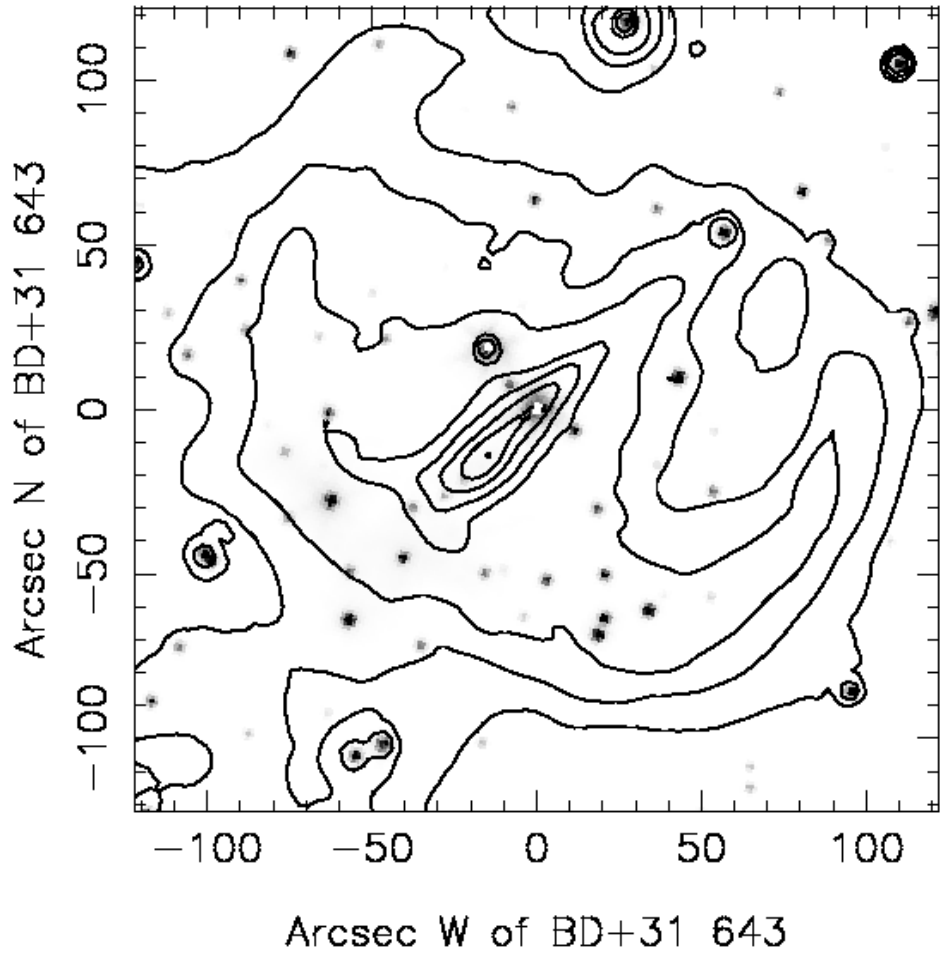}
\caption{Region of IC 348 centered on the debris disk candidate
BD+31$\arcdeg$643 (HD 281159).  The IRAC 3.6 $\mu$m image (see J06)
is shown in grayscale, overlaid by contours of MIPS-24 $\mu$m
emission.  The contour  intervals are 62, 74, 91, 116, 150, 198,
267, 363, and 500 MJy sr$^{-1}$.}
\label{fig:bd+31}
\end{figure*}

\begin{figure*}[tbp]
\plotone{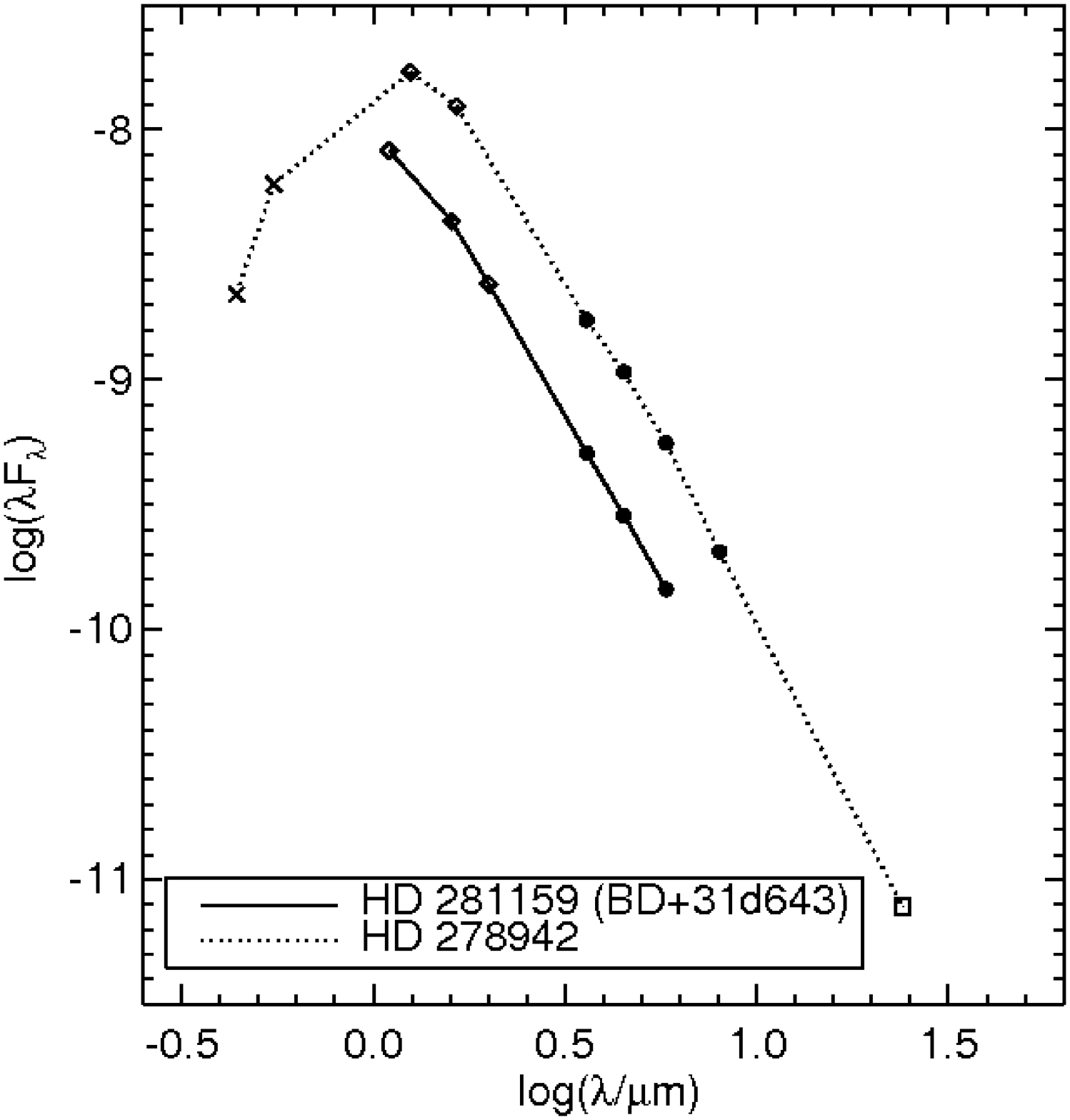}
\caption{SEDs for HD 281159 (BD+31$\arcdeg$643; solid line) and
HD 278942 (dashed line).  Notation is similar to earlier figures;
$\times$ are optical data, diamonds are 2MASS, dots are IRAC,
and square is MIPS-24.  Note that BD+31$\arcdeg$643 has an upper
limit at MIPS-24 that is comparable to the detection for HD
278942. Both stars have a long-wavelength SED completely
consistent with a Rayleigh-Jeans slope.}
\label{fig:ringstarsed}
\end{figure*}

HD 281159 (BD+31$\arcdeg$643) is a B5 star primarily responsible
for illuminating the IC 348 reflection nebula (Witt and Schild
1986), and is a 0.47$\arcsec$ binary (Alzner 1998).  Optical
coronagraphic images by Kalas and Jewitt (1997) detected a
bright linear feature within the IC 348 nebulosity, centered on
HD 281159 and with diameter 40$\arcsec$, and suggested this
might be a debris disk seen in scattered light. However, the
extent of the linear nebulosity (10,000 AU for $d$= 250 pc)  is
much larger than the diameters of known debris disks; the
nearest IRAS source, IRAS 03414+3200, is offset 28$\arcsec$ SE
of the stellar position; and the disk has not been confirmed in
subsequent Hubble Space Telescope imaging, perhaps because of
low surface brightness (Kalas, unpublished; HST GO program
7414).

Spitzer's view of the region surrounding HD 281159 is shown in
Figure~\ref{fig:bd+31}.  The region has a complex combination of
stellar point sources  and extended emission, with the stars
best seen in the 3.6 \mum\ IRAC image and the extended emission
being especially prominent at 24 \mum.  At lower surface
brightness levels, the extended  emission has the appearance of
an incomplete spherical shell centered on  HD 281159, $\sim
200\arcsec$ (50,000 AU) in diameter and open toward the
northwest. Within the shell, there is a narrow spike of very
bright 24 \mum\ emission extending from southeast to northwest
across the star.  This spike is the highest surface  brightness
feature in the entire Perseus cloud at 24 \mum, and follows the
position angle of the disk-like nebulosity found by Kalas and
Jewitt (1997). However, the surface brightness distribution
along this spike is highly asymmetric about the star.  Along the
spike, the surface brightness peak is found 22$\arcsec$
southeast of the star; at the same distance on the opposite side
of the star, the spike feature is a factor of $\sim$4 times
fainter. At low surface brightness levels, the spike appears to
project as far as 80$\arcsec$ northwest of the star, but on the
southeast side it terminates only 55$\arcsec$ away, near the
apparent shell wall.  The thickness of the spike is spatially
resolved at 24 \mum, with a FWHM 14$\arcsec$ (3500 AU).

Previous work had suggested an infrared excess in HD 281159. The
Spitzer SED (Fig.\ \ref{fig:ringstarsed}) is consistent with a bare
stellar photosphere out to 8 \mum.  There is no point source at 24
\mum, to an upper limit of 50 mJy determined by PSF planting (e.g.,
adding an artificial point source to the MIPS image at the stellar
position, and determining the smallest flux normalization at which it
remains clearly visible). The source is undetected at 70 \mum\ or in
the submillimeter continuum (Enoch \etal\ 2006).  Previous ISO-SWS
observations made in 1997 (Mer\'in 2004) showed a non-negligible
infrared excess starting at $\sim$4 \mum. To compare that observation
with the  Spitzer measurements, we convolved the SWS spectrum from the
ISO  Data Archive (TDT\#65201414, pipeline version 10.1) with the
Spitzer  IRAC and MIPS filter passbands.  The ISO flux densities are
consistent with  the Spitzer ones except at 8 and 24 \mum, where the
ISO ones are much  higher.  We attribute the discrepancy to the bright
extended emission around  the source at 8 and 24 \mum\
(Figure~\ref{fig:bd+31}), which should  have strongly contaminated the
ISO measurements in the large $14\times20\arcsec$ entrance aperture
of SWS.

The infrared properties of the region around HD 281159 are very
difficult to reconcile with a circumstellar debris disk.  The
star has no clear infrared excess.  For a central luminosity of
1600 \lsun,  24 \mum\ thermal emission from a disk should peak
at radii near 200 AU -- i.e., within 1$\arcsec$ of the star. 
While a spike-like feature  corresponding to the proposed disk
structure is present at 24 \mum,  its strong asymmetry is
inconsistent with a centrally heated circumstellar disk.  The
spike feature's length at 24 \mum\ is $>$ 3 times larger than
its reported optical size, exacerbating the size discrepancy
between it and known debris disks.  To the southeast, the resolved
thickness of the spike corresponds to an opening angle of
35$\arcdeg$ as seen from the star -- much larger than would be
expected for a disk close to edge-on.

What is the distribution of material in the region around HD
281159?  An extended disk appears very unlikely in the light of
the Spitzer results.  Nevertheless, the star is still likely to
be the illuminating source of  the surrounding region in both
scattered light and 24 \mum\ emission. HD 281159 is likely
another example of a Pleiades-like phenomenon, where an
interstellar cloud illuminated by a high luminosity star mimics
some properties of a circumstellar disk.  Kalas \etal\ (2002)
report six similar objects.  In this case, the star appears to
be clearing a cavity in surrounding molecular cloud material.
The spike feature would appear to be some kind of filamentary
interstellar cloud structure crossing behind  or in front of HD
281159, and passing closest to the star on its southeast-projected
side.  Filamentary structures are expected to form during 
magnetically-mediated collapse of interstellar clouds, and can
be seen at lower surface brightness levels in other parts of
Perseus (such as around HD 278942; see
Fig.~\ref{fig:ringstarimg}). In the case of HD 281159, such a
structure is located very near a luminous star, and
coincidentally projected across it.

\subsection{HD 278942}
\label{sec:hd278942}

\begin{figure*}[tbp]
\plotone{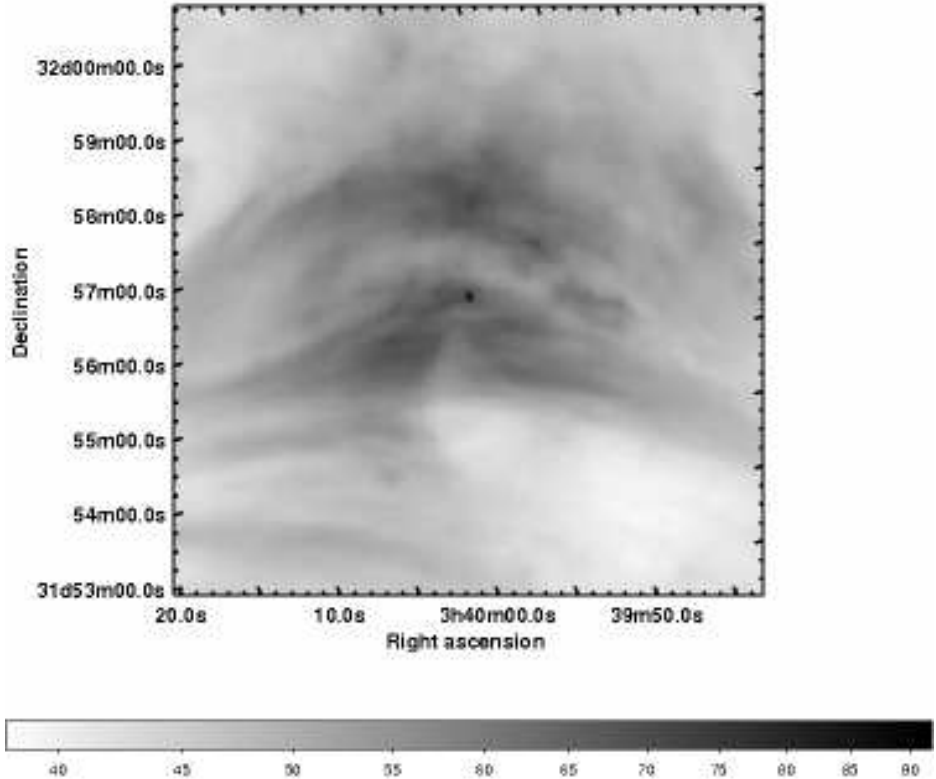}
\caption{Image of HD 278942 at 24 \mum, shown in linear stretch.  The
star appears as a relatively isolated point source, suggesting that
it is detached from the material it is illuminating.}
\label{fig:ringstarimg}
\end{figure*}



HD 278942 (IRAS 03367+3145) is a B0 V star about 1$\arcdeg$ West
of IC 348.  It lies at the center of a large, bright ring of
extended mid-infrared emission first identified by IRAS.  The
ring structure is clearly seen in our 24 \mum\ image mosaic, and
extends well outside the \av=2 extinction contour used to define
the c2d map region (see
Fig.~\ref{fig:24mosaic}-\ref{fig:3color}, Fig.~\ref{fig:160ext},
and Sec.~\ref{sec:extemiss}).  Studies of the ring by Andersson
\etal\ (2000) and Ridge \etal\ (2006) suggest that it may
represent a weak H II region excited by HD 278942, and located
just behind the Perseus molecular clouds.  

The 24 \mum\ image of the region (see
Figure~\ref{fig:ringstarimg}) shows that HD 278942 lies at the
center of a bright striated region  of extended emission more
than 0.5$\arcdeg$ in diameter.  The emission surrounding this
source appears extremely ``blue'' to MIPS (see
Fig.~\ref{fig:3color}), meaning that relative to the rest of the
Perseus extended emission, this location is much brighter at 24
\mum\ than at 70 or 160 \mum. The star stands out as an isolated
point source against this mid-infrared  nebulosity; there is no
sharp peak in the nebula surface brightness  around the stellar
position.  This ``detached'' relationship between the star and
nebulosity indicates that extended warm dust is not distributed
within a filled circumstellar volume.  Instead, it appears that
the illuminated dust represents a cloud surface that is offset a
considerable distance from the star, consistent  with previous
arguments that the HD 278942 lies behind the main Perseus cloud
complex.

The bright extended emission surrounding HD 278942 makes it
difficult to isolate the stellar fluxes in lower resolution
mid-infrared datasets such as IRAS and MSX.  The IRAS data
suggested a strong 25 \mum\ excess, and led Andersson \etal\
(2000) to suggest that the star possessed a circumstellar debris
disk.  Spitzer flux density measurements, obtained using PSF
fitting photometry, are shown as a spectral energy distribution
in Figure~\ref{fig:ringstarsed}.  The star is undetected at 70
and 160 \mum, and shows a Rayleigh-Jeans spectrum between 3.5
and 24 \mum.  There is no evidence for an infrared excess from
circumstellar dust.

\subsection{BD+30$\arcdeg$540 and Environs}
\label{sec:bd+30d540}

\begin{figure*}[tbp]
\plotone{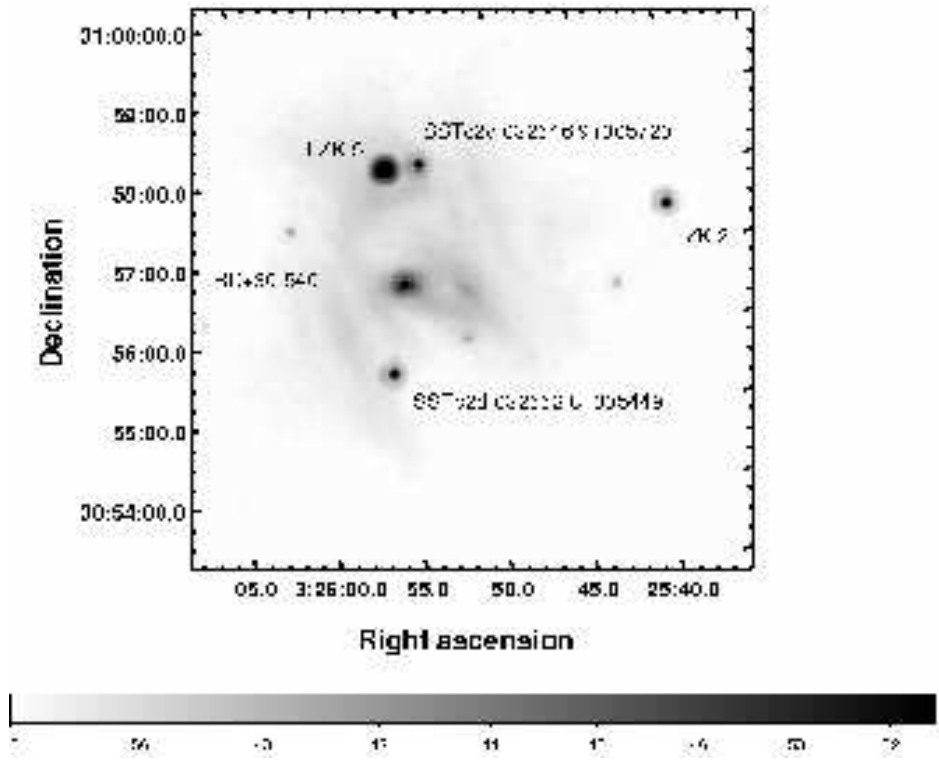}
\caption{Spitzer image of BD+30 540 at 24 microns, shown in log
stretch.  The source is nebulous, and illuminates an extended
emission nebula.  The surface brightness units are in MJy sr$^{-1}$.  
The vertical image axis corresponds to position angle 341.3$\arcdeg$.}
\label{fig:bd30d540}
\end{figure*}

BD+30$\arcdeg$540 (SAO 56444; IRAS 03227+3045) is a B8 V star which
illuminates the reflection nebula Van den Bergh 13, $10\arcmin$ North
of L1448.   This object is covered by our Spitzer MIPS scans, but
falls outside the region of IRAC coverage (see Fig.~\ref{fig:where}). 
As can be seen in Figure~\ref{fig:bd30d540}, the 24 \mum\ image of the
region shows that the region around the star is nebulous at this
wavelength and immersed in patchy mid-infrared nebulosity extending to
distances of  2$\arcmin$ (30,000 AU at 250 pc).  Several other point
sources are present in the field within $\sim5\arcmin$ that have
infrared excess\footnote{We note for completeness that we do not
consider this grouping of stars an aggregate of the sort discussed
above, because it covers a larger area and has larger typical
nearest-neighbor distances.}: the emission line stars LZK 5 (SSTc2d
032548.9+305725, 1.5$\arcmin$ NNW of BD+30) and LZK 2 (SSTc2d
032533.2+305544, 3.6$\arcmin$ West of BD+30); SSTc2d 032546.9+305720,
28$\arcsec$ West of LZK 5;  and SSTc2d 032552.8+305449, about
1$\arcmin$ southeast of BD+30$\arcdeg$540. BD+30$\arcdeg$540, LZK 2,
and LZK 5 are also detected at 70 \mum. At 24 \mum, the star itself is
situated on the West side of a compact, elliptical patch of nebulosity
that is elongated at PA 80$\arcdeg$.  Its 24 \mum\ flux density is
$\sim$0.1 Jy, consistent with the IRAS 25 \mum\ upper limit.  Unlike
the four nearby stars with infrared excess,  which show Class II
spectral energy distributions, BD+30$\arcdeg$540 is a Class I source
-- one of the reddest objects in Fig.~\ref{fig:k_k70} with
\ks$-$[70]$>$15 mag.  The 70 \mum\ point source has a flux density of
1.2 Jy, more than a factor of five smaller than the IRAS 60 \mum\ flux
density.  The rest of the 70 \mum\ flux density is distributed in
bright nebulosity extending 1.6$\arcmin$ to the East of the star at PA
80$\arcdeg$; when this is accounted for, the Spitzer 70 \mum\ and IRAS
60 \mum\ flux densities come into rough agreement.  Unfortunately the
70 \mum\ nebulosity near BD+30$\arcdeg$540 is crossed by a strong scan
artifact, which  prevents any detailed interpretation of its
structure.


\subsection{HH 211 Outflow and Source}
\label{sec:hh211}

\begin{figure*}[tbp]
\plotone{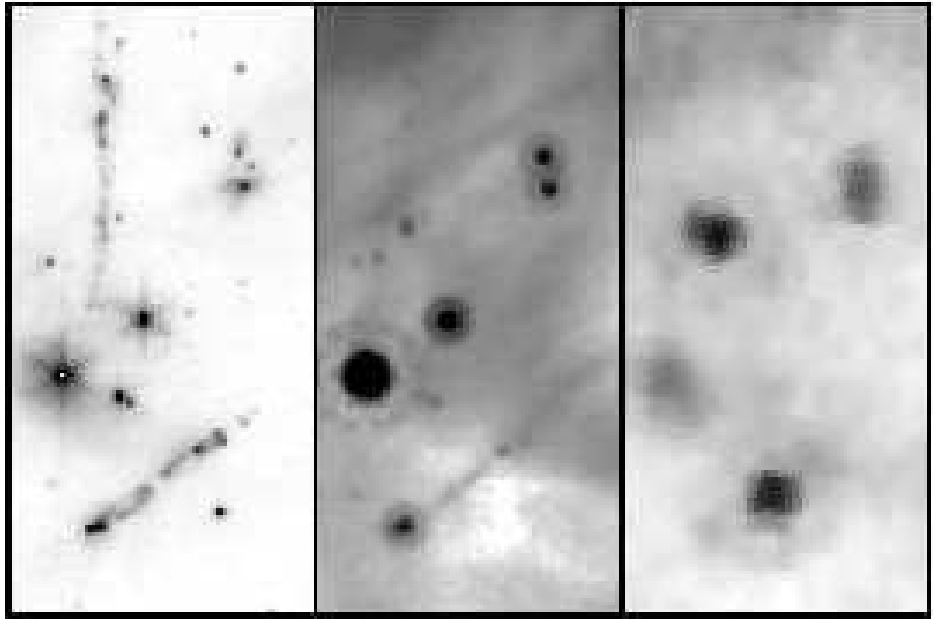}
\caption{Spitzer images of the HH 211 field at 4.5, 24, and 70
\mum, left to right.  This uses a logarithmic stretch for the
4.5 \mum\ image and a linear stretch for the 24 and 70 \mum. 
The two outflow driving sources are extremely red.  The HH 211
jet is detected in the broadband 24 \mum\ image, and its two
bowshocks are seen at both 24 and 70 \mum.}
\label{fig:hh211bandw}
\end{figure*}

\begin{figure*}[tbp]
\plotone{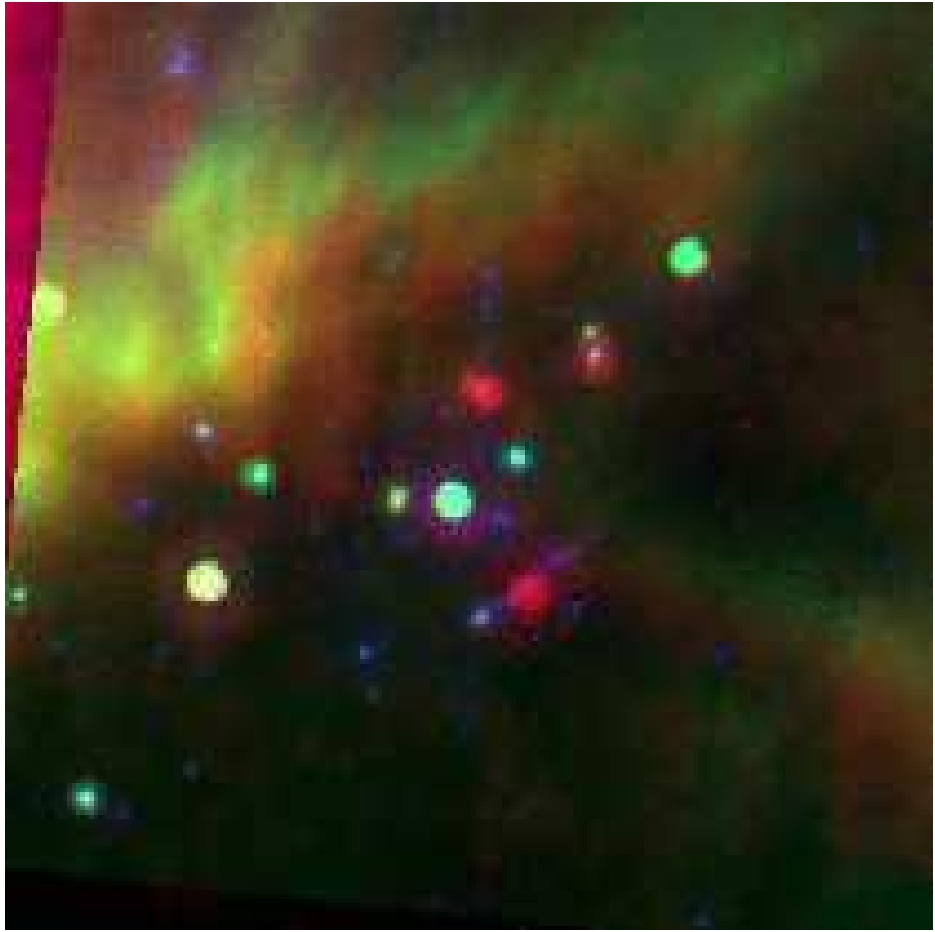}
\caption{Spitzer 3-color image of the HH 211 field at 4.5
(blue), 24 (green), and 70 (red) \mum.  This uses a logarithmic
stretch for the 4.5 \mum\ image and a linear stretch for the 24
and 70 \mum. }
\label{fig:hh211color}
\end{figure*}

\begin{figure*}[tbp]
\plotone{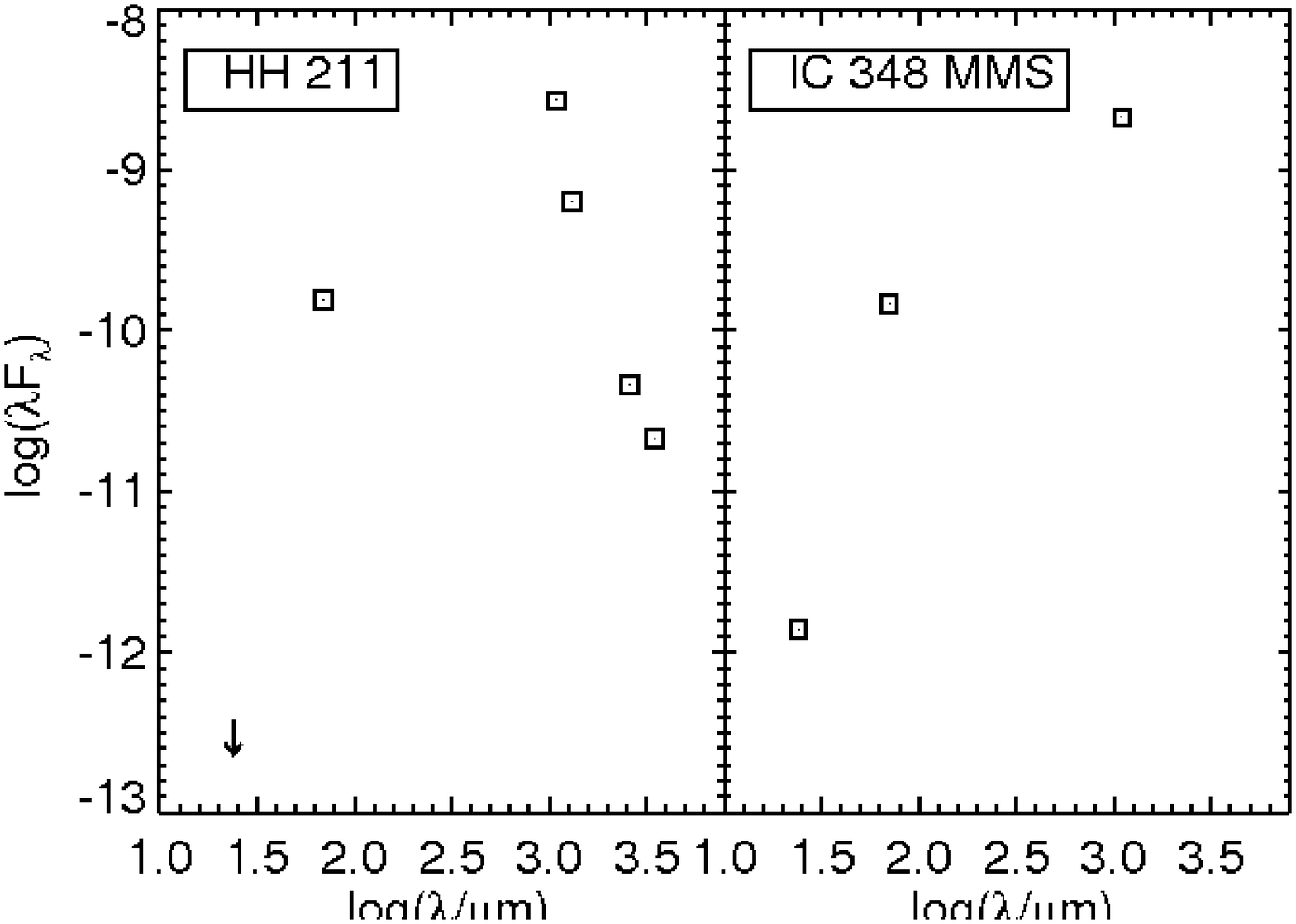}
\caption{SED for HH 211 and IC 348 MMS.  Note that most of the
points are longer than Spitzer wavelengths; only the two
shortest-wavelength points in HH 211 and the three shortest in IC
348 MMS are Spitzer points.}
\label{fig:hh211sed}
\end{figure*}

Herbig-Haro 211 is a bipolar outflow on the southwest side of IC
348, originally discovered in near-infrared H$_2$ emission by
McCaughrean \etal\ (1994).  Gueth and Guilloteau (1999) mapped
the corresponding CO outflow at high resolution, finding it to
be highly collimated. They also detected the central outflow
source in the millimeter continuum, and derived a circumstellar
mass of 0.2 M$_{\odot}$. Near-infrared spectroscopy of the flow
is reported by O'Connell \etal\ (2005).  A wide-field H$_2$
imaging survey of the region by Eisl\"offel \etal\ (2003)
identified a new North-South oriented outflow a few arcmin from
HH 211, and suggested that a second bright mm continuum source
along its axis (``IC 348 MMS'') was the driving source.  Both of
these outflow sources are detected in the centimeter continuum
(Avila \etal\ 2001). Neither has been detected in the
near-infrared, and source confusion in the complex IC 348 field
prevented their detection by IRAS. MIPS observations, by virtue
of their improved sensitivity  and spatial resolution over
previous far-infrared studies, offer  a new opportunity to
characterize these two Class 0 outflow sources.

Spitzer images of the HH 211 field at 4.5, 24, and 70 \mum\ are
shown in Figure~\ref{fig:hh211bandw}, and a wider field of view
3-color image is shown in Figure~\ref{fig:hh211color}.  The
outflows appear prominently in the IRAC 4.5 \mum\ band, which
includes  the S(10) and S(9) pure rotational transitions of
H$_2$.  The mid-infrared morphology of both flows appears very
similar to their near-infrared vibrational emission (Eisl\"offel
\etal\ 2003).  Neither driving source is detected at any of the
IRAC bands, unlike the Class 0 objects discussed in J06. At 24
\mum, the HH 211 source is still undetected, but several flow
features are visible: the narrow bipolar jets; a central gap in
the narrow jet at the position of the outflow source; bright
emission from the outermost southeast bowshock (knot j, in the
nomenclature of McCaughrean \etal\ 1994); and fainter emission
from the northwest bowshocks (knots b/c, and e/d).  In the other
North-South outflow, the driving source is just detected with a
flux density of 11 mJy, and faint indications of the North
outflow emission are seen.  At  70 \mum, both driving sources
become very bright (F${\nu}\approx$3.5 Jy).  Both are slightly
extended: Gaussian fitting, followed by subtraction in 
quadrature of the nominal PSF FWHM, finds that the HH 211 source
is elongated by 11$\arcsec$ (2750 AU) at PA 52$\arcdeg$, and
that IC 348 MMS is elongated by 13$\arcsec$  (3250 AU) at PA
125$\arcdeg$.  Significantly,  both the southeast and northwest
bowshocks in HH 211 are detected at 70 \mum\ -- the first time
individual HH knots have been clearly detected in MIPS 70 \mum\
images.  Neither of the two outflow sources can be identified at
160 \mum, due to saturation and confusion of bright extended
emission  in the region. Spectral energy distributions for both
sources are shown  in Figure~\ref{fig:hh211sed}.

What is the nature of the far-IR emission from the HH 211
bowshocks?   Spectroscopic studies found that continuum
emission was negligible near 24 \mum\ in HH 47 (Noriega-Crespo
\etal\  2004), HH 7 (Noriega-Crespo \etal\  2000; Neufeld \etal\
2006), and HH 54 (Neufeld \etal\ 2006).  An emission line such as
[\ion{S}{1}] 25.2 \mum\ might account for the visibility of the
flow features in the MIPS 24 \mum\ filter.  If so, the line flux
corresponding to the 40 mJy flux density measured for the HH 211
SE bowshock would be $1.3\times10^{-15}$ W m$^{-2}$, which
equates to a luminosity of $2\times10^{-4}$ $L_{\odot}$ at 250
pc distance.  The 70 \mum\ flux  densities of the southeast and
northwest bowshocks, measured after subtracting the PSF of 
central source, are 80 and 20 mJy respectively.  The
[\ion{O}{1}] line at 63.18 \mum\ is known to be bright in HH
flows (Moro-Martin et al. 2001), and the MIPS 70 \mum filter has
70\% transmission at this wavelegnth.  If all the 70 \mum\ emission
from the southeast bowshock originated in this transition,  the
implied line flux is 8.9$\times10^{-16}$ W m$^{-2}$ (about 10\%
the value measured in the Cep E bowshocks by Moro-Martin \etal\
2001), and the values inferred for the line luminosity would
again be $\sim2\times10^{-4}$  $L_{\odot}$.  In this view, the
total power emitted in the [\ion{S}{1}] and [\ion{O}{1}]  lines
would be comparable to that in 1-0 S(1) line of H$_2$ at 2.12
\mum\  (McCaughrean \etal\ 1994).  A more speculative
possibility is that we have detected dust continuum emission
from the bowshock region.  In this view, the warm dust would be
located in shocked ambient material, and in the pre-shock medium
heated by a radiative precursor.  The color temperature implied
by the measured flux densities is 105 K; a blackbody at this
temperature would require an emitting area of $\sim 10^{24}$
m$^{2}$, which is comfortably smaller than the MIPS 24 \mum\
beam.  The source luminosity would be comparable to the emission
line case.  Spectroscopic observations will be needed to
distinguish between these two possibilities in the specific case
of HH 211.  The high implied luminosity of the HH 211 bowshocks,
the absence of outer flow features beyond the known bowshocks
in the wide-field IRAC maps, and the extremely red driving
sources all point toward HH 211 being an extremely young
outflow.

The Spitzer 70 \mum\ flux densities of the two outflow driving
sources are consistent with previously reported ISO 60 \mum\
measurements  (Froebrich \etal\ 2003).  The new 24 \mum\
measurements provide a strong new constraint on the bolometric
temperatures of both objects.   Both are among the reddest
objects in the entire Perseus cloud: with [24]$-$[70] color $>$7
mag. A 30 K blackbody is consistent with the observations of IC
348 MMS,  and corresponds to an upper limit temperature for HH
211.  The low temperatures of the two sources arise from both
their dense envelopes and the fact that both are viewed from
near the equator plane of their developing circumstellar disks. 
At 70 \mum, the HH 211 source is elongated along the same
position angle as in the 1.1 mm continuum (Enoch \etal\ 2006),
60$\arcdeg$ from the flow axis and 30$\arcdeg$ misaligned from
the major axis of the circumstellar disk (Palau \etal\ 2006).

\section{Conclusions}
\label{sec:concl}

We have presented here the MIPS observations of $\sim$10.5
square degrees of the Perseus molecular cloud at 24, 70, and 160
\mum.  In this paper, we have selected just a few items to
highlight from this incredibly rich data set.  

Bright extended emission is present at all three wavelengths
throughout the MIPS Perseus mosaics.  The majority of the
extended structures we see have direct counterparts in the IRAS
maps.  At 24 and 70 \mum, there is more extended emission on the
East side of the complex, where several B stars act as
illuminating sources.  The distribution of 160 \mum emission closely 
follows the contours of \av\ extinction on the West side of the
complex.  The large 1.4$\arcdeg$ diameter IRAS ring illuminated by
HD 278942 does not follow the \av\ contours.  The structure
of the nebulosity immediately surrounding HD 278942 is consistent
with a cloud wall illuminated by a star immediately behind it,
as suggested for the IRAS ring by Ridge et al. (2006).

The source counts found at MIPS 24 and 70 are a combination of
background galaxies, background and foreground stars, and true
cluster members.  At 24 \mum, in the IC 348 and NGC 1333
clusters, there is a clear excess of cloud member sources for
flux densities $>$2-3 mJy.  Across the much larger ``rest of the
cloud'' region, an excess of on-cloud sources is seen for flux
densities $>$10 mJy.  At 70 \mum, there is a clear excess above
SWIRE of sources obtained at all flux densities $>$200 mJy in
the ``rest of the cloud'' and in IC 348; in NGC 1333, an excess
in number counts is found for objects brighter than about 65
mJy.  The brightest measurable objects in Perseus are found in
NGC 1333, where there is a clear excess of these relative to
other parts of Perseus.

As a result of observational constraints, the MIPS coverage is
about 3 times that of the IRAC coverage.  In the region that is
covered by IRAC, 92\% of the MIPS-24 objects have an IRAC match
at some band. The relatively shallow NIR \jhk\ data from 2MASS
provides some additional data, but only about 30\% of the
MIPS-24 sources have \ks-band counterparts.  Nonetheless, a
third of those have MIR excesses, making them potentially cloud
members.  Many interesting objects throughout this complex 
have well-defined SEDs.

By making color-magnitude plots using combinations of \ks, 24,
and 70 \mum, we can make an attempt at separating likely
YSOs from likely galaxies such as those found in SWIRE.   Many
candidate young objects are found throughout the cloud, not just
in the well-known clusters; in sheer numbers, there are as many
young object candidates outside the clusters as in the
clusters.  In contrast to J06, we find a much lower
frequency of Class I and flat spectrum sources outside the
clusters.  The difference can be traced to the selection
criteria; by requiring \ks\ detections, we are implicitly
limiting our selection to the brighter objects, and J06 includes
fainter objects.  The MIPS data show that the abundance of {\it
bright} Class I and flat spectrum sources is not higher in the
rest of the cloud versus the NGC 1333 and IC 348 clusters. We
find that NGC 1333 is on average younger than IC 348, consistent
with expectations.  

We find objects that are redder in \ks$-$[24], \ks$-$[70], and
[24]$-$[70] than any extragalactic source found in SWIRE at
these flux levels; we consider these to be potentially the most
embedded young objects, and they are found both in the clusters
and in the rest of the cloud.  

There are tightly clustered clumps, arcs, and strings of
apparently young objects bright at 24 \mum\ located throughout
the Perseus cloud.  We have presented SEDs and images of the
components of a new aggregate called Per 6 and two objects that
IRAS thought were single point sources but that break into
pieces when viewed by MIPS.  Each of these groupings of stars
has a diversity of SEDs despite being closely co-located (within
$\sim$0.1 pc), similar to another aggregate found in the
Galactic First Look Survey in L1228 (Padgett \etal\ 2004).  This
suggests that either  (a) all of the objects started with
similar properties but at different times such that there is a
real age spread and their circumstellar environments today
reflect different stages in a process of uniform evolution, or
(b) all of the objects are roughly coeval and there is a real
diversity of evolutionary states and/or or initial conditions. 
If the former is the case, then some stars are $\lesssim0.1$ Myr
old and some are $\gtrsim10$ Myr old.  If the latter is the
case, the evolution of circumstellar matter must be such that
the timescale to evolve from Class 0 to Class I is comparable to
the shortest possible time to evolve from Class 0 to Class III,
and it is perhaps strongly influenced by conditions specific to
each star. 

There is no real variability of any of the 24 \mum\ sources in
our field on timescales of $\sim$6 hours to $\sim$10\%.  

We used the MIPS data to assess how well the IRAS survey did in
identifying point sources in a region where complex extended
emission is present, posing a significant source of confusion to
the large aperture IRAS measurements. Most of the 12 and 25
\mum\ highest-quality IRAS PSC point sources are either
retrieved or it is clear why a point source was reported for
that location.  There are many more PSC ``point sources''
reported at 60 and 100 \mum\ (than at 12 and 25 \mum); a much
lower fraction of these objects are recovered.   The IRAS FSC is
much more robust for studies of stellar sources in regions with
complex extended emission, and should be used in preference to
the IRAS PSC unless MIPS data are available.

The 16 objects detected at all three MIPS bands throughout the
cloud are expected to be among the most interesting cloud
members; they all have colors consistent with embedded young
objects, though one on the edge of the field is likely to be a
background galaxy (based on POSS observations).  We suggest that
at least one of these objects may be a new Class 0 object. 

Two debris disk candidates, HD 281159 (BD+31$\arcdeg$643 in IC 348) 
and HD 278942, are found not to have infrared excess; IRAS measurements
suggesting an excess were contaminated by bright extended emission.
Two jets in IC 348, HH211 and IC 348 MMS, are seen at 24 \mum;
bowshocks are detected at 24 and 70 \mum\ -- the first time
individual HH knots have been clearly detected in MIPS 70 \mum\
images.

A complete list of candidate young objects in Perseus, along with
discussions of clustering and evolutionary timescales, will appear
in S.-P.\ Lai, in preparation.

\acknowledgements 

We wish to thank D.\ Shupe for helpful conversations in regards
to differential source counts, and S.\ Strom, J.\ Muzerolle, and
J.\ Najita for helpful conversations regarding disk lifetimes.
We thank R.\ D.\ Blum for making a plot of \ks\ vs.\ks$-$[24] using the
LMC data from Blum \etal\ (2006) and not making us load his
extremely large catalog and do it ourselves.
We thank the anonymous referee for a thorough and thoughtful report. 
L.\ M.\ R.\ wishes to acknowledge funding from the Spitzer
Science Center to allow her to take ``science retreats'' to work
intensively on this paper.   Most of the support for this work,
part of the Spitzer Space Telescope Legacy Science Program, was
provided by NASA through contracts 1224608, 1230782, and
1230779, issued by the Jet Propulsion Laboratory, California
Institute of Technology under NASA contract 1407.  We thank the
Lorentz Center in Leiden for hosting several meetings that
contributed to this paper.  B.\ M.\ acknowledges the Fundaci\'on
Ram\'on Areces for financial support. Support for J.\ K.\ J.\
and P.\ C.\ M.\ was provided in part by a NASA Origins grant,
NAG5-13050.  Astrochemistry in Leiden is supported by a NWO
Spinoza grant and a NOVA grant.   K.\ E.\ Y.\ was supported by
NASA under Grant No.\ NGT5-50401 issued through the Office of
Space Science. This research has made use of NASA's Astrophysics
Data System (ADS) Abstract Service, and of the SIMBAD database,
operated at CDS, Strasbourg, France.  This research has made use
of data products from the Two Micron All-Sky Survey (2MASS),
which is a joint project of the University of Massachusetts and
the Infrared Processing and Analysis Center, funded by the
National Aeronautics and Space Administration and the National
Science Foundation.  These data were served by the NASA/IPAC
Infrared Science Archive, which is operated by the Jet
Propulsion Laboratory, California Institute of Technology, under
contract with the National Aeronautics and Space
Administration.  The research described in this paper was
partially carried out at the Jet Propulsion Laboratory,
California Institute of Technology, under contract with the
National Aeronautics and Space Administration.

\appendix
\section{IRAS sources not recovered by MIPS}

As discussed in Section~\ref{sec:iras}, we compared the MIPS 24
and 70 \mum\ images and source catalogs with the IRAS Point
Source  (PSC) and Faint Source (FSC) catalogs.  A significant
number of IRAS detections in Perseus are not recovered by
Spitzer/MIPS.  The following tables were constructed by
overlaying the PSC and FSC source positions on the MIPS 24 and
70 \mum\ images.  If an object (or more than one object) 
appeared at the IRAS catalog position, it was marked as having
been recovered, even if the MIPS detection was saturated (which
happened frequently).  If an IRAS object was located in a region
of bright extended emission in the MIPS image with no
immediately obvious point source(s), it was noted that IRAS was
likely to have been confused by nebulosity.  If no MIPS point
source was readily discernible at the IRAS catalog position, it
was marked as missing from the MIPS data.   All unconfused IRAS
25 \mum\ sources should be recovered in the MIPS 24 \mum\ band,
given its greater sensitivity.  Similarly, all unconfused IRAS
60 \mum\ sources should be recovered in the MIPS 70 \mum\
band.   IRAS sources that were listed in the IRAS catalogs (PSC
and FSC) as clear detections (data quality flag of 3) but that
are missing in one or both MIPS bands or confused by nebulosity
are given in the following Tables.  

Comparing to Table~3, the objects included in the Tables below are (a)
the 13 PSC objects listed as clear detections at 25 \mum\ but that
MIPS-24 reveals are likely to have been simply confused by the
nebulosity; (b) the 8 PSC objects listed as clear detections at 60
\mum\ but missing entirely from the 70 \mum\ map; (c) the 36 PSC
objects listed as clear detections at 60 \mum\ but that MIPS-70
reveals as clumps of nebulosity; (d) the one FSC object listed as a
clear detection at 25 \mum\ but that MIPS-24 does not recover at all;
and (e) the 8 objects listed as clear detections at 25 \mum\ but that
MIPS-24 resolves into clumps of nebulosity.  No PSC objects are listed
as clear IRAS detections at 25 \mum\ but are completely missing at 24
\mum, and no FSC objects are listed as clear IRAS detections at 60
\mum.  For completeness, we note that one PSC object clearly detected
at 12 \mum\ was not recovered at 24 \mum\ (or at 70 \mum); it is
03324+3020.  A source is present at 24 \mum, but offset $\sim1\arcmin$
West of the IRAS position.  It is reasonably likely that these sources
should be matched, but as the positional offset was larger than
expected, we have listed it as missing.

\begin{deluxetable}{lll}
\tablecaption{IRAS PSC detections not recovered by Spitzer/MIPS}
\tablewidth{0pt}
\tablehead{
\colhead{PSC name} &  \colhead{\ldots from 24 \mum} &
\colhead{\ldots from 70 \mum}}
\startdata
03265+3014&missing (listed as upper limit at 12,25) & missing
(listed as detection at 60)\\ 
03292+3124&confused by nebulosity&missing (listed as detection at 60)\\ 
03303+3108&confused by nebulosity&confused by nebulosity\\ 
03310+3026&confused by nebulosity?&missing (listed as detection at 60)\\ 
03313+3117&confused by nebulosity?&confused by nebulosity\\ 
03319+3044&confused by nebulosity?&confused by nebulosity\\ 
03323+3049&confused by nebulosity?&confused by nebulosity?\\ 
03326+3055&confused by nebulosity?&confused by nebulosity?\\ 
03334+3042&confused by nebulosity&confused by nebulosity\\ 
03338+3123&missing (listed as upper limit at 12,25)&confused by nebulosity\\ 
03346+3116&confused by nebulosity&confused by nebulosity\\ 
03349+3117&off edge&confused by nebulosity\\ 
03354+3114&confused by nebulosity&confused by nebulosity\\ 
03356+3121&confused by nebulosity&confused by nebulosity\\ 
03358+3138&confused by nebulosity&confused by nebulosity\\ 
03363+3200&confused by nebulosity&missing (listed as detection at 60)\\ 
03366+3105&confused by nebulosity&missing (listed as detection at 60)\\ 
03367+3145&confused by nebulosity&confused by nebulosity\\ 
03367+3147&confused by nebulosity&confused by nebulosity\\ 
03367+3147&confused by nebulosity&confused by nebulosity\\ 
03370+3155&confused by nebulosity&confused by nebulosity\\ 
03371+3103&confused by nebulosity&confused by nebulosity\\ 
03371+3135&confused by nebulosity&confused by nebulosity\\ 
03372+3107&confused by nebulosity&confused by nebulosity\\ 
03372+3112&confused by nebulosity&confused by nebulosity\\ 
03374+3056&confused by nebulosity&missing (listed as detection at 60)\\ 
03380+3114&confused by nebulosity&confused by nebulosity\\ 
03380+3140&confused by nebulosity&confused by nebulosity\\ 
03380+3143&confused by nebulosity&confused by nebulosity\\ 
03382+3145&confused by nebulosity&confused by nebulosity\\ 
03385+3109&confused by nebulosity&confused by nebulosity\\ 
03385+3149&confused by nebulosity&confused by nebulosity\\ 
03386+3206&confused by nebulosity&confused by nebulosity\\ 
03391+3223&confused by nebulosity&confused by nebulosity\\ 
03391+3227&missing (listed as upper limit at 12,25)&confused by nebulosity\\ 
03404+3156&confused by nebulosity&confused by nebulosity\\ 
03406+3144&confused by nebulosity&confused by nebulosity\\ 
03410+3204&confused by nebulosity&confused by nebulosity\\ 
03411+3155&confused by nebulosity&confused by nebulosity\\ 
03411+3235&confused by nebulosity?&confused by nebulosity\\ 
03414+3200&confused by nebulosity&confused by nebulosity\\ 
03415+3121&missing (listed as upper limit at 12,25)&confused by nebulosity\\ 
03417+3159&confused by nebulosity&confused by nebulosity\\ 
03417+3207&confused by nebulosity&confused by nebulosity\\ 
03418+3242&confused by nebulosity?&missing (listed as detection at 60)\\ 
03424+3234&missing (listed as upper limit at 12,25)&confused by nebulosity\\ 
03427+3206&confused by nebulosity?&confused by nebulosity\\ 
03429+3237&confused by nebulosity?&confused by nebulosity\\ 
03434+3235&confused by nebulosity?&confused by nebulosity\\ 
03437+3219&confused by nebulosity&confused by nebulosity\\ 
03439+3131&off edge&confused by nebulosity\\ 
03448+3302&confused by nebulosity&missing (listed as detection at 60)\\ 
03449+3240&confused by nebulosity&confused by nebulosity\\ 
03450+3223&broken into pieces&confused by nebulosity\\ 
03452+3245&missing (listed as upper limit at 12,25)&confused by nebulosity\\ 
03454+3230&missing (listed as upper limit at 12,25)&confused by nebulosity\\ 
03455+3242&confused by nebulosity?&confused by nebulosity\\ 
\enddata
\end{deluxetable}

\begin{deluxetable}{lll}
\tablecaption{IRAS FSC detections not recovered by Spitzer/MIPS}
\tabletypesize{\footnotesize}
\tablewidth{0pt}
\tablehead{
\colhead{FSC name} &  \colhead{\ldots from 24 \mum} &
\colhead{\ldots from 70 \mum}}
\startdata
F03345+3116&confused by nebulosity&off edge\\ 
F03368+3147&confused by nebulosity&confused by nebulosity\\ 
F03399+3134&confused by nebulosity&confused by nebulosity\\ 
F03414+3200&confused by nebulosity&confused by nebulosity\\ 
F03415+3210&confused by nebulosity&confused by nebulosity\\ 
F03416+3158&confused by nebulosity&confused by nebulosity\\ 
F03416+3206&confused by nebulosity&confused by nebulosity\\ 
F03419+3209&confused by nebulosity&confused by nebulosity\\ 
F03226+3059&missing (listed as upper limit at 12, solid detection
at 25)&missing (listed as weak detection at 60)\\ 
\enddata
\end{deluxetable}

\end{document}